\documentclass[11pt]{article}
\usepackage{graphicx}
\usepackage{longtable,lscape}
\usepackage{amsfonts}
\newcommand{\compl}{{\mathbb C}}
\newcommand{\real}{{\mathbb R}}
\newcommand{\captionfonts}{\small}
\makeatletter  
\long\def\@makecaption#1#2{%
  \vskip\abovecaptionskip
  \sbox\@tempboxa{{\captionfonts #1: #2}}%
  \ifdim \wd\@tempboxa >\hsize
    {\captionfonts #1: #2\par}
  \else
    \hbox to\hsize{\hfil\box\@tempboxa\hfil}%
  \fi
  \vskip\belowcaptionskip}
\makeatother   
\begin{document}

\title{Quantum Structure in Cognition}
\author{Diederik Aerts\\
        \normalsize\itshape
        Center Leo Apostel for Interdisciplinary Studies \\
         \normalsize\itshape
         Department of Mathematics and Department of Psychology\\
        \normalsize\itshape
        Vrije Universiteit Brussel, 1160 Brussels, 
       Belgium \\
        \normalsize
        E-Mail: \textsf{diraerts@vub.ac.be}
		}
\date{}
\maketitle

\begin{abstract}
\noindent
The broader scope of our investigations is the search for the way in which concepts and their combinations carry and influence meaning and what this implies for human thought. More specifically, we examine the use of the mathematical formalism of quantum mechanics as a modeling instrument and propose a general mathematical modeling scheme for the combinations of concepts. We point out that quantum mechanical principles, such as superposition and interference, are at the origin of specific effects in cognition related to concept combinations, such as the guppy effect and the overextension and underextension of membership weights of items. We work out a concrete quantum mechanical model for a large set of experimental data of membership weights with overextension and underextension of items with respect to the conjunction and disjunction of pairs of concepts, and show that no classical model is possible for these data. We put forward an explanation by linking the presence of quantum aspects that model concept combinations to the basic process of concept formation. We investigate the implications of our quantum modeling scheme for the structure of human thought, and show the presence of a two-layer structure consisting of a classical logical layer and a quantum conceptual layer. We consider connections between our findings and phenomena such as the disjunction effect and the conjunction fallacy in decision theory, violations of the sure thing principle, and the Allais and Elsberg paradoxes in economics.
\end{abstract} 

\medskip
\begin{quotation}
\end{quotation}

\medskip
\section*{Introduction}

To understand the mechanism of how concepts combine to form sentences and texts and carry and communicate meaning between human minds is one of the major challenges facing the study of human thought. In this article we will further elaborate the theory about the combination of concepts that was initiated in Gabora and Aerts (2002a,b) and Aerts and Gabora (2005a,b), and continued in Aerts (2007a,b). In this approach, the influence of a context on a concept is an intrinsic part of the theory. Concepts `change continuously under influence of context', and this change is described as a `change of the state of the concept'. Our theory is essentially a contextual theory, which is one of the reasons why we can model the concepts in the way a quantum entity is described by the mathematical formalism of quantum mechanics, which is a contextual physical theory describing physical entities whose states change under influence of contexts of measurement. In other words, we use the formalism of quantum theory for the mathematical modeling of the concepts in our theory. We put forward a number of new insights including the surprising  one that the structure of quantum field theory, which we introduced in our modeling scheme in Aerts (2007b), plays an essential role. This allowed us to propose a specific hypothesis about the structure of human thought, viz. the hypothesis that we can identify within human thought a superposition of two layers whose structure follows from our quantum-based model of a large set of experimental data on the combination of concepts (Hampton 1988a,b). The layered structure of human thought is directly related to the quantum field structure of our scheme, more specifically to the use of Fock space in our modeling of these data (Aerts 2007a,b). We will illustrate these findings by working out in detail a relatively simple and concrete mathematical quantum model for this large collection of experimental data of Hampton (1988a,b).

The experiments in Hampton (1988ab) measure the deviation from classical set theoretic membership weights of exemplars or items with respect to pairs of concepts and their conjunction and disjunction. The reason for this focus is that Hampton's investigation was inspired by the so called `guppy effect' in concept conjunction found by Osherson and Smith (1981). Osherson and Smith considered the concepts {\it Pet} and {\it Fish} and their conjunction {\it Pet-Fish}, and observed that, while an exemplar or item such as {\it Guppy} was a very typical example of {\it Pet-Fish}, it was neither a very typical example of {\it Pet} nor of {\it Fish}. This demonstrates that the typicality of a specific item with respect to the conjunction of concepts can show unexpected behavior. As a result of the work of Osherson and Smith, the problem is often referred to as the `pet-fish problem' and the effect is usually called the `guppy effect'. Hampton identified a guppy-like effect for the membership weights of items with respect to pairs of concepts and their conjunction (Hampton, 1988a), and equally so for the membership weights of items with respect to pairs of concepts and their disjunction (Hampton, 1988b). Many experiments and analyses of effects due to combining concepts in general have since been conducted (Hampton, 1987, 1988a,b, 1991, 1996, 1997a,b; Osherson \& Smith, 1981, 1982; Rips, 1995; Smith \& Osherson, 1984; Smith, Osherson, Rips \& Keane, 1988; Springer \& Murphy, 1992; Storms, De Boeck, Van Mechelen \& Geeraerts, 1993; Storms, De Boeck, Hampton \& van Mechelen, 1999). However, none of the currently existing concept theories provides a satisfactory description and/or explanation of such effect for concept combinations. 

It is important to explain why we specifically want to investigate the modeling of the experimental data of Hampton (1988a,b). Our search for the modeling of the combination of concepts is not only a search for good models for specific sets of data. Indeed, we have come to suspect that `the way concepts combine and how they carry and communicate meaning' is governed by the presence of quantum structure in cognition. There is a well-established corpus of literature in theoretical physics describing methods to prove the presence of quantum structure by `only looking at experimental data', and it is irrelevant to the validity of these methods whether the data are the result of experiments in the area of physics or in any other domain of science (Aerts \& Aerts 2008). Theoretical physicists who are familiar with these approaches also know that `data showing deviations from set theoretic rules' are a major indication of the presence of quantum structure. This is why, the moment we became aware of the experimental results of Hampton (1988a,b), we assumed they might be the right data to prove the presence of quantum structure by making use of the techniques and methods developed in theoretical physics. This is exactly what we are doing in section \ref{classnonclassdata}, where we derive inequalities that characterize classical data, and hence show that the data of Hampton (1988a,b) are non-classical `in the same sense that the quantum mechanical data in physics are non-classical'. However, it is only by also working out an explicit quantum modeling of these data, which is what we do in Aerts (2007,ab) and in the present article, that the presence of quantum structure is fully proved: There are experimental data in cognition that `cannot be modeled by means of a classical theory' and for which `a quantum model does exist'.

Apart from this theoretical motivation of our modeling, namely to prove the existence of genuine quantum structure in cognition, we are also interested in the pure modeling power of the quantum modeling scheme we put forward. This raises the question in which sense successful modeling of the large set of data of Hampton (1988a,b) provides evidence of a broad validity of our modeling scheme. The literature on concept combinations contains numerous examples of effects of different types. Moreover, because existing theories have such great difficulties to model even simple combinations of `two' concepts, the ultimate aim of modeling sentences, texts, books, i.e. `all kinds of collections of combinations of concepts', has almost gone out of sight altogether. The general modeling power of our theory is based on different aspects, two of which in particular break with existing approaches and theories: (i) our theory is intrinsically contextual, and (ii) we explicitly introduce the notion of `state of a concept', and it is this state which can change under the influence of context. Here are some examples. If {\it Kitchen} is combined with {\it Island} to form {\it Kitchen Island}, it becomes very improbable for such a principal feature of {\it Island} as {\it Surrounded by Water} to apply. If {\it Stone} is combined with {\it Lion} to form {\it Stone Lion}, it becomes definitely not true for such a principal feature of {\it Lion} as {\it Is A Living Being} to apply. In the example of {\it Stone Lion}, the concept {\it Stone} provides a context for {\it Lion}, which changes the state of {\it Lion} in such a way that the feature {\it Is a Living Being} no longer applies. {\it Is a Living Being} is considered a principal feature of {\it Lion} because this feature applies to most states of the concept {\it Lion}, which does not mean that there is no state where it does not apply, and a context that transforms its state to exactly such a state. This is what the context {\it Stone} does. And hence this is the way the combination {\it Stone Lion} is modeled in our theory. The example of {\it Kitchen Island} is modeled in a similar way in our theory, and examples of greater complexity are worked out in detail in Aerts and Gabora (2005a,b).

We also proposed a detailed model for the concept {\it Pet-Fish} in Aerts (2005,b), where the guppy effect is modeled, and {\it Pet-Fish} appears as a specific state of {\it Pet} under the influence of the context {\it The Pet is a Fish}, and also as a specific state of {\it Fish} under the context {\it The Fish is a Pet}. Why then still pay special attention to the guppy effect, as we started doing in Aerts (2007,a,b), if this effect can be modeled as induced by context? The answer is that Hampton (1988,a,b) experiments made clear that more can be done and also more can be said about modeling than what we worked out in Aerts and Gabora (2005a,b). The mathematical formalism of quantum mechanics proves to allow modeling not only the influence of context in concept combinations -- as we did in Aerts and Gabora (2005a,b) -- but also `the emergence of new states'. This additional possibility is due to the `superposition principle' of quantum mechanics. In this article we give an explanation of the role of this emergent effect and the contextual effects and of how they give rise to a general quantum modeling scheme based on the subtle joint action of different quantum effects within the mathematical structure of quantum field theory. We have not proven that our theory enables the modeling of all possible combinations of large collections of concepts, but we will, in subsection \ref{largecollection} of this article, provide a scheme, a roadmap if you like, of how to work out in a general way the modeling of combinations of large collections of concepts.

Our modeling is less concerned with specific pure linguistic structures than it is aimed at the `meaning aspects' of concepts and their combinations, intending to uncover more and more the way meaning flows and interacts in the combination of concepts. This approach has consequences both for the nature of our modeling and for its potential bearing on other issues, theories and disciplines. In this sense, it is linked to the traditional problem of artificial intelligence, and we believe that one of the reasons that so little progress has been made in this field is partly due to the poor understanding of how meaning flows and interacts within concept combinations. The non-classical effects we investigate, i.e the `guppy effect' and the `over- and underextension' in membership weights, are not linked to peculiar effects of a linguistic nature either. They are related quite directly to non-classical ways of human decision-making, revealed in situations such as the conjunction fallacy (Tversky \& Kahneman, 1982) and the disjunction effect (Tversky \& Shafir,1992). Economics is yet another scientific domain where the same effects have been identified. Indeed, historically it has been the first of all. Savage's `sure thing principle' was formulated in 1944, and violations of this principle, which are in fact direct examples of the disjunction effect in decision theory, and underextension for the disjunction in concept theory, were identified and reported as early as 1953 (Allais 1953), and subsequently on quite a number of occasions (Elsberg, 1961). In this sense, it is not a coincidence that the disjunction effect as well as the conjunction fallacy have been studied in approaches where quantum aspects are similarly used in the modeling of these effects (Busemeyer, Matthew \& Wang, 2006; Franco, 2007; Khrennikov, 2008), and that also in economics quantum mechanics has been used for modeling purposes (Schaden, 2002; Baaquie, 2004; Haven, 2005; Khrennikov, 2009). We have provided more details of these and other connections in subsection \ref{decisiontheoryeconomics}. 

A next remark we want to make is that quantum structures are different from classical structures in more than one respect. In our study of applying quantum to cognition we have identified five main aspects that play a fundamental role and that are specific to quantum structures as compared to classical structures. They are (i) contextual influence, (ii) emergence due to superposition, (iii) interference, (iv) entanglement and (v) quantum field theoretic aspects.

In Gabora and Aerts (2002) and Aerts and Gabora (2005a,b), we focused on `contextual influence'. Indeed, unlike classical structures, quantum structures serve to model contextual influence. More specifically, we used the mathematical structure of a State Context Property System or SCOP (Aerts 2002; Gabora \& Aerts 2002; Aerts \& Gabora 2005a,b, Nelson \& McEvoy 2007; Gabora, Rosch \& Aerts 2008; Hettel, Flender \& Barros 2008; Flender, Kitto \& Bruza 2009), which is a generalization of the traditional Hilbert space of standard quantum mechanics. Such a SCOP describes concepts by means of their states, their properties, and the contexts that are relevant to their change. This makes it possible to model `contextual influence', one of the above five quantum aspects, which can be experimentally tested by considering weights related to typicality of exemplars and weights related to applicability of features, and how they change under the influence of a context.

In Aerts (2007a,b), we focused on how `emergence due to superposition', `interference' and also `specific quantum field theoretic aspects' could be used to model the type of deviation in concept combinations that have been identified in the guppy effect and in the membership effects measured by Hampton (1988a,b) in case of disjunction and conjunction, but also in a variety of other effects due to concept combinations. Although we have not worked out the concrete modeling for many of these situations, we have grounds to believe that the theory developed in Aerts (2007a,b) is generally applicable.

There is one limitation to what we have done so far, which we will explicitly point out here. To experimentally test the modeling of concepts and combinations of concepts, one has considered different quantities, including typicality, membership, applicability, etc \ldots, where one class of quantities is linked to exemplars -- sometimes also called `items' or `instantiations' -- of the considered concepts and a second class of quantities is linked to features of these concepts. In Aerts and Gabora (2005a), we developed the SCOP model attributing equal attention to the feature-linked quantities as to the exemplar-linked quantities, hence modeling the influence of context for both classes of quantities. When constructing an explicit Hilbert space quantum model for the experimental data testing contextual influence in Aerts and Gabora (2005a), i.e. making the SCOP model more concrete in a mathematical way, we largely shifted our attention to the modeling of the exemplar-linked quantities, namely the typicality of exemplars with respect to a concept. If, however, the Hilbert space model for the typicality of exemplars is considered in detail, it can be inferred that also the feature-linked quantities, e.g. applicability of features, can be modeled in a similar way in this Hilbert space. The quantum models elaborated in Aerts (2007a,b) focus only on exemplar-linked experimental quantities, namely the membership weights of exemplars of the considered concepts measured in Hampton (1988a,b). To our knowledge, neither Hampton nor any others have systematically investigated deviations with respect to conjunction and disjunction of feature-linked experimental quantities. To resolve this limitation, experimental data will need to be collected with respect to feature-linked quantities, accompanied by an assessment of whether the modeling developed in Aerts (2007a,b) can successfully be applied to these quantities as well.

As we have already hinted at, if we interpret the quantum representation that we built in Aerts (2007a,b), where the modeling centers on `emergence due to superposition', `interference' and `specific quantum field theoretic aspects', we can derive a specific structure for human thought. What we propose is that human thought comprises two layers, the one superposed with the other, which we have called the `classical logical layer' and the `quantum conceptual layer', respectively. The thought process within the classical logical layer is given form by an underlying classical logical conceptual process. The thought process within the quantum conceptual layer is given form under the influence of the totality of the surrounding conceptual landscape, where the different concepts figure as individual entities, also when they are combinations of other concepts, contrary to the classical logical layer, where combinations of concepts figure as classical combinations of entities and not as individual entities. In this sense, one can speak of a phenomenon of `conceptual emergence' taking place in this quantum conceptual layer, certainly so for combinations of concepts. The quantum conceptual thought process is indeterministic in essence, and since all concepts of the interconnected web that forms the landscape of concepts and combinations of them attribute as individual entities to the influences reigning in this landscape, the nature of quantum conceptual thought contains aspects that we strongly identify as holistic and synthetic. However, the quantum conceptual thought process is not unorganized or irrational. Quantum conceptual thought is as firmly structured as classical logical thought but in a very different way. We believe that science has hardly uncovered the structure of quantum conceptual thought because it has been believed to be intuitive, associative, irrational, etc... -- in other words, `rather unstructured'. Its structure has not been sought for because it has always been believed to be hardly existent in the first place. An idealized version of this quantum conceptual thought process, or a substantial part of it, can be modeled as a quantum mechanical process. Hence we believe that important aspects of the basic structure of quantum conceptual thought can be uncovered based on the quantum structure modeling developed in Aerts (2007a,b), and the simpler model that we will work out explicitly in the remainder of this article.

\section{A General Scheme for Quantum Modeling} \label{generalscheme}
In this section we will explain the general scheme for quantum modeling worked out in Aerts (2007a,b) and our earlier work. 
We will first explain Hampton's experiments and introduce some of his data because this is the main experimental material of our discussion.

\subsection{The Guppy Effect for Membership}
Since the work of Eleanor Rosch and collaborators (Rosch, 1973a, 1973b), cognitive scientists view membership of an item for a specific concept category usually not as a `yes-or-no' notion, but a graded or fuzzy notion. This means that we can characterize the item by assigning it a membership weight, which is a number between 0 and 1, both inclusive, where 1 corresponds to membership of the concept category, 0 corresponds to non-membership of the concept category, and values between 1 and 0 indicate a graded or fuzzy degree of membership of the item with respect to the considered concept. Following this approach of graded membership, Hampton (1988a,b) experimentally identified an effect similar to the guppy effect for typicality with respect to the conjunction and disjunction of concepts.

More concretely, Hampton (1988a) considered, for example, the concepts {\it Bird} and {\it Pet} and their conjunction {\it Bird and Pet}. He then conducted tests to measure how subjects rated the membership weights of different items. In the case of the item {\it Cuckoo} for the concept {\it Bird}, the outcome was 1, while the rating of the membership weight of {\it Cuckoo} for the concept {\it Pet} was 0.575. When subjects were asked to rate the membership weight of the item {\it Cuckoo} for the combination {\it Bird and Pet}, the outcome was 0.842. This means that subjects found {\it Cuckoo} to be `more strongly a member of the conjunction {\it Bird and Pet}' than they found it to be a member of the concept {\it Pet} on its own. If we consider the `logical' meaning of a conjunction intuitively, we must say that this is a strange effect. Indeed, if somebody finds that {\it Cuckoo} is a {\it Bird and a Pet}, they may be expected equally to agree with the statement that {\it Cuckoo} is a {\it Pet} if the conjunction of concepts behaved in a way similar to the conjunction of logical propositions. Hampton (1988a) called this deviation from what one would expect according to a standard classical interpretation of conjunctions of concepts `overextension'.

Hampton (1988b) considered the disjunction of concepts, for example, the concepts {\it Home Furnishings} and {\it Furniture} and their disjunction {\it Home Furnishings or Furniture}. With respect to this pair, Hampton considered the item {\it Ashtray}. Subjects rated the membership weight of {\it Ashtray} for the concept {\it Home Furnishings} as 0.7 and the membership weight of the item {\it Ashtray} for the concept {\it Furniture} as 0.3. However, the membership weight of {\it Ashtray} with respect to the disjunction {\it Home Furnishings or Furniture} was rated as only 0.25, i.e. less than either of the weights assigned for both concepts apart. This means that subjects found {\it Ashtray} to be `less strongly a member of the disjunction {\it Home Furnishings or Furniture}' than they found it to be a member of the concept {\it Home Furnishings} alone or a member of the concept {\it Furniture} alone. If one thinks intuitively of the `logical' meaning of a disjunction, this is an unexpected result. Indeed, if somebody finds that {\it Ashtray} belongs to {\it Home Furnishings}, they would be expected to also believe that {\it Ashtray} belongs to {\it Home Furnishings or Furniture}. The same holds for {\it Ashtray} and {\it Furniture}. Hampton (1988b) called this deviation from what one would expect according to a standard classical interpretation of the disjunction `underextension'. 

To be more specific about the nature of this guppy effect for conjunction and disjunction, we will consider two concepts, concept $A$ and concept $B$, the conjunction of these two concepts, denoted as `$A$ and $B$', and the disjunction of these concepts, denoted as `$A$ or $B$'. Furthermore, we will consider different items $X$, and for each of these items $X$, its membership weight $\mu(A)$ with respect to concept $A$, its membership weight $\mu(B)$ with respect to concept $B$, its membership weight $\mu(A\ {\rm and}\ B)$ with respect to $A$ and $B$, and its membership weight $\mu(A\ {\rm or}\ B)$ with respect to $A$ or $B$. 

A typical experiment testing the guppy effect, such as the experiments considered in Hampton (1988a,b), proceeds as follows. The tested subjects are asked to choose a number from the following set: $\{-3,-2, -1,0,$ $+1,+2,+3\}$, where the positive numbers +1, +2 or +3 mean that they consider `the item to be a member of the concept' and the typicality of the membership increases with an increasing number. Hence +3 means that the subject who attributes this number considers the item to be a very typical member, and +1 means that he or she considers the item to be a not so typical member. The negative numbers indicate non-membership, again in increasing order, i.e. -3 indicates strong non-membership, and -1 represents weak non-membership. Choosing 0 means the subject is indecisive about the membership or non-membership of the item. Table 2 and 3 represent the items and pairs of concepts that Hampton (1988a,b) tested for the guppy effect with respect to the conjunction and the disjunction. In both experiments -- the one testing the guppy effect for conjunction and the one testing it for disjunction -- subjects were asked to repeat the procedure for all the items and concepts considered. Membership weights were then calculated by dividing the number of positive ratings by the number of non-zero ratings.

The validity of a `graded structure approach' to concept modeling was criticized for `being unstable' by Barsalou (1987). If we look at the `elements of instability' that Barsalou analyzes, we can see that they are the very elements that we, in our Aerts and Gabora (2005a,b) approach, put forward as `elements that provoke a change of the state of the concept'. This means that the effects that Barsalou (1987) qualified as unstable, are captured by the notion of `state of a concept' in our Aerts and Gabora (2005a,b) approach. These effects are what we have called `contextual effects' in Aerts and Gabora (2005a,b), while Gabora, Rosch and Aerts (2008) examined their `ecological aspects'.

If we typify membership of an item for a concept by means of weights to explicitly account for the graded and fuzzy structure of the membership notion, from a mathematical point of view, we can then represent a concept by a set if this set is a fuzzy set in the sense introduced in Zadeh (1965). In fuzzy-set theory, the common rule for conjunction is the minimum rule and the common rule for disjunction is the maximum rule. More concretely, following this rule, the membership weight with respect to the conjunction of two concepts equals the smallest of the two membership weights with respect to the constituent concepts, and the membership weight for the disjunction of two concepts equals the greatest of the two membership weights with respect to the constituent concepts. Osherson and Smith (1981) showed how the situation of the pet-fish problem conflicts with the minimum rule of fuzzy-set theory for the conjunction. We now introduce the `observed weight of the conjunction concept - minimum weight of both concepts' and the `maximum weight of both concepts - observed weight of the disjunction concept'
\begin{equation} \label{deltaconjdisj}
\Delta_c=\mu(A\ {\rm and}\ B) - \min(\mu(A), \mu(B)) \quad \Delta_d=\max(\mu(A), \mu(B))-\mu(A\ {\rm or}\ B)
\end{equation}
and call $\Delta_c$ the `conjunction minimum rule deviation' and $\Delta_d$ the `disjunction maximum rule deviation'.
In Table 1 and 2 we can see how individual items deviate from the minimum rule for the conjunction and for maximum rule for the disjunction respectively. The complete set of conjunction and disjunction data of Hampton (1988a,b) can be found in Tables 3 and 4.

\subsection{Classical and Non Classical Data} \label{classnonclassdata}
Before putting forward our general scheme for quantum modeling, we will analyze in greater detail -- as has been done by Hampton and others -- the deviation from what one would expect in classical terms for conjunction and disjunction data measured, as we explained in the previous section. We will do this first of all to make clear from a mathematical point of view `which are the situations that cannot be modeled within a classical set theoretic setting', and secondly, such that we can show in a systematic way `what is the reason that a quantum mechanical setting allows for a modeling of these non-classical situations'. Although the commonest `conjunction rule' in fuzzy-set theory is the `minimum rule' and the commonest `disjunction rule' in fuzzy-set theory is the `maximum rule', it is possible to carry out a more in-depth analysis of `classical conjunction data' and `classical disjunction data'.

We can define classical conjunction data for the situation of an item $X$ with respect to concepts $A$ and $B$ and their conjunction `$A$ and $B$' and classical disjunction data for the situation of an item $X$ with respect to the concepts $A$ and $B$ and their disjunction `$A$ or $B$' as data that can be modeled within a measure theoretical or Kolmogorovian probability structure. An explanation of such a measure or probability structure follows.

\bigskip
\noindent
\noindent {\it Definition of `Measure and Kolmogorovian Probability':
A measure $P$ is a function defined on a $\sigma$-algebra (pronounced sigma-algebra) $\sigma(\Omega)$ over a set $\Omega$ and taking values in the extended interval $[0,\infty]$ such that the following three conditions are satisfied: (i) The empty set has measure zero; (ii) Countable additivity or $\sigma$-additivity: if  $E_1$, $E_2$, $E_3$, $\dots$ is a countable sequence of pairwise disjoint sets in $\sigma(\Omega)$, the measure of the union of all the $E_i$ is equal to the sum of the measures of each $E_i$; (iii) The triple $(\Omega,\sigma(\Omega),P)$ satisfying (i) and (ii) is then called a measure space, and the members of $\sigma(\Omega)$ are called measurable sets.
A Kolmogorovian probability is a measure with total measure one. A Kolmogorovian probability space $(\Omega,\sigma(\Omega),P)$ is a measure space $(\Omega,\sigma(\Omega),P)$ such that $P$ is a Kolmogorovian probability. The three conditions expressed in a mathematical way are
\begin{equation} \label{KolmogorovianMeasure}
P(\emptyset)=0 \quad P(\bigcup_{i=1}^\infty E_i)=\sum_{i=1}^\infty P(E_i) \quad P(\Omega)=1
\end{equation}
}
\noindent
We now need to explain what is a $\sigma$-algebra over a set to understand the above definition.

\bigskip
\noindent
{\it Definition of `$\sigma$-algebra over a set': A $\sigma$-algebra over a set $\Omega$ is a non-empty collection $\sigma(\Omega)$ of subsets of $\Omega$ that is closed under complementation and countable unions of its members. It is a Boolean algebra, completed to include countably infinite operations.}

\bigskip
\noindent
Measure structures are the most general classical structures devised by mathematicians and physicists to structure weights. A Kolmogorovian probability is such a measure applied to statistical data. It is called `Kolmogorovian', because Andrey Kolmogorov was the first to axiomatize probability theory in this manner (Kolmogorov, 1977).
 
If we analyze this general idea for classical conjunction and disjunction data, i.e. that they are modeled by a classical measure structure, we can show that `conjunction data for which the minimum rule of fuzzy-set theory is valid are classical' and `disjunction data for which the maximum rule of fuzzy-set theory is valid are classical'. This means that the items that Hampton and others classified as problematic, i.e. with overextension for the conjunction and underextension for the disjunction, remain problematic in our way of defining classical conjunction and disjunction data, i.e. they are non-classical. However, the rather limited minimum rule for the conjunction and maximum rule for the disjunction need not necessarily be valid for conjunction and disjunction data to be classical. Hence there do exist conjunction data and disjunction data in Hampton's collection that are classical and for which the minimum rule and maximum rule of fuzzy-set theory, respectively, are not satisfied. On the other hand, next to overextension for the conjunction and underextension for the disjunction, there is something else, not explicitly identified by Hampton and others, which can make conjunction and disjunction data non classical. Some of Hampton's data are problematic even if they do not show overextension for the conjunction or underextension for the disjunction. Overextension is not the only problematic aspect of the conjunction data collected in experiments on concepts and their conjunctions, and underextension is not the only problematic aspect of the disjunction data collected in experiments on concepts and their disjunctions.

\subsection{Classical and Non Classical Conjunction Data} \label{conjunctiondata}
To make all this specific, we will first concentrate on the case of conjunction and in the subsequent subsection focus on that of disjunction.

\bigskip
\noindent {\it Definition of `Classical Conjunction Data':
We say that data that are the weights $\mu(A)$, $\mu(B)$ and $\mu(A\ {\rm and}\ B)$ of an item $X$ with respect to a pair of concepts $A$ and $B$ and their conjunction `$A$ and $B$' are `classical conjunction data' with respect to these concepts if there exists a Kolmogorovian probability space $(\Omega,\sigma(\Omega),P)$ and events $E_A, E_B \in \sigma(\Omega)$ of the events algebra $\sigma(\Omega)$ such that 
\begin{equation}
P(E_A) = \mu(A) \quad P(E_B) = \mu(B) \quad {\rm and} \quad P(E_A \cap E_B) = \mu(A\ {\rm and}\ B)
\end{equation}}
We can prove an interesting theorem that makes it possible to characterize classical conjunction data in an easy way.

\bigskip
\noindent {\bf Theorem 1:} {\it The membership weights $\mu(A), \mu(B)$ and $\mu(A\ {\rm and}\ B)$ of an item $X$ with respect to concepts $A$ and $B$ and their conjunction `$A$ and $B$' are classical conjunction data if and only if they satisfy the following inequalities}
\begin{eqnarray} \label{ineq01}
&0 \le \mu(A\ {\rm and}\ B) \le \mu(A) \le 1 \\ \label{ineq02}
&0 \le \mu(A\ {\rm and}\ B) \le \mu(B) \le 1 \\ \label{ineq03}
& \mu(A) + \mu(B) - \mu(A\ {\rm and}\ B) \le 1
\end{eqnarray}
Proof: See Appendix A

\bigskip
\noindent
Inequalities (\ref{ineq01}) and (\ref{ineq02}) can be verified by looking at the quantity $\Delta_c$ defined in (\ref{deltaconjdisj}). We have $\Delta_c \le 0$ and hence $\mu(A\ {\rm and}\ B) \le {\rm min}(\mu(A), \mu(B))$ if and only if both inequalities (\ref{ineq01}) and (\ref{ineq02}) are satisfied. On the other hand, if $\Delta_c > 0$, a situation named `overextension' by Hampton (1988a), at least one of the inequalities (\ref{ineq01}) and (\ref{ineq02}) is not satisfied. This makes it possible to put forward another theorem.

\bigskip
\noindent {\bf Theorem 2:} {\it Consider the situation of an item $X$ such that the membership weight $\mu(A\ {\rm and}\ B)$ with respect to the conjunction of two concepts $A$ and $B$ is `overextended' with respect to the membership weights $\mu(A)$ and $\mu(B)$ of the item $X$ with respect to the individual concepts $A$ and $B$. The weights $\mu(A)$, $\mu(B)$ and $\mu(A\ {\rm and}\ B)$ are then non-classical conjunction data, i.e. they cannot be modeled by a Kolmogorovian probability space.}

\bigskip
\noindent
The fact that there is no overextension, however, is not sufficient for $\mu(A)$, $\mu(B)$ and $\mu(A\ {\rm and}\ B)$ to be able to be modeled within a Kolmogorovian probability space. Also inequality (\ref{ineq03}) needs to be satisfied. For this we introduce a new quantity
\begin{equation} \label{Kconj}
k_c = 1 -\mu(A)-\mu(B)+\mu(A\ {\rm and}\ B)
\end{equation}
using the letter $k$ for Kolmogorov, and we call it the `Kolmogorovian conjunction factor'. If $0 \le k_c$ then (\ref{ineq03}) is satisfied, and if $k_c < 0$ then (\ref{ineq03}) is not satisfied. This makes it possible to formulate the following theorem.

\bigskip
\noindent {\bf Theorem 3:} {\it The membership weights $\mu(A)$, $\mu(B)$ and $\mu(A\ {\rm and}\ B)$ of an item $X$ with respect to concepts $A$, $B$ and the conjunction of $A$ and $B$ are classical conjunction data, i.e. they can be modeled by means of a Kolmogorovian probability space, if and only if $\Delta_c \le 0$, i.e. there is no `overextension', and $0 \le k_c$, i.e. the Kolmogorovian conjunction factor is not negative.}

\bigskip
\noindent
In Table 1 the quantities $\Delta_c$ and $k_c$ are given for some of the items tested by Hampton (1988a). Most items that cannot be modeled within a Kolmogorovian probability space are overextended items, i.e. items for which $0 < \Delta_c$. We have labeled these items by means of the letter $\Delta$. These are the items that Hampton (1988a), following the guppy-effect analysis of Osherson and Smith (1981), already observed to be the problematic ones. We have labeled the classical items by means of the letter $c$. There are a few items only where it is the other inequality (\ref{ineq03}) that is violated, and which for this reason cannot be modeled by a Kolmogorovian space either. We have labeled these items by means of the letter $k$ and they can be found in the complete list of items tested by Hampton (1988a) in Table 4. In the next section we will see that for Hampton's disjunction experiment many more items are non-classical of the $k$-type, hence of the non-Guppy type. Let us analyze first the situation of the disjunction and see how the Kolmogorovian factor needs to be defined for this situation.

\subsection{Classical and Non Classical Disjunction Data}

We will first explicitly define what are classical disjunction data based on the general idea we put forward.

\bigskip
\noindent {\it Definition of `Classical Disjunction Data':
We say that data that are the weights $\mu(A)$, $\mu(B)$ and $\mu(A\ {\rm or}\ B)$ of an item $X$ with respect to a pair of concepts $A$ and $B$ and their disjunction `$A$ or $B$' are `classical disjunction data' with respect to these concepts if there exists a Kolmogorovian probability space $(\Omega, \sigma(\Omega), P)$ and events $E_A, E_B \in \sigma(\Omega)$ of the events algebra $\sigma(\Omega)$ such that 
\begin{equation}
P(E_A) = \mu(A) \quad P(E_B) = \mu(B) \quad {\rm and} \quad P(E_A \cup E_B) = \mu(A\ {\rm or}\ B)
\end{equation}}
We will now prove the analogue of theorem 1 for the disjunction.

\bigskip
\noindent
{\bf Theorem 4:} {\it The membership weights $\mu(A), \mu(B)$ and $\mu(A\ {\rm or}\ B)$ of an item $X$ with respect to concepts $A$ and $B$ and their disjunction `$A$ or $B$' are classical disjunction data if and only if they satisfy the following inequalities}
\begin{eqnarray} \label{disjunctionineq01}
&0 \le \mu(A) \le \mu(A\ {\rm or}\ B) \le 1 \\ \label{disjunctionineq02}
&0 \le \mu(B) \le \mu(A\ {\rm or}\ B) \le 1 \\ \label{disjunctionineq03}
&0 \le \mu(A) + \mu(B) - \mu(A\ {\rm or}\ B)
\end{eqnarray}
Proof: See Appendix B

\bigskip
\noindent
Inequalities (\ref{disjunctionineq01}) and (\ref{disjunctionineq02}) can be verified by looking at the quantity $\Delta_d$ defined in (\ref{deltaconjdisj}). We have $\Delta_d \le 0$ and hence $\mu(A\ {\rm and}\ B) \le {\rm min}(\mu(A), \mu(B))$ if and only if both inequalities (\ref{disjunctionineq01}) and (\ref{disjunctionineq02}) are satisfied. On the other hand, if $\Delta_d > 0$, a situation named `underextension' by Hampton (1988b), at least one of the inequalities (\ref{disjunctionineq01}) and (\ref{disjunctionineq02}) is not satisfied. This makes it possible to put forward another theorem.

\bigskip
\noindent
{\bf Theorem 5:} {\it Consider the situation of an item $X$ such that the membership weight $\mu(A\ {\rm or}\ B)$ with respect to the disjunction of two concepts $A$ and $B$ is `underextended' with respect to the membership weights $\mu(A)$ and $\mu(B)$ of the item $X$ with respect to the individual concepts $A$ and $B$. The weights $\mu(A)$, $\mu(B)$ and $\mu(A\ {\rm or}\ B)$ are then non-classical disjunction data, i.e. they cannot be modeled by a Kolmogorovian probability space.}

\bigskip
\noindent
The fact that there is no underextension, however, is not sufficient for $\mu(A)$, $\mu(B)$ and $\mu(A\ {\rm or}\ B)$ to be able to be modeled within a Kolmogorovian probability space. Also inequality (\ref{disjunctionineq03}) needs to be satisfied. We therefore introduce the quantity
\begin{equation} \label{Kdisj}
k_d = \mu(A)+\mu(B)-\mu(A\ {\rm or}\ B)
\end{equation}
and we call it the `Kolmogorovian disjunction factor'. If $0 \le k_d$ then (\ref{disjunctionineq03}) is satisfied, and if $k_d < 0$ then (\ref{disjunctionineq03}) is not satisfied. This makes it possible to formulate the following theorem.

\bigskip
\noindent {\bf Theorem 6:} {\it The membership weights $\mu(A)$, $\mu(B)$ and $\mu(A\ {\rm or}\ B)$ of an item $X$ with respect to concepts $A$, $B$ and the disjunction of $A$ and $B$ are classical disjunction data, i.e. they can be modeled by means of a Kolmogorovian probability space, if and only if $\Delta_d \le 0$, i.e. there is no `underextension', and $0 \le k_d$, i.e. the Kolmogorovian disjunction factor is not negative.}

\bigskip
\noindent
In Table 2 the quantities $\Delta_d$ and $k_d$ are given for some of the items tested by Hampton (1988b). Underextension is the commonest form of non-classicality in the case of disjunction, and the underextended items are labeled by means of the letter $\Delta$. These are the items that Hampton (1988b) already observed to be the problematic ones. The classical items are again labeled by the letter $c$. Many more items are non-classical than in the case of conjunction in the sense that the Kolmogorovian disjunction factor $k_d$ is negative, as can be seen in the complete list of all items in Table 3, and we have labeled these items using the letter $k$.

Aerts, Aerts \& Gabora (2009) present a simple geometric way, making use of polytopes, to distinguish between classical and non-classical experimental data. It also links our analysis to the work of Itamar Pitowsky on correlation polytopes (Pitowsky 1989).

\subsection{Presenting the Quantum Modeling Scheme}
\label{QuantumModelingScheme}

In quantum mechanics, a state of a quantum entity is described by a vector of length equal to 1. The Hilbert space of quantum mechanics is essentially the set of these vectors, with each vector representing the state of the quantum entity under consideration, and equipped with some additional structure. We denote vectors using the bra-ket notation introduced by Paul Adrien Dirac, one of the founding fathers of quantum mechanics (Dirac, 1958), i.e. $\left| A \right\rangle $, $\left| B \right\rangle $. Vectors denoted in this way are called `kets', to distinguish them from another type of vectors, denoted as $\left\langle A \right|$, $\left\langle B \right|$ and called `bras', which we will introduce later.

A state of a quantum entity is described by a ket vector, and by analogy we will describe the state of a concept by a ket vector. More concretely, consider the concept $A$, then the state of concept $A$ is represented by ket vector $\left| A \right\rangle$. We introduced the notion of `state of a concept' in detail in Aerts \& Gabora (2005a), and this is also the way we use it in the present article. The `state of a concept' represents `what the concept stands for with respect to its relevant features and contexts'.

The additional structure of a Hilbert space, as compared to being a vector space, is meant to express the notions of length, orthogonality and weight. This is achieved by introducing a product between a bra vector, for example $\langle A|$, and a ket vector, for example $|B\rangle $, denoted as $\langle A|B\rangle$ and called a bra-ket. A bra-ket is always a complex number, and the absolute value of complex number $\langle A|B\rangle$ is equal to the length of $|A\rangle$ times the length of $|B\rangle$ times the cosine of the angle between vectors $|A\rangle$ and $|B\rangle$. From this it follows that we have a definition of the length of a ket and bra vector $\||A\rangle\|=\|\langle A|\|=\sqrt{\langle A|A\rangle}$. In quantum mechanics a state of the quantum entity is represented by means of a ket vector of length 1. Hence we can now specify this requirement for the vectors concerning concepts $A$ and $B$. Vectors $|A\rangle$ and $|B\rangle$ are such that $\langle A|A\rangle=\langle B|B\rangle=1$. We said that $\langle A|B\rangle$ is a complex number whose absolute value equals the length of $|A\rangle$ times the length of $|B\rangle$ times the cosine of the angle between $|A\rangle$ and $|B\rangle$. This means that $|A\rangle$ and $|B\rangle$ are orthogonal, in the sense that the angle between both vectors is $90^\circ$, if $\langle A|B\rangle= 0$. We denote this as $|A\rangle \perp |B\rangle$. We said that $\langle A|B\rangle$ is a complex number; additionally, in the quantum formalism $\langle A|B\rangle$ is the complex conjugate of $\langle B|A\rangle$. Hence $\langle B|A\rangle^*=\langle A|B\rangle$.
Further, the operation bra-ket $\langle \cdot | \cdot \rangle$ is linear in the ket and anti-linear in the bra. Hence $\langle A|(x|B\rangle+y|C\rangle)=x\langle A|B\rangle+y\langle A|C\rangle$ and $(a\langle A|+b\langle B|)|C\rangle=a^*\langle A|C\rangle+b^*\langle B|C\rangle$. The absolute value of a complex number is defined as the square root of the product of this complex number and its complex conjugate. Hence we have $|\langle A|B\rangle|=\sqrt{\langle A|B\rangle \langle B|A\rangle}$.
  
An orthogonal projection $M$ is a linear function on the Hilbert space, hence $M: {\cal H} \rightarrow {\cal H}, |A\rangle \mapsto M|A\rangle$, which is Hermitian and idempotent, which means that for $|A\rangle, |B\rangle \in {\cal H}$ and $x, y \in \compl$ we have (i) $M(z|A\rangle+t|B\rangle)=zM|A\rangle+tM|B\rangle$ (linearity); (ii) $\langle A|M|B\rangle=\langle B|M|A\rangle$ (hermiticity); and (iii) $M \cdot M=M$ (idempotenty).

Measurable quantities, often called observables in quantum mechanics, are represented by means of Hermitian linear functions on the Hilbert space, and for two valued observables these Hermitian functions are orthogonal projections. This is why we can describe the decision measurement of `being a member of' or `not being a member of' with respect to a concept by means of an orthogonal projection on the Hilbert space. Concretely, let us consider an item $X$, then the decision measurement `being a member of' with respect to a concept is represented by means of the orthogonal projection $M$. And the probability $\mu(A)$ for a test subject to decide `in favor of membership' of item $X$ with respect to concept $A$ is given in quantum mechanics by the following equation $\mu(A)=\langle A|M|A\rangle$.

We have all tools at hand now to present our quantum modeling scheme. Let us therefore consider the two concepts $A$ and $B$. Both $A$ and $B$ are described quantum mechanically in a Hilbert space ${\cal H}$, so that they are represented by states $|A\rangle$ and $|B\rangle$ of ${\cal H}$, respectively. We describe concept `$A$ or $B$' by means of the normalized superposition state ${1 \over \sqrt{2}}(|A\rangle+|B\rangle)$, and also suppose that $|A\rangle$ and $|B\rangle$ are orthogonal, hence $\langle A|B\rangle=0$. An experiment considered in Hampton (1988a,b) consists in a test aimed to ascertain whether a specific item $X$ is `a member of' or `not a member of' a concept. We represent this experiment by means of a projection operator $M$ on this Hilbert space ${\cal H}$. This experiment is applied to concept $A$, to concept $B$, and to concept `$A$ or $B$', respectively, yielding specific probabilities $\mu(A)$, $\mu(B)$ and $\mu(A\ {\rm or}\ B)$. These probabilities represent the degrees to which a subject is likely to choose $X$ to be a member of $A$, $B$ and `$A$ or $B$'. In accordance with the quantum rules, these probabilities are given by \begin{equation}
\mu(A)=\langle A|M|A\rangle \quad \mu(B)=\langle B|M|B\rangle \quad \mu(A\ {\rm or}\ B)={1 \over 2}(\langle A|+\langle B|)M(|A\rangle+|B\rangle)
\end{equation}
Applying the linearity of Hilbert space and taking into account that $\langle B|M|A\rangle^*=\langle A|M|B\rangle$, we have
\begin{equation} \label{interference}
\mu(A\ {\rm or}\ B)={1 \over 2}(\langle A|M|A\rangle+\langle A|M|B\rangle+\langle B|M|A\rangle+\langle B|M|B\rangle)={\mu(A,X)+\mu(B,X) \over 2}+\Re\langle A|M|B\rangle
\end{equation}
where $\Re\langle A|M|B\rangle$ is the real part of the complex number $\langle A|M|B\rangle$. This is called the `interference term' in quantum mechanics. Its presence produces a deviation from the average value ${1 \over 2}(\mu(A)+\mu(B))$, which would be the outcome in the absence of interference. Note that thus far we have applied two of the quantum elements discussed, namely `superposition', in taking ${1 \over \sqrt{2}}(|A\rangle+|B\rangle)$ to represent `$A$ or $B$', and `interference', as the effect appearing in equation (\ref{interference}).

This `quantum model based on superposition and interference' can be realized in a three-dimensional complex Hilbert space $\compl^3$. Rather than presenting a detailed analysis as can be found in Aerts (2007a,b), we focus on this $\compl^3$ realization in this article. We suppose that $\mu(A)\not=0$, $\mu(B)\not=0$, $\mu(A)\not=1$ and $\mu(B)\not=1$, because the cases where one of the membership weights is 0 or 1 call for a specific approach, as can be found in Aerts (2007a,b). We also remark that, for a given $\mu(A)$ and $\mu(B)$, the situation is always such that one of the two quantities $\mu(A)+\mu(B)$ or $(1-\mu(A)+(1-\mu(B)$ is greater than or equal to 1 and the other is smaller than or equal to 1. In case $1 \le \mu(A)+\mu(B)$, we put $a=\mu(A)$, $b=\mu(B)$ and in case $1 \le (1-\mu(A)+(1-\mu(B)$, we put $a=1-\mu(A)$ and $b=1-\mu(B)$. We take $M(\compl^3)$, the subspace of $\compl^3$ spanned by vectors $(1,0,0)$ and $(0,1,0)$, and choose
\begin{eqnarray} \label{vectorA}
|A\rangle&=&(\sqrt{a},0,\sqrt{1-a}) \\ \label{vectorB}
|B\rangle&=&e^{i\beta}(\sqrt{(1-a)(1-b) \over a},\sqrt{a+b-1 \over a},-\sqrt{1-b}) \\ \label{anglebeta}
\beta&=&\arccos({2\mu(A\ {\rm or}\ B)-\mu(A)-\mu(B) \over 2\sqrt{(1-a)(1-b)}})
\end{eqnarray}
This gives rise to a quantum mechanical description of the situation with probability weights $\mu(A), \mu(B)$ and $\mu(A\ {\rm or}\ B)$. Let us verify this. We have $\langle A|A\rangle=a+1-a=1$, $\langle B|B\rangle={(1-a)(1-b) \over a}+{a+b-1 \over a}+1-b=1$, which shows that both vectors $|A\rangle$ and $|B\rangle$ are unit vectors. We have $\langle A|B\rangle=\sqrt{(1-a)(1-b)}e^{i\beta}-\sqrt{(1-a)(1-b)}e^{i\beta}=0$, which shows that $|A\rangle$ and $|B\rangle$ are orthogonal. Furthermore, we have $\langle A|M|B\rangle=\sqrt{(1-a)(1-b)}e^{i\beta}$ and hence $\Re\langle A|M|B\rangle=\sqrt{(1-a)(1-b)}\cos\beta={1 \over 2}(2\mu(A\ {\rm or}\ B)-\mu(A)-\mu(B))$. Applying (\ref{anglebeta}) this gives $\mu(A\ {\rm or}\ B)={1 \over 2}(\mu(A)+\mu(B))+\Re\langle A|M|B\rangle$, which corresponds to (\ref{interference}), which shows that, given the values of $\mu(A)$ and $\mu(B)$, the correct value for $\mu(A\ {\rm or}\ B)$ is obtained in this quantum model.

Let us work out some examples. Consider the item {\it Pencil Eraser} with respect to the pair of concepts {\it Instruments} and {\it Tools} and their disjunction {\it Instruments or Tools}. Hampton (1988b) measured $\mu(A)=0.4$, $\mu(B)=0.7$ and $\mu(A\ {\rm or}\ B)=0.45$. This means that this situation does not allow a classical model, since $\mu(A\ {\rm or}\ B)<\mu(B)$. Let us construct the $\compl^3$ quantum model for this item. We have $\mu(A)+\mu(B)=1.1$, and hence put $a=\mu(A)=0.4$ and $b=\mu(B)=0.7$. After making the calculations of equations (\ref{vectorA}), (\ref{vectorB}) and (\ref{anglebeta}), we find $|A\rangle=(0.6325, 0, 0.7746)$, $|B\rangle=e^{i\beta}(0.6708, 0.5, -0.5477)$ and $\beta=103.6330^\circ$. As a second example we consider the item {\it Ashtray} with respect to the pair of concepts {\it House Furnishings} and {\it Furniture} and their disjunction {\it House Furnishings or Furniture}. Hampton (1988b) measured $\mu(A)=0.7$, $\mu(B)=0.3$ and $\mu(A\ {\rm or}\ B)=0.25$. Again, this situation does not allow a classical model since $\mu(A\ {\rm or}\ B)<\mu(A)$ and $\mu(A\ {\rm or}\ B)<\mu(B)$. For the $\compl^3$ realization we find $|A\rangle=(0.8367, 0, 0.5477)$, $|B\rangle=e^{i\beta}(0.5477, 0, -0.8367)$ and $\beta=123.0619^\circ$. Our final example concerns the item {\it Field Mouse} with respect to the pair of concepts {\it Pets} and {\it Farmyard Animals} and their disjunction {\it Pets or Farmyard Animals}. For this item, the values $\mu(A)=0.1$, $\mu(B)=0.7$ and $\mu(A\ {\rm or}\ B)=0.4$. This situation does not allow a classical model either, since $\mu(A\ {\rm or}\ B)<\mu(B)$. For the $\compl^3$ realization we find $|A\rangle=(0.9487, 0, 0.3162)$, $|B\rangle=e^{i\beta}(0.2789, 0.4714, -0.8367)$ and $\beta=90^\circ$.

In this $\compl^3$ model, only $e^{i\beta}$ appears as `not a real number' in the vector $|B\rangle$. For two values of $\beta$, namely $\beta=0^\circ$ and $\beta=180^\circ$, $e^{i\beta}$ is a real number, which means that for these two values of $\beta$ we can make a graphical representation of the situation in $\real^3$. The interference effect is present for these two values of $\beta$ but can take only two values. The role of the complex numbers is to allow it to obtain any value in between these two values. In Table 2 we have calculated the vectors $|A\rangle$ and $|B\rangle$ and the angle $\beta$ for a number of Hampton (1988b)'s experimental data. For those items where no vector is shown in Table 2 it means that the $\compl^3$ model does not exist. This is one of the main reason to extend our modeling to Fock space, and we will explain how we do this in subsection \ref{Fockspace}. The content of Table 1 will be explained later, since we first need to understand our quantum modeling scheme for the conjunction for which we need to introduce Fock space.

\subsection{Quantum Field Theory and Two Modes of Human Thought} \label{QuantumFieldTheory}

Our use of vector ${1 \over \sqrt{2}}(|A\rangle+|B\rangle)$ to model the `$A$ or $B$' concept reflects the modeling of the archetypical `double-slit type of situation' in quantum mechanics. Quantum mechanics describes the situation where both slits are open using the wave function which is the normalized superposition of the wave functions that describe the situations where only one of the two slits is open. In Aerts (2007a,b), introducing the Feynman integral version of quantum mechanics for the situation of the description of concepts and their disjunction, we analyzed in detail how first quantum principles yield the choice of vector ${1 \over \sqrt{2}}(|A\rangle+|B\rangle)$ to model the `$A$ or $B$' concept. However, the question that we are now concerned with is: ``Does ${1 \over \sqrt{2}}(|A\rangle+|B\rangle)$ really model the `or' situation (for example in a double-slit situation), and, by analogy, the `$A$ or $B$' concept?". To find the right answer to this question, let us discuss a number of elements that point to a potential problem. The first element is that for the `classical limit situation' in quantum mechanics, i.e. the situation with no interference, the value of $\mu(A\ {\rm or}\ B)$ reduces to ${1 \over 2}(\mu(A)+\mu(B))$. This is neither $\max(\mu(A),\mu(B))$, expected from a fuzzy set perspective of the disjunction, nor $\mu(A)+\mu(B)-\mu(A)\mu(B)$, expected from a Kolmogorovian approach. Moreover, the value of $\mu(A\ {\rm or}\ B)={1 \over 2}(\mu(A)+\mu(B))$ will in general `not satisfy the inequalities that we have derived for what we have called classical disjunction data', i.e. inequalities (\ref{disjunctionineq01}), (\ref{disjunctionineq02}) and (\ref{disjunctionineq03}). However, if we consider the double-slit situation in a classical mechanics setting, and one particle -- a classical particle -- is fired with both slits open, the probability of its detection on a screen behind both slits is indeed the average of the probabilities of its detection on the same screen in case only one of the slits is open. In other words, the equation $\mu(A\ {\rm or}\ B)={1 \over 2}(\mu(A)+\mu(B))$ correctly represents the double-slit situation for a classical particle and a pair of classical slits. Furthermore, the interference taking place when a quantum particle is fired in the case of a double slit, is accounted for by the interference term $\Re\langle A|M|B\rangle$ contained in equation (\ref{interference}), where $\mu(A\ {\rm or}\ B)={1 \over 2}(\mu(A)+\mu(B))+\Re\langle A|M|B\rangle$ equals this average plus this interference term. So what is the problem? Well, there seems to be a fundamental issue that we have not yet fully understood. And, surprisingly, it is quantum field theory we have to turn to in order to shed light on the problem.

\scriptsize
\setlongtables 
\begin{longtable}{|llllllll|} 
\caption{Part of the concepts and items and their membership weights measured in of experiment 4 in Hampton (1988a). $\mu(A)$, $\mu(B)$ and $\mu(A{\rm{\ and\ }}B)$ are the membership weights of concepts $A$, $B$ and the conjunction `$A$ and $B$', respectively, for the considered item. $\Delta_c$ is the `conjunction minimum rule deviation' and $k_c$ the `Kolmogorovian factor'. We use label $c$ for a classical item and $\Delta$ for a $\Delta$-type non classical item. $|A\rangle$ and $|B\rangle$ are the vectors representing $A$ and $B$ in the $\compl^3$ quantum modeling, and $m^2$ and $n^2$ are the weight in Fock space.
} \\
\hline 
& {\it label} & $\mu(A)$ & $\mu(B)$ & $\mu(A{\rm{\ and\ }}B)$ & $\Delta_c$ & $k_c$ & $|A\rangle+|B\rangle$; $m^2\mu(A)\mu(B)$+$n^2{1 \over 2}(\mu(A)+\mu(B))$ \\ 
\endfirsthead 
\hline 
& {\it label} & $\mu(A)$ & $\mu(B)$ & $\mu(A{\rm{\ and\ }}B)$ & $\Delta_c$ & $k_c$ & $|A\rangle+|B\rangle$; $m^2\mu(A)\mu(B)$+$n^2{1 \over 2}(\mu(A)+\mu(B))$ \\
\hline \hline
\endhead 
\hline \hline  
\multicolumn{8}{|l|}{\it $A$=Furniture, $B$=Household Appliances} \\
\hline
{\it Desk Lamp} & $\Delta$ & 0.725 & 0.825 & 0.825 & 0.1 & 0.275 & (0.8515, 0, 0.5244)+$e^{i76.8253^\circ}$(0.2576, 0.8710, -0.4183) \\
{\it Coffee Table} & $\Delta$ & 1 & 0.15 & 0.3846 & 0.2346 & 0.2346 & 0.4480(0.15)+0.5520(0.575) \\
{\it Painting} & $\Delta$ & 0.6154 & 0.0513 & 0.1053 & 0.0540 & 0.4386 & 0.7558(0.0316)+0.2442(0.3333) \\
\hline
\multicolumn{8}{|l|}{\it $A$=Food, $B$=Plant} \\
\hline
{\it Peppercorn} & $\Delta$ & 0.875 & 0.6207 & 0.7586 & 0.1379 & 0.2629 & (0.9354, 0, 0.3536)+$e^{i87.1634^\circ}$(0.2328, 0.7527, -0.6159)  \\
{\it Sponge} & $\Delta$ & 0.0263 & 0.3421 & 0.0882 & 0.0619 & 0.7198 & 0.5478(0.0090)+0.4522(0.1842) \\
\hline
\multicolumn{8}{|l|}{\it $A$=Weapon, $B$=Tool} \\
\hline
{\it Toothbrush} & $c$ & 0 & 0.55 & 0 & 0 & 0.45 & 1(0)+0(0.275) \\
{\it Chisel} & $\Delta$ & 0.4 & 0.975 & 0.6410 & 0.2410 & 0.2660 & (0.6325, 0, 0.7746)+$e^{i112.3003^\circ}$(0.1936, 0.9682, -0.1581) \\
\hline
\multicolumn{8}{|l|}{\it $A$=Building, $B$=Dwelling} \\
\hline
{\it Cave} & c & 0.2821 & 0.95 & 0.2821 & 0 & 0.05 & 0.9595(0.2679)+0.0405(0.6160) \\
{\it Tree House} & c & 0.5 & 0.9 & 0.95 & -0.05 & 0.45 & (0.8771, 0, 0.4804)+$e^{i77.0244^\circ}$(0.2148, 0.8944, -0.3922)  \\
\hline
\multicolumn{8}{|l|}{\it $A$=Machine, $B$=Vehicle} \\
\hline
{\it Dogsled} & $\Delta$ & 0.1795 & 0.925 & 0.275 & 0.0955 & 0.1705 & 0.7178(0.1660)+0.2822(0.5522) \\
{\it Course liner} & $\Delta$ & 0.875 & 0.875 & 0.95 & 0.075 & 0.2 & (0.9354, 0, 0.3536)+$e^{i53.1301^\circ}$(0.1336, 0.9258, -0.3536) \\
\hline
\multicolumn{8}{|l|}{\it $A$=Bird, $B$=Pet} \\
\hline
{\it Lark} & $\Delta$ & 1 & 0.275 & 0.4872 & 0.2122 & 0.2122 & 0.4147(0.275)+0.5853(0.6375) \\
{\it Elephant} & $c$ & 0 & 0.25 & 0 & 0 & 0.75 & 1(0)+0(0.125) \\
\hline
\end{longtable}

\noindent
\normalsize

\scriptsize
\setlongtables 
\begin{longtable}{|llllllll|} 
\caption{Part of the concepts and items and their membership weights measured in experiment 2 of Hampton (1988b). $\mu(A)$, $\mu(B)$ and $\mu(A{\rm{\ or\ }}B)$ are the membership weights of concepts $A$, $B$ and the disjunction `$A$ or $B$', respectively, for the considered item. $\Delta_d$ is the `disjunction maximum rule deviation' and $k_d$ the `Kolmogorovian factor'. We use label $c$ for a classical item, $\Delta$ for a $\Delta$-type and $k$ for a $k$-type non classical item. $|A\rangle$ and $|B\rangle$ are the vectors representing $A$ and $B$ in the $\compl^3$ quantum modeling, and $m^2$ and $n^2$ are the weight in Fock space.} \\
\hline 
& {\it label} & $\mu(A)$ & $\mu(B)$ & $\mu(A{\rm{\ or\ }}B)$ & $\Delta_d$ & $k_d$ & $|A\rangle\!+\!|B\rangle$; $m^2\!(\!\mu(A)\!+\!\mu(B)\!-\!\mu(A)\mu(B)\!)$+$n^2\!{1 \over 2}(\!\mu(A)\!+\!\mu(B)\!)$ \\ 
\endfirsthead 
\hline 
& {\it label} & $\mu(A)$ & $\mu(B)$ & $\mu(A{\rm{\ or\ }}B)$ & $\Delta_d$ & $k_d$ & $|A\rangle\!+\!|B\rangle$; $m^2\!(\!\mu(A)\!+\!\mu(B)\!-\!\mu(A)\mu(B)\!)$+$n^2\!{1 \over 2}(\!\mu(A)\!+\!\mu(B)\!)$  \\
\hline \hline
\endhead 
\hline \hline  
\multicolumn{8}{|l|}{\it $A$=House Furnishings, $B$=Furniture} \\
\hline
{\it Wall-Hanging} & c & 0.9 & 0.4 & 0.95 & -0.05 & 0.35 & 1(0.94)+0(0.65) \\
{\it Door Bell} & c & 0.5 & 0.1 & 0.55 & -0.05 & 0.05 & 1(0.55)+0(0.3) \\
{\it Ashtray} & $\Delta$ & 0.7 & 0.3 & 0.25 & 0.45 & 0.75 & (0.8367, 0, 0.5477)+$e^{i123.0619^\circ}$(0.5477, 0, -0.8367) \\
{\it Sink Unit} & $\Delta$ & 0.9 & 0.6 & 0.6 & 0.3 & 0.9 & (0.9487, 0, 0.3162)+$e^{i138.5904^\circ}$(0.2108, 0.7454,-0.6325) \\
\hline
\multicolumn{8}{|l|}{\it $A$=Hobbies, $B$=Games} \\
\hline
{\it Gardening} & c & 1 & 0 & 1 & 0 & 0 & 1(1)+0(0.5) \\
{\it Beer Drinking} & $\Delta$ & 0.8 & 0.2 & 0.575 & 0.225 & 0.425 & (0.4472, 0, 0.8944)+$e^{i79.1931^\circ}$(0.8421, 0.3015, -0.4472)  \\
{\it Stamp Collection} & c & 1 & 0.1 & 1 & 0 & 0.1 & 1(1)+0(0.55) \\
{\it Wrestling} & $\Delta$ & 0.9 & 0.6 & 0.625 & 0.275 & 0.875 & (0.9487, 0, 0.3162)+$e^{i126.8699^\circ}$(0.2108, 0.7454, -0.6325)   \\
\hline
\multicolumn{8}{|l|}{\it $A$=Pets, $B$=Farmyard Animals} \\
\hline
{\it Collie Dog} & c & 1 & 0.7 & 1 & 0 & 0.7 & 1(1)+0(0.85) \\
{\it Rat} & $\Delta$ & 0.5 & 0.7 & 0.4 & 0.3 & 0.8 &  (0.7071, 0, 0.7071)+$e^{i121.0909^\circ}$(0.5477, 0.6325, -0.5477) \\
{\it Field Mouse} & $\Delta$ & 0.1 & 0.7 & 0.4 & 0.3 & 0.4 & (0.9487, 0, 0.3162)+$e^{i90^\circ}$(0.2789, 0.4714, -0.8367) \\
\hline 
\multicolumn{8}{|l|}{\it $A$=Spices, $B$=Herbs} \\
\hline
{\it Molasses} & c & 0.4 & 0.05 & 0.425 & -0.025 & 0.025 & 0.9756(0.43)+0.0244(0.225) \\
{\it Salt} & $\Delta$ & 0.75 & 0.1 & 0.6 & 0.15 & 0.25 & (0.5, 0, 0.8660)+$e^{i50.2820^\circ}$(0.5477, 0.7746, -0.3162) \\
{\it Curry} & $\Delta$ & 0.9 & 0.4 & 0.75 & 0.15 & 0.55 & (0.9487, 0, 0.3162)+$e^{i65.9052^\circ}$(0.2582, 0.5774, -0.7746) \\
{\it Parsley} & c & 0.5 & 0.9 & 0.95 & -0.05 & 0.45 & 1(0.95)+0(0.7) \\
\hline
\multicolumn{8}{|l|}{\it $A$=Instruments, $B$=Tool} \\
\hline
{\it Pencil Eraser} & $\Delta$ & 0.4 & 0.7 & 0.45 & 0.25 & 0.65 & (0.6325, 0, 0.7746)+$e^{i103.6330^\circ}$(0.6708, 0.5, -0.5477) \\
{\it Computer} & $\Delta$ & 0.6 & 0.8 & 0.6 & 0.2 & 0.8 & (0.7746, 0, 0.6325)+$e^{i110.7048^\circ}$(0.3651, 0.8165, -0.4472)\\
{\it Spoon} & $\Delta$ & 0.65 & 0.9 & 0.7 & 0.2 & 0.85 & (0.8185, 0, 0.5745)+$e^{i117.8987^\circ}$(0.2219, 0.9224, -0.3162) \\
{\it Pliers} & c & 0.8 & 1 & 1 & 0 & 0.8 & 1(1)+0(0.9) \\
\hline
\multicolumn{8}{|l|}{\it $A$=Sportswear, $B$=Sports Equipment} \\
\hline
{\it Sunglasses} & $\Delta$ & 0.4 & 0.2 & 0.1 & 0.3 & 0.5 & (0.7746, 0, 0.6325)+$e^{i135^\circ}$(0.3651, 0.8165, -0.4472) \\
{\it Golf Ball} & c & 0.1 & 1 & 1 & 0 & 0.1 & 1(1)+0(0.55) \\
{\it Sailing Life Jacket} & c & 1 & 0.8 & 1 & 0 & 0.8 & 1(1)+0(0.9) \\
{\it Tennis Racket} & c & 0.2 & 1 & 1 & 0 & 0.2 & 1(1)+0(0.6) \\
\hline
\multicolumn{8}{|l|}{\it $A$=Household Appliances, $B$=Kitchen Utensils} \\
\hline
{\it Cake Tin} & c & 0.4 & 0.7 & 0.95 & -0.25 & 0.15 & (0.7071, 0, 0.7071)+$e^{i53.1301^\circ}$(0.7071, 0, -0.7071) \\
{\it Cooking Stove} & c & 1 & 0.5 & 1 & 0 & 0.5 & 1(1)+0(0.75) \\
{\it Rubbish Bin} & c & 0.5 & 0.5 & 0.8 & -0.3 & 0.2 & (0.6325, 0, 0.7746)+$e^{i19.4712^\circ}$(0.6708, 0.5, -0.5477) \\
{\it Spatula} & c & 0.55 & 0.9 & 0.95 & -0.05 & 0.5 & 0.9783(0.955)+0.0217(0.725) \\
\hline
\multicolumn{8}{|l|}{\it $A$=Fruits, $B$=Vegetables} \\
\hline
{\it Apple} & c & 1 & 0 & 1 & 0 & 0 & 1(1)+0(0.5) \\
{\it Chili Pepper} & c & 0.05 & 0.5 & 0.5 & 0 & 0.05 & 0.9(0.525)+0.1(0.275) \\
{\it Raisin} & $\Delta$ & 1 & 0 & 0.9 & 0.1 & 0.1 & 0.8(1)+0.2(0.5) \\
{\it Tomato} & c & 0.7 & 0.7 & 1 & -0.3 & 0.4 & (0.7348, 0, 0.6782)+$e^{i121.8967^\circ}$(0.6052, 0.4513, -0.6557) \\
{\it Peanut} & c & 0.3 & 0.1 & 0.4 & -0.1 & 0 & 1(0.37)+0(0.2) \\
{\it Elderberry} & $\Delta$ & 1 & 0 & 0.8 & 0.2 & 0.2 & 0.6(1)+0.4(0.5) \\
\hline
\end{longtable}

\noindent
\normalsize
What we have not yet considered is the fact that in quantum mechanics and in classical mechanics, as well as in the situation of the concepts considered above, there are fundamentally two deeply differing cases involved with respect to `a situation of disjunction'. We can gain further insight into the root of these two cases by considering them in everyday language. Let us do this for concepts. The situation is that of `an item $X$' and a pair of concepts $A$ and $B$, and a question about membership of this item $X$ with respect to $A$ `or' $B$. There are two ways we can approach this situation. The first is to ask ourselves whether $X$ is a member of the concept `$A$ or $B$'. The second is to ask ourselves whether this item $X$ is a member of $A$ `or' whether this item $X$ is a member of $B$. In both approaches the possible outcomes are `yes' or `no' with respect to the membership question of an item, and hence probabilities exist with respect to these outcomes. However, these are two fundamentally different ways of approach, yielding different probabilities for the outcomes `yes' and `no'. Before we proceed, we will introduce names for these two ways, and subsequently explain why we have chosen these names. We call the first the `one-particle way' and the second the `two-particle way'. These two ways exist also for the double-slit situation, as we will show. The `one-particle way' for the double-slit situation consists in firing one particle at the two open slits, electing a spot $X$ on the screen behind the two slits and considering the probability that the particle hits this spot $X$. This probability is the average of the probabilities that the particle hits the spot in case only one of both slits is open. The `two-particle way' for the double-slit situation consists in firing two identical particles at the two slits, such that one particle passes through one slit and the other particle passes through the other slit, and considering the probability that one of these particles hits spot $X$. If the probability of detection at spot $X$ for one particle passing through one slit is $\mu(A)$, and the probability of detection at spot $X$ for the other particle passing through the other slit is $\mu(B)$, then the probability of detection of one of these particles at spot $X$ is $\mu(A)+\mu(B)-\mu(A)\mu(B)$. Indeed, the probability that one of the particles, the one passing through slit $A$, is `not' detected at spot $X$ is given by $1-\mu(A)$, and the probability that the other identical particle, the one passing through the other slit $B$, is `not' detected at spot $X$ is given by $1-\mu(B)$. Hence, the probability that `no particle is detected at spot $X$ is given by $(1-\mu(A))(1-\mu(B)$. This means that the probability that (at least) one particle is detected at spot $X$ is given by $1-(1-\mu(A))(1-\mu(B)=\mu(A)+\mu(B)-\mu(A)\mu(B)$. This is a very different formula for $\mu(A\ {\rm or}\ B)$ than the average ${1 \over 2}(\mu(A)+\mu(B))$, which should come as no surprise since it describes a very different type of `or' situation. 

One of the ways to consider the difference between both `or' situations is that the first involves `one particle' or, in the case of concepts, `one item', while the second involves `two identical particles' or, in the case of concepts, `two identical items'. All other aspects of both situations are very similar. Two slits are involved in the double-slit situation, and two concepts are involved in the concept situation. Hence the difference is determined by what happens with the particles in passing through the slits. Either there is one particle passing through one of both slits or there are two identical particles such that each passes through one of the slits. Although structurally identical, the difference is a lot more subtle in the case of concepts. Indeed, for the double-slit and the `one-particle way' or the `two-particle way', one can, anyhow if it is about classical particles, distinguish, in the sense that in one case there is only one particle involved, while in the other case there are two particles involved. In the concept situation, `one item' is involved in the case of the `one-particle way' and `two identical items' are involved in the case of the `two-particle way'. To identify this difference within a human thought-process is much more complicated, however, also since `the making of an identical item starting from a considered item within human thought is a process taking place all the time'.

Let us consider the concrete situation of a subject participating in one of the experiments described in Hampton (1988b). We can reflect intuitively on the `human thought process' taking place in his or her mind and imagine that both ways can take place. 
We will illustrate this by means of an example. Suppose a subject is asked to answer with `yes' or `no' the question whether {\it Almond} is a member of {\it Fruits or Vegetables}. Following only the first way -- the `one-particle way' --, the subject would consider `{\it Fruits or Vegetables}' a wholly new concept and answer the question of whether or not {\it Almond} is a member of this new concept `{\it Fruits or Vegetable}'. Following only the second way -- the `two-particle way' --, the subject would consider two identical items {\it Almond} in turn, deciding for the one whether it is a member of {\it Fruits} and for the other whether it is a member of {\it Vegetables}. While doing so the subject makes a complicated confrontation of these two decision possibilities with the notion of `or' in the background and decides in this way about `yes' or `no' with respect to membership for the item {\it Almond}. This complicated confrontation is what has been organized in truth tables by logicians within their discipline called `logic', because indeed it comes to deciding `yes' for membership in case `yes, yes', `yes, no' or `no, yes' would have been decided for membership with respect to the concepts apart.

 What Hampton (1988b) experiments show is that `both ways take place'. After carefully analyzing the experimental data, which we will do in subsection \ref{Fockspace}, we conclude that these two ways occur in superposition. The `superposition' of both ways, the one-particle way and the two-particle way, is exactly what quantum-fields theory offers to meet our modeling need, and the mathematical space it uses for this is Fock space. That is why in Aerts (2007a,b) we elaborated a Fock space model for the disjunction of two concepts.

The following examples taken from the Hampton experiments further illustrate the mechanism that we propose here. They also serve to help explain what we mean by our hypothesis that `human thought comprises two superposed layers'. Let us consider the item {\it Apple} with respect to the pair of concepts {\it Fruits} and {\it Vegetables} and their disjunction {\it Fruits or Vegetables}. A subject following the `two-particle way' will go about answering the question more or less as follows: ``An apple is certainly a fruit and it is definitely not a vegetable". In this part of the thought process, the subject splits up the item {\it Apple} into two identical items, confronting them with the concepts {\it Fruits} or {\it Vegetables}, respectively. The subject's thought process then continues: ``But since it is a fruit, it must necessarily also be a `fruit or a vegetable'". In this part of the thought process, the subject follows a logical line of reasoning, combining the two confrontations with the concepts. Let us now suppose that he the subject follows the `one-particle way' for resolving the {\it Apple} and `{\it Fruits or Vegetables}' question. In this case, the subject would consider `{\it Fruits or Vegetables}' to be a wholly new concept and determine whether or not {\it Apple} is a member of this new concept. {\it Apple} being a very archetypical {\it Fruit}, it is very unlikely for the subject to regard {\it Apple} as a strong member of the new {\it Fruit or Vegetable} concept, because this is an overall concept for all items belonging to {\it Fruits} or {\it Vegetables}. In other words, while in this situation the `two-particle way' should yield $\mu(A\ {\rm or}\ B)=1$ for this situation, the `one-particle way' should yield $\mu(A\ {\rm or}\ B)={1 \over 2}$. The experiments described in Hampton (1988b) concerning {\it Apple} with respect to {\it Fruits}, {\it Vegetables} and {\it Fruits or Vegetables} give membership weights $\mu(A)=1$, $\mu(B)=0$ and $\mu(A\ {\rm or}\ B)=1$. This shows that this particular comparison between {\it Apple} and `{\it Fruits or Vegetables}' is very much dominated by the two-particle way. 

Let us now return to the example of {\it Almond}. The experimental data about membership weights of {\it Almond} with respect to the pair of concepts {\it Fruits} and {\it Vegetables} and their disjunction described in Hampton (1988b) are $\mu(A)=0.2$, $\mu(B)=0.1$ and $\mu(A\ {\rm or}\ B)=0.425$, respectively. Hence $k_d=\mu(A)+\mu(B)-\mu(A\ {\rm or}\ B)=-0.125\le0$, which shows that {\it Almond} is a $k$-type non-classical item. We can see that, for {\it Almond}, the second way, the `one-particle way', is strongly dominant. While {\it Almond} was assigned hardly any weight as a member of both {\it Fruits} and {\it Vegetables}, it was assigned considerable weight as a member `{\it Fruits or Vegetables}'. Apparently, {\it Almond} was regarded as one of those items that raise doubts as to whether they are {\it Fruits} or {\it Vegetables}, neither of the two categories offering a satisfactory typification individually. This makes it a fairly good member of the new concept `{\it Fruits or Vegetables}'.

Analogous to `one-particle way' and `two-particle way', we will use the terms `one train of thought way' and `two trains of thought way' in dealing with concepts. Indeed, one can imagine the `two-particle way' as two parallel trains of thought taking place in the subject's mind. One train of thought is aimed at deciding about the membership of item $X$ with respect to concept $A$ and the second train of thought is aimed at deciding about the membership weight of item $X$ with respect to concept $B$. The two trains of thought take place in parallel, and either of them confirming membership of item $X$ with respect to one of the concepts, say concept $A$, is sufficient for membership of item $X$ with respect to `$A$ or $B$' to be confirmed as well. Once one of the trains of thought has established membership, the outcome of the other train of thought has become irrelevant. Only if both trains of thought deny membership with respect to both concepts, will membership with respect to the disjunction be denied as well. Since this way of reasoning is in line with classical logic with respect to the disjunction of two propositions, we have called it the `classical logical way'. If this is the `only' way of reasoning in the subject's mind, the data will come out as classical data, i.e. data that can be modeled within a classical measure or probability structure. An item for which the `classical logical' way is very dominant will experimentally show itself as what we have called a classical item. {\it Apple} with respect to {\it Fruits}, {\it Vegetables} and `{\it Fruits or vegetables}' is such an example. What we have called the `one-particle way' or `one train of thought way' introduces a way of thought that is very different from this classical logical thought. We will call it `quantum conceptual thought'. A subject who follows only the quantum conceptual thought way, focuses on one and only one train of thought with respect to `$A$ or $B$', and hence directly wonders whether the item $X$ is a member or not a member of `$A$ or $B$'. In `quantum conceptual thought' the subject considers $A$ or $B$ as a new concept, hence the emergence of the concept `$A$ or $B$'. Quantum mechanics as a mathematical formalism is suited to describe `quantum conceptual thought' since the mathematical operation of `superposition' produces a `new state with new features', and classical theories do not entail such a possibility.

The distinction of two modes of thought has been proposed by many and in many different ways. Sigmund Freud, in his seminal work `The interpretation of dreams', already proposed to consider thought as consisting of two processes, which he called primary and secondary (Freud, 1899) and were to become popularly known as the conscious and the subconscious. William James subsequently introduced the idea of `two legs of thought', specifying the one as `conceptual', i.e. exclusive, static, classical and following the rules of logic, and the other as `perceptual', i.e. intuitive and penetrating. He expressed the opinion that `just as we need two legs to walk, we also need both conceptual and perceptual modes to think' (James, 1910). More recently, Jerome Bruner introduced the `paradigmatic mode of thought', transcending particularities to achieve systematic categorical cognition where propositions are linked by logical operators, and the `narrative mode thought', engaging in sequential, action-oriented, detail-driven thought, where thinking takes the form of stories and `gripping drama' (Bruner, 1990). Aspects of different modes of thought and the influence of their presence on human cognitive evolution were also proposed in Gabora and Aerts (2009). Another point that we think is relevant to our structure of two superposed layers -- subsection \ref{decisiontheoryeconomics} explains why -- is how Tversky \& Kahneman (1982) introduced the notion of `representative heuristic' to describe the decision process that lies at the basis of what happens during `disjunction effect situations' or `conjunction fallacy situations' and other similar situations. We believe that all these examples, and others, are related but also different from the superposed layers we introduce in this paper. We intend to dedicate future research to these relations and differences, continuing along the lines of Aerts and D'Hooghe (2009).

\subsection{Fock Space and What About Conjunction}
\label{Fockspace}

As said, our general scheme makes use, next to `superposition' and `interference', of a `quantum field theoretic aspect'. In quantum field theory the entity that is described, i.e. the field, consists of superpositions of different configurations of many quantum particles. This is why the mathematical space used to describe this quantum field is a Fock space, which is a direct sum -- this is the superposition part -- of different Hilbert spaces, where each Hilbert space represents a certain number of quantum particles. In the case where we consider two concepts $A$ and $B$, a field theoretic model consists of the direct sum of two Hilbert spaces. One Hilbert space ${\cal H}$ describes the `one-particle situation', i.e. the situation where `$A$ or $B$' is considered to be a new concept, and its state in this Hilbert space is ${1 \over \sqrt{2}}(|A\rangle+|B\rangle)$, while the experiment to determine membership of an item $X$ is described by the orthogonal projection $M$. It is the description that we have presented in section \ref{QuantumModelingScheme}.

A second Hilbert space describes the `two-particle way', which is the tensor product ${\cal H} \otimes {\cal H}$. The state of the concepts $A$ and $B$ is represented by a vector $|A\rangle \otimes |B\rangle$ of this tensor product Hilbert space. To know how to describe the `decision measurement' for this `two-particle way' quantum-mechanically, let us analyze in detail the decision process. We will do this by modeling the situation where a subject is asked to decide on the membership of an item $X$ with respect to concept `$A$ or $B$'. To answer this question, the subject will consider `two identical items $X$', pondering on the membership of one of the two identical items $X$ with respect to $A$ `and' the membership of the other one of the two identical items $X$ with respect to $B$. The outcome will therefore be one of the following answers: (i) `yes, yes', which means that the subject decides that one of the items $X$ is a member of $A$ and the other identical item $X$ is a member of $B$; (ii) `yes, no', which means that the subject decides that one of the items $X$ is not a member of $A$ and the other is a member of $B$; (iii) `no, yes', which means that the subject decides that one of the items $X$ is not a member of $A$ and the other is a member of $B$; and finally (iv) `no, no', which means that the subject decides that one of the items $X$ is not a member of $A$ and the other item $X$ is not a member of $B$. The subject will affirm membership of $X$ with respect to `$A$ or $B$' if the outcome is `yes, yes', `yes, no' or `no, yes'.
The decision experiment with respect to membership of item $X$ is described by the orthogonal projection $M \otimes M$, which is a linear operator on the tensor product Hilbert space ${\cal H} \otimes {\cal H}$. Following the above analysis, membership weight for the disjunction is given by
\begin{equation}
\mu(A\ {\rm or}\ B)=1-(\langle A|\otimes \langle B|)(1-M) \otimes (1-M)|(|A\rangle \otimes |B\rangle)
\end{equation}
where indeed a `yes' for the disjunction means that one of the outcomes `yes, yes', `yes, no' or `no, yes' is obtained, which is equivalent to the outcome `no, no' not being obtained.

In sections \ref{QuantumModelingScheme} and \ref{QuantumFieldTheory} we discussed the disjunction of two concepts $A$ and $B$ at great length. We will now look at the conjunction of two concepts $A$ and $B$. Now that we have understood that ${1 \over \sqrt{2}}(|A\rangle+|B\rangle)$ represents a new concept `$A$ or $B$', and not the logical construction `concept $A$' or `concept $B$', it may well be that with the conjunction also correspond a `one-particle way' and a `two-particle way'. There is yet another fact that points to this. Note, for example, that `underextension', which is the commonest effect measured by Hampton (1988b) for disjunction, produces a value for $\mu(A\ {\rm or}\ B)$ that deviates from $\max(\mu(A),\mu(B))$ `in the direction of the average ${1 \over 2}(\mu(A)+\mu(B))$'. More specifically, the average ${1 \over 2}(\mu(A)+\mu(B))$ is in general `underextended' itself. Overextension, which is the commonest effect measured by Hampton (1988a) for conjunction, produces a value for $\mu(A\ {\rm and}\ B)$ that deviates from $\min(\mu(A),\mu(B))$ `also in the direction of the average ${1 \over 2}(\mu(A)+\mu(B))$'. And the average ${1 \over 2}(\mu(A)+\mu(B))$ is in general `overextended' itself. This suggests that `underextension for the disjunction' and `overextension for the conjunction' could well be caused by the same effect. And indeed, in our explanatory scheme, they are caused by the same effect, namely by the effect of the presence of the emergence of a new concept. In the case of the conjunction, this is the concept `$A$ and $B$'. We can now explain how we have modeled the conjunction data in Table 1, i.e. using the same quantum description, for example the one worked out in $\compl^3$ for the disjunction. We refer to Aerts (2007b) for a detailed elaboration of this modeling.

Finally we have all the material to explain the modeling in Fock space, and, therefore, the motivation for the simple quantum model presented in section \ref{simplemodel}. Fock space ${\cal F}$ is the direct sum of both Hilbert spaces, the `two-particle way' Hilbert space and the `one-particle way' Hilbert space, hence ${\cal F}=({\cal H} \otimes {\cal H}) \oplus {\cal H}$, and the state $\psi(A,B)$ of the concepts $A$ and $B$ in Fock space is described by a normalized linear combination of the `two-particle state' and the `one-particle state'. Concretely, this leads to
\begin{eqnarray}
\psi(A,B)=me^{i\theta}|A\rangle \otimes |B\rangle + {ne^{i\phi}\over \sqrt{2}}(|A\rangle+|B\rangle) \quad m^2+n^2=1
\end{eqnarray}
The experiment testing whether an item $X$ `is' or `is not' a member of the concept `$A$ and $B$' is described by the projection operator $M \otimes M \oplus M$ working on this Fock space. The probability that the outcome is `yes', i.e. that a subject decides in favor of membership of the item $X$ with respect to the concept `$A$ and $B$', is given by
\begin{eqnarray}
\mu(A\ {\rm and}\ B)&=& (me^{i\theta}\langle A| \otimes \langle B| + {ne^{i\phi}\over \sqrt{2}}(\langle A|+\langle B|))M \otimes M \oplus M(me^{i\theta}|A\rangle \otimes |B\rangle + {ne^{i\phi}\over \sqrt{2}}(|A\rangle+|B\rangle))\\
&=&m^2(\langle A| \otimes \langle B|)M \otimes M|(|A\rangle \otimes |B\rangle)+{n^2 \over 2}(\langle A|+\langle B|)M(|A\rangle+|B\rangle) \\
&=&m^2\langle A|M|A\rangle \langle B|M|B\rangle+{n^2 \over 2}(\langle A|M|A\rangle+\langle B|M|B\rangle+\langle A|M|B\rangle+\langle B|M|A\rangle) \\ \label{fockspaceconjunction}
&=&m^2\mu(A)\mu(B)+n^2({\mu(A)+\mu(B) \over 2}+\Re\langle A|M|B\rangle)
\end{eqnarray}
Let us mention that our quantum modeling of the conjunction differs from the quantum model of the conjunction presented by Franco (2007), in that he employs conditional probability and we make use of how the conjunction appears in Fock space. Let us construct the Fock space probability for the disjunction now.

The probability $\mu(A\ {\rm or}\ B)$ that a subject decides in favor of membership of the item $X$ with respect to the concept `$A$ or $B$' is given by 1 minus the probability of a decision against membership of the item $X$ with respect to the concept `$A$ and $B$'. This means that
\begin{eqnarray}
\mu(A\ {\rm or}\ B)&=&1-m^2(1-\mu(A))(1-\mu(B)-n^2({1-\mu(A)+1-\mu(B) \over 2}+\Re\langle A|1-M|B\rangle) \\
&=&m^2+n^2- m^2(1-\mu(A)-\mu(B)+\mu(A)\mu(B))-n^2({2-\mu(A)-\mu(B) \over 2}-\Re\langle A|M|B\rangle) \\ \label{fockspacedisjunction}
&=&m^2(\mu(A)+\mu(B)-\mu(A)\mu(B))+n^2({\mu(A)+\mu(B) \over 2}+\Re\langle A|M|B\rangle)
\end{eqnarray}
By means of equations (\ref{fockspaceconjunction}) and (\ref{fockspacedisjunction}) we can show how the experimental data of Hampton (1988a,b) make a strong case for the introduction of Fock space as a generalization of a single Hilbert space. For example, consider the item {\it Cave} with respect to the pair of concepts {\it Building} and {\it Dwelling} and their conjunction {\it Building and Dwelling}. As we can see in Table 1, we have $\mu(A)=0.2821$ and $\mu(B)=0.95$ and $\mu(A\ {\rm and}\ B)=0.2821$, which shows that {\it Cave} is a classical item. But it cannot be modeled by means of quantum interference only, i.e. in the `one-particle way'. We can understand why quantum interference alone cannot model {\it Cave}. The `one-particle way' produces deviations from the average ${1 \over 2}(\mu(A)+\mu(B))$ due to interference. However, if one of the weights $\mu(A)$ or $\mu(B)$ is close to zero, and the other is close to 1, like in the case of {\it Cave}, the average is a number far removed from $\mu(A)$ and from $\mu(B)$. The size of the interference is proportional to the smallest term of $\sqrt{\mu(A)\mu(B)}$ or $\sqrt{(1-\mu(A))(1-\mu(B)}$, which means that interference is small if $\mu(A)$ and $\mu(B)$ are close to 0 or 1. If then additionally $\mu(A\ {\rm and}\ B)$ is close to 0 or 1, the value of  $\mu(A\ {\rm and}\ B)$ will not be reached by adding or subtracting the interference from the average. This is what happens for {\it Cave} where $\mu(A)$ and $\mu(A\ {\rm and}\ B)$ are both close to zero, and $\mu(B)$ is close to 1. If the item is classical, and one of the weights is close to zero and the other is close to 1, the weight of the conjunction will be close to the smallest of $\mu(A)$ and $\mu(B)$, resulting in the situation which does not allow modeling by quantum interference. Table 1 gives examples of items in this situation for the conjunction, more specifically {\it Coffee Table} and {\it Painting} for to the pair of concepts {\it Furniture} and {\it Household Appliances}; {\it Sponge} for the pair of concepts {\it Food} and {\it Plant}; {\it Toothbrush} for the pair of concepts {\it Weapon} and {\it Tool}; {\it Cave} for the pair of concepts {\it Building} and {\it Dwelling}; {\it Dogsled} for the pair of concepts {\it Machine} and {\it Vehicle} and {\it Lark} and {\it Elephant} for the pair of concepts {\it Bird} and {\it Pet}, and Table 2 shows a list of items where for the disjunction no quantum model with only interference exist, more specifically {\it Wall-Hanging} and {\it Door Bell} for the pair of concepts {\it House Furnishings} and {\it Furniture}; {\it Gardening} and {\it Stamp Collection} for the pair of concepts {\it Hobbies} and {\it Games}; {\it Collie Dog} for the pair of concepts {\it Pets} and {\it Farmyard Animals}; {\it Molasses} and {\it Parsley} for the pair of concepts {\it Spices} and {\it Herbs}; {\it Pliers} for the pair of concepts {\it Instruments} and {\it Tools}; {\it Golf Ball}, {\it Sailing Life Jacket} and {\it Tennis Racket} for the pair of concepts {\it Sportswear} and {\it Sports Equipment}; {\it Cooking Stove} and {\it Spatula} for the pair of concepts {\it Household Appliances} and{\it Kitchen Utensils}; {\it Apple}, {\it Chili Pepper}, {\it Raisin}, {\it Peanut} and {\it Elderberry} for the pair of concepts {\it Fruits} and {\it Vegetables}. The others items in Table 1 and Table 2 are presented there as quantum modeled with interference alone by making use of the $\compl^3$ model of section \ref{QuantumModelingScheme}.

The items that don't allow a quantum model with interference alone can usually be modeled in Fock space with equations (\ref{fockspaceconjunction}) and (\ref{fockspacedisjunction}) for conjunction and disjunction. For example, consider equation (\ref{fockspaceconjunction}) and suppose that we neglect the interference term. Then we have $\mu(A\ {\rm and}\ B)=m^2\mu(A)\mu(B)+n^2{1 \over 2}(\mu(A)+\mu(B))$, which is the convex combination of the product and the average of both weights $\mu(A)$ and $\mu(B)$. Table 1 shows how each one of the items that cannot be modeled by quantum interference alone for the conjunction can be modeled in Fock space, and it gives the convex weights $m^2$ and $n^2$ for each item. In an analogous way, we show how the items that cannot be modeled by quantum interference alone for the disjunction are modeled in Fock space by giving the value of the convex weights $m^2$ and $n^2$ using equation (\ref{fockspacedisjunction}) with zero interference. This allows us to put forward the following statement.

\bigskip
\noindent
{\bf Motivation for Fock space:} {\it Classical items with one of their weights close to 0 and the other close to 1 cannot in general be modeled by a quantum model where only quantum interference is introduced as quantum effect. They can usually be modeled in Fock space.}

\subsection{Application to Decision Theory, Economics and Other Domains} \label{decisiontheoryeconomics}
In this section we will analyze the relevance of the general quantum modeling scheme to decision theory and economics, explaining how our hypothesis of the superposed layered structure of classical logical and quantum conceptual thought can shed light on situations there. Next to the modeling within our quantum modeling scheme, we will also suggest, and attempt to prove, that our concept analysis can help explain the effects that we will consider in these domains. We will also mention other domains of science where we have identified connections and analogies, such as semantic analysis and artificial intelligence, and point out domains where other scientists have applied quantum structures.

The disjunction effect in decision theory is an example of a situation that can be described in our general quantum modeling scheme (Tversky \& Shafir,1992). This effect, in its game theoretic version, has already been modeled successfully within a quantum game theoretic scheme in Busemeyer, Matthew and Wang (2006). Let us shortly describe how it is modeled within our scheme. The disjunction effect occurs when test subjects prefer option $x$ (to $y$) if they know that both event $A$ and event $B$ will occur (usually $B$ is taken to be not $A$, but that is not essential), but refuse option $x$ (or prefer option $y$) if they know neither whether $A$ will occur nor whether $B$ will occur. The best-known example of this disjunction effect is the so-called {\it Hawaii problem}, which also originated its analysis, and which is about the following two situations (Tversky \& Shafir,1992).

{\it Disjunctive version:}
Imagine that you have just taken a tough qualifying examination. It is the end of the fall quarter, you feel tired and run-down, and you are not sure that you passed the exam. In case you failed you have to take the exam again in a couple of months after the Christmas holidays. You now have an opportunity to buy a very attractive 5-day Christmas vacation package to Hawaii at an exceptionally low price. The special offer expires tomorrow, while the exam grade will not be available until the following day. Would you: $x$ buy the vacation package; $y$ not buy the vacation package; $z$ pay a \$5 non-refundable fee in order to retain the rights to buy the vacation package at the same exceptional price the day after tomorrow after you find out whether or not you passed the exam.

{\it Pass/fail version:}
Imagine that you have just taken a tough qualifying examination. It is the end of the fall quarter, you feel tired and run-down, and you find out that you passed the exam (failed the exam. You will have to take it again in a couple of month after the Christmas holidays). You now have an opportunity to buy a very attractive 5-day Christmas vacation package to Hawaii at an exceptionally low price. The special offer expires tomorrow. Would you: $x$ buy the vacation package; $y$ not buy the vacation package: $z$ pay a \$5 non-refundable fee in order to retain the rights to buy the vacation package at the same exceptional price the day after tomorrow.

In the Hawaii problem, more than half of the subjects chose option $x$ (buy the vacation package) if they knew the outcome of the exam (54\% in the pass condition and 57\% in the fail condition), but only 32\% did so if they did not know the outcome of the exam. Let us explicitly construct the quantum model for the Hawaii problem, and use the $\compl^3$ realization that we introduced in subsection \ref{QuantumModelingScheme}. The states $|A\rangle$ and $|B\rangle$ of the Hilbert space now represent situation $A$, `having passed the exam', and situation $B$, `not having passed the exam'. The decision to be taken is `to buy the vacation package' or `not buy the vacation package'. This decision is described in the Hilbert space by means of projection operator $M$. The probability for an outcome `yes' (buy the package) in the `pass' situation (state $|A\rangle$) is $0.54$, i.e. in our notation $\mu(A)=0.54$. The probability for an outcome `yes' (buy the package) in the `fail' situation (state $|B\rangle$) is $0.57$, i.e. in our notation $\mu(B)=0.57$. The probability for an outcome `yes' (buy the package) in the `outcome unknown' situation (state ${1 \over \sqrt{2}}(|A\rangle+|B\rangle)$) is $0.32$, i.e. in our notation $\mu(A\ {\rm or}\ B)=0.32$. Applying the $\compl^3$ model introduced in \ref{QuantumModelingScheme}, we have to verify first whether $\mu(A)+\mu(B)=1.11$ is smaller or greater than 1, and since it is greater, we put $a=0.54$, $b=0.57$ and apply equations (\ref{vectorA}), (\ref{vectorB}) and (\ref{anglebeta}) to find
\begin{equation}
|A\rangle=(0.7348, 0, 0.6782) \quad |B\rangle=e^{i\beta}(0.6052, 0.4513, -0.6557) \quad \beta=121.8967^\circ 
\end{equation}
which completes the $\compl^3$ quantum model. We can indeed verify that, using these values for $|A\rangle$, $|B\rangle$ and $\theta$, and choosing $M(\compl^3)$ as the plane spanned by $(1, 0, 0)$ and $(0, 1, 0)$, i.e. $M$ being the orthogonal projector on this plane, we get
\begin{eqnarray}
\mu(A)&=&\langle A|M|A\rangle=(0.7348)^2=0.54 \\
\mu(B)&=&\langle B|M|B\rangle=e^{-i\beta}e^{i\beta}((0.6052)^2+(0.4513)^2)=(0.6052)^2+(0.4513)^2=0.57 \\
\mu(A\ {\rm or}\ B)&=&{1 \over 2}(\langle A|+\langle B|)M(|A\rangle+|B\rangle)={1 \over 2}(\mu(A)+\mu(B))+\langle A|M|B\rangle+\langle B|M|A\rangle) \\
&=&{1 \over 2}(0.54+0.57+(0.7348)(0.6052)e^{i121.8967^\circ}+(0.6052)(0.7348)e^{-i121.8967^\circ}) \\
&=&0.555+0.4447\cos121.8967^\circ=0.555-0.235=0.32
\end{eqnarray}
which are the values found experimentally in the Hawaii experiment. 

Our quantum modeling scheme enables us to explain this Hawaii problem as follows. The whole landscape of concepts and their combinations play a role in determining the weights that are connected with the decision process. If we would model this conceptual landscape within Fock space, we would be able to calculate the weights as predicted by quantum calculation in Fock space. Of course, constructing the Fock space states of this conceptual landscape requires carrying out a large number of experiments to analyze the situation and gather sufficient knowledge on the preparation of such states. To illustrate this, let us suppose that the Hawaii dilemma is not about buying or not buying a vacation package but about saying yes or no to a free Sauna and Jacuzzi relaxation weekend. In this case, the outcome may well be the opposite, with a larger than expected number of subjects saying yes even without knowing their exam results. We believe that in the Hawaii situation, the underextension is due to the fact that the idea of making a trip is appealing only if the outcome of the results is known, whether pass or fail, whereas a free relaxation weekend is much more likely to be appealing even if the exam results have not been released yet. This situation is very similar to, for example, {\it Almond} being an overextended item with respect to {\it Fruits or Vegetables}, because it fits well with the uncertainty `whether it is a Fruit or a Vegetable'. This kind of reflections is discussed in greater detail in Aerts and D'Hooghe (2009).

Busemeyer, Matthew and Wang (2006) model the game theoretic variant of the disjunction effect on a quantum game theoretic model, and, in a very interesting way, use the Schr\"odinger equation to describe the dynamics of the decision process. Their model is part of a general operational approach of comparing classical stochastic models with quantum dynamic models, and deciding by comparison with experimental data which of both models has most predictive power (Busemeyer, Wang \& Townsend, 2006). Also Andrei Khrennikov has presented a quantum model for decision making (Khrennikov, 2008). By means of an algorithm probabilistic data of any origin are represented by complex probability amplitudes. The algorithm is used to model the Prisoners Dilemma and the disjunction effect.

The model of the Hawaii situation which we present here is very similar -- at least with regard to the structure of state space, i.e. Hilbert space -- to what can be found in Busemeyer, Matthew and Wang (2006) and in Khrennikov (20080, since we use here only the `one-particle way'. The situation changes, however, if we add the `two-particle way' for the disjunction, working in Fock space, like we did for our `disjunction of concepts model' in Aerts (2007a,b) and in the present article. Detailed testing with many different alternatives in choice for the `disjunction effect experimental situation' is necessary to find out whether a Fock space model is needed. If the type of explanation for the effect that we put forward here, namely that the whole conceptual landscape plays a role, is correct, most probably a Fock space model will be needed to complement the single Hilbert space model if different experimental data pertaining to the same disjunction effect situation are attempted to be modeled. There are some recent experiments on the disjunction effect that point in this direction (Bagassi \& Macchi, 2007), and we have analyzed them in some detail in Aerts and D'Hooghe (2009).

Economics is another domain of science where we can identify situations involving effects that can be described using our quantum modeling scheme. Here are some examples. One of the basic principles of economics, playing a fundamental role in the von Neumann-Morgenstern theory, for instance, is Savage's `sure thing principle' (Savage, 1944). Savage introduced this principle in the following story: {\it A businessman contemplates buying a certain piece of property. He considers the outcome of the next presidential election relevant. So, to clarify the matter to himself, he asks whether he would buy if he knew that the Democratic candidate were going to win, and decides that he would. Similarly, he considers whether he would buy if he knew that the Republican candidate were going to win, and again finds that he would. Seeing that he would buy in either event, he decides that he should buy, even though he does not know which event obtains, or will obtain, as we would ordinarily say.}

The `sure thing principle' is equivalent to the independence axiom of expected utility theory: `independence' here means that if a person is indifferent in his or her choice between simple lotteries $L_1$ and $L_2$, he or she will also be indifferent in choosing between $L_1$ mixed with an arbitrary simple lottery $L_3$ with probability $p$ and $L_2$ mixed with $L_3$ with the same probability $p$.
Problems in economics such as the Allais paradox (Allais, 1953) and the Ellsberg paradox (Ellsberg, 1961) show an inconsistency with the predictions of the expected utility hypothesis, namely a violation of the `sure thing principle'.

We can readily identify an item for a pair of concepts measured by Hampton where the `sure thing principle' is violated if we would transpose it to the situation of concepts and decision for membership of items. For example, let us consider the item {\it Diving Mask} with respect to the pair of concepts {\it Sportswear} and {\it Sports Equipment} and their disjunction {\it Sportswear or Sports Equipment}. Hampton (1988b) measured $\mu(A)=1$, $\mu(B)=1$ and $\mu(A\ {\rm or}\ B)=0.95$. These values for $\mu(A)$, $\mu(B)$ and $\mu(A\ {\rm or}\ B)$ are modeled by our quantum model, the vector is given in Table 3, and equals $x=(0.0664, 0, 0, 0, 0.9978, 0, 0, 0)$, with angles $\theta=107^\circ$ and $\phi=12.95^\circ$. Hence, we can also use our quantum model to model these values if they are collected experimentally in the situation of the `sure thing principle'. More concretely for the situation originally proposed by Savage: (i) the businessman will buy the property with certainty if the Democratic candidate were to win, i.e. $\mu(A)=1$; (ii) the businessman will buy the property with certainty if the Republican candidate were to win, i.e. $\mu(B)=1$; (iii) the businessman will buy the property `not' with certainty if he does not know the outcome of the elections, i.e. if the situation is such that the Democratic `or' the Republican candidate were to win, i.e. $\mu(A\ {\rm or}\ B)=0.95$. Our quantum model also allows modeling of much lower values than $0.95$ for $\mu(A\ {\rm or}\ B)$ with $\mu(A)=\mu(B)=1$ for an experimentally tested `sure thing principle situation' to be modeled.

We have investigated correspondences between our approach and existing cognitive science models of knowledge representation. In this way, we were able to prove that modern approaches to semantic analysis (Deerwester, Dumais \& Harshman, 1990; Berry, Dumais \& O'Brien, 1995; Lund \& Burgess, 1996), if reformulated as Hilbert-space problems, reveal quantum structures similar to those we employ in our quantum modeling scheme (Aerts and Czachor, 2004), and we have worked out a concrete comparison with Latent Semantic Analysis (Landauer, Foltz \& Laham, 1998). In Aerts and Czachor (2004), we could prove a similar correspondence with distributed representations of cognitive structures developed for the purposes of neural networks (Smolensky, 1990).

After connecting our approach to ongoing semantic analysis and artificial intelligence investigations, we also looked into its relation with other cognitive science approaches. More specifically, G\"ardenfors (2004) comes to mind. Although there are certainly interesting connections to point out, and we plan to do so concretely in future work, there are also quite some differences between the foundations of his approach and ours. We already said that the modeling of linguistic structures is not our primary concern, since we focus on the semantic content, on the `flow and interaction of meaning'. There is, however, something more to be said on this point. Although modeling, and specifically the modeling of large collections of data, is the ultimate test for a theory, we must admit that `modeling' is not our primary concern either. Modeling is a necessary and very important test for the theory, but our primary concern is `explanation'. Perhaps this can be better understood by considering our first publication on the subject, Aerts and Aerts (1994). The idea of applying quantum statistics to model a human decision process was then inspired by the fact that in such a decision process the `possible outcomes of the decision do not exist prior to the decision being taken'. Classical statistics is a formalization of a situation where these possible outcomes `do exist' but `we lack knowledge about them'. Quantum statistics is exactly the opposite in this respect, for it describes processes where indeterminism appears `not due to a lack of knowledge of an already existing set of events', but `due to probability connected to potential (and hence not yet existing) events'. This is the reason and the origin of the quantum decision model we proposed in Aerts and Aerts (1994). At this moment, we are still driven by a very similar idea, namely the idea that human decision-making is structured in a quantum way, not only due to a situation of indeterminism that is not of the `lack of knowledge' type, but also due to the existence of effects, i.e. overextension, guppy effect, disjunction effect, conjunction fallacy, violation of the Savage `sure thing principle', that are connected with the way we have conceptually structured our world, by means of concepts and combinations of concepts. We indeed believe that this quantum structure percolates to the realm of `how meaning is carried by concepts'. This idea gains considerable plausibility thanks to the fact that the `disjunction effect' and the `conjunction fallacy' in decision theory and the violation of Savage's `sure thing principle' are completely similar effects to the `underextension' and `overextension' effects related to disjunction and conjunction in concept combination, and also that concept formation is largely a combinatory process of disjunctions and conjunctions (see section \ref{conceptformation}).

Returning to G\"ardenfors (2004), we can say that in his approach the focus is strongly on modeling, and hence a use of geometric structures that is much broader than what our quantum modeling scheme has to offer is possible. In this sense, we feel greater affinity with the attempts to model human thought within logical structures, started by Aristotle and continued by Boole and numerous logicians after them, except that the quantum Fock space connected structures describe a process of thought that is much more complex and where classical logical thought appears as one of the branches of a superposition in Fock space.

Quantum mechanics has also been used in the domain of information retrieval (Widdows, 2003, 2006; Widdows \& Peters, 2003; Van Rijsbergen, 2004), and here it is more specifically the quantum logic structure underlying quantum mechanics that has proved effective in producing theoretical models for information retrieval. Widdows and Peters (2003) demonstrated that the use of the quantum logic connective for negation, i.e. orthogonality, is a powerful tool for exploring and analyzing word meanings that has distinct advantages over Boolean operators in document retrieval experiments. Van Rijsbergen (2004) introduced a general theory for information retrieval based on quantum logic structures. He showed how three keystone models used in information retrieval, a vector space model, a probabilistic model and a logical model, can be described in Hilbert space, where a document can be represented by a vector and relevance by a Hermitian Operator. Both approaches bear correspondence to our approach.

In the same period of time, the formalization of context effects in relation to concepts and the study of operational issues related to such effects considering quantum logic was undertaken in Bruza and Cole (2005), inspiring further study of quantum structures in this respect (Bruza, Widdows \& Woods (in press)). Another subject that has been investigated is that of quantum structures in language, more specifically the entanglement of words in human semantic space, resulting in proposals of possible violations of Bell inequalities (Aerts, Czachor \& D'Hooghe, 2006; Bruza, Kitto, Nelson \& McEvoy, 2008; Bruza, Kitto, Nelson \& McEvoy, 2009).

\section{A Simple Quantum Model Illustrating the General Scheme} \label{simplemodel}

In the foregoing section we have given all elements of our general scheme of quantum modeling to explain what is the nature of the simple quantum model that we present in this section. The quantum model we present here is constructed in an eight dimensional real vector space $\real^8$, consisting of the direct sum of two four dimensional real vector spaces, hence $\real^8=\real^4 \oplus \real^4$, where the first $\real^4$ plays the role of what we have called the `two particle way' in the foregoing section. We will show that indeed the `classical items' and the `classical parts of a general item' can be represented in it. In the second $\real^4$ we represent the analogue of the `one particle way' in foregoing section.

What is now the specific value of this model compared to the model elaborated in Aerts (2007a,b)? This model is such that all the items being tested by Hampton with respect to one pair of concepts and its disjunction or conjunction can be modeled giving also a fixed representation to their conjunction and disjunction. For the standard interference quantum model, the one put forward in Aerts (2007a,b), and in other approaches where quantum mechanics is used in cognition and economy, this is not the case. More specifically the superposition vector ${1 \over \sqrt{2}}(|A\rangle+|B\rangle)$, if $|A\rangle$ and $|B\rangle$ are chosen from specific fixed rays in Hilbert space, does not necessarily corresponds to one specific ray in Hilbert space, because a lot of different rays can be reached by choosing different phases for $|A\rangle$ and $|B\rangle$. This is a well known phenomenon in quantum mechanics and even at the origin of the continuous set of values  that is reached by interference. The `real vector space model' presented in the remainder of this article shows that `superposition and interference without complex numbers' plus all the other aspects of quantum mechanics, is sufficient to produce a model for the set of cognitive data of Hampton (1988a,b). Such `superposition and interference without complex numbers' allows the representation of the conjunction and disjunction by means of a definite geometric structure, namely a two dimensional subspace, for each pair of concepts. Next to this the model shows in a simple way how the `two particle way' and `one particle way' play out. We believe however that the model built in Aerts (2007a,b), the one we have presented in section \ref{QuantumModelingScheme}, will turn out to be the more fundamental one, certainly if it comes to developing a generalized theory for combinations of large collections of concepts, as put forward in subsection \ref{largecollection}. In the next subsection we illustrate how the classical data can be represented in the first $\real^4$.

\subsection{Quantum Modeling of the Classical Items} \label{quantumrule}

We model the classical items in a 4-dimensional vector space over the real numbers, hence $\real^4$. We consider the item {\it Sailboat} of the pair of concepts {\it Machine} and {\it Vehicle} and their conjunction {\it Machine and Vehicle} to illustrate the modeling. Subjects attributed membership weights $\mu(A)=0.5641$, $\mu(B)=0.8$ and $\mu(A\ {\rm and}\ B)=0.4211$ respectively with respect to the concepts {\it Machine} and {\it Vehicle} and their conjunction {\it Machine and Vehicle}. The `conjunction minimum rule deviation' equals -0.1430, and the `Kolmogorovian conjunction factor' equals 0.0570, which makes the item is `classical', following our analysis of section \ref{conjunctiondata}. We need to specify (i) how we will represent the concepts $A$, $B$ and their conjunction `$A$ and $B$'; (ii) how we will represent the item $X$; and (iii) what are the quantum rules for calculating the weights.

Consider the canonical orthonormal base $\{e_{AB}, e_{AB'}, e_{A'B}, e_{A'B'}\}$ where $e_{AB}=(1,0,0,0), e_{AB'}=(0,1,0,0), e_{A'B}=(0,0,1,0)$ and $e_{A'B'}=(0,0,0,1)$. The concept $A$ is represented by the subspace ${\cal A}$ generated by the vectors $e_{AB}$ and $e_{AB'}$, and the concept $B$ is represented by the subspace ${\cal B}$ generated by the vectors $e_{AB}$ and $e_{A'B}$. Hence we have
\begin{equation}
{\cal A}=\{(x_{AB}, x_{AB'}, 0, 0)\ \vert\ x_{AB},x_{AB'} \in \real\} \quad {\cal B}=\{(x_{AB}, 0, x_{A'B}, 0)\ \vert\ x_{AB},x_{AB'}\in \real\}
\end{equation}
The conjunction `$A$ and $B$' is represented by the 1-dimensional subspace generated by the vector $e_{AB}$, which is equal to the intersection of ${\cal A}$ and ${\cal B}$. The disjunction `$A$ or $B$' is represented by the 3-dimensional subspace generated by $e_{AB}, e_{AB'}$ and $e_{A'B}$, which is equal to the sum of ${\cal A}$ and ${\cal B}$. Hence we have
\begin{equation}
{\cal A \cap B}=\{(x_{AB}, 0, 0, 0)\ \vert\ x_{AB}\in \real\} \quad
{\cal A}+{\cal B}=\{(x_{AB}, x_{AB'}, x_{A'B}, 0)\ \vert\ x_{AB},x_{AB'},x_{A'B}\in\real\}
\end{equation}
The item $X$ is represented by a unit vector $x \in \real^4$. Let us introduce the quantum rule for calculating the weights. 

\bigskip
\noindent {\it {\bf Quantum Rule:} The membership weight of an item $X$ with respect to a concept $A$ is given by the square of the length of the orthogonal projection of the vector $x$ representing the item $X$ on the subspace ${\cal A}$ representing the concept $A$.}

\bigskip
\noindent
Consider the unit vector $x$ representing the item $X$ is $(x_{AB}, x_{AB'}, x_{A'B}, x_{A'B'}) \in \real^4$. Hence, $x_{AB},$ $x_{AB'},$ $x_{A'B},$ $x_{A'B'} \in \real$, and $x_{AB}^2+x_{AB'}^2+x_{A'B}^2+x_{A'B'}^2=1$. The orthogonal projection on ${\cal A}$ of $x$ is the vector $(x_{AB}, x_{AB'}, 0, 0)$, and the square of its length is $x_{AB}^2+x_{AB'}^2$. And so following the quantum rules, the membership weight $\mu(A)$ of item $X$, represented by unit vector $(x_{AB}, x_{AB'}, x_{A'B}, x_{A'B'})$, with respect to concept $A$, represented by subspace ${\cal A}$, is given by $x_{AB}^2+x_{AB'}^2$. This means that, generally, for unit vector $(x_{AB}, x_{AB'}, x_{A'B}, x_{A'B'})$ representing item $X$, we have
\begin{eqnarray} \label{eqclass01}
&\mu(A)=x_{AB}^2+x_{AB'}^2 \\ \label{eqclass02}
&\mu(B)=x_{AB}^2+x_{A'B}^2 \\ \label{eqclass03}
&\mu(A\ {\rm and}\ B)=x_{AB}^2 \\ \label{eqclass04}
&\mu(A\ {\rm or}\ B)=x_{AB}^2+x_{AB'}^2+x_{A'B}^2
\end{eqnarray}
{\bf Theorem 7:} {\it Suppose we have an item $X$ with membership weights $\mu(A)$, $\mu(B)$ and $\mu(A\ {\rm and}\ B)$ with respect to the concepts $A$, $B$ and the conjunction `$A$ and $B$'. Suppose that inequalities \ref{ineq01}, \ref{ineq02} and \ref{ineq03} are satisfied, such that $X$ is a classical conjunction item. If we choose $x=(x_{AB},x_{AB'}, x_{A'B},x_{A'B'})$ such that
\begin{eqnarray} \label{eqclass05}
&x_{AB}=\pm\sqrt{\mu(A\ {\rm and}\ B)} \\ \label{eqclass06}
&x_{AB'}=\pm\sqrt{\mu(A)-\mu(A\ {\rm and}\ B)} \\ \label{eqclass07}
&x_{A'B}=\pm\sqrt{\mu(B)-\mu(A\ {\rm and}\ B)} \\ \label{eqclass08}
&x_{A'B'}=\pm\sqrt{1-\mu(A)-\mu(B)+\mu(A\ {\rm and}\ B)}=\pm\sqrt{k_c}
\end{eqnarray}
then $x$ is a unit vector of $\real^4$, and (\ref{eqclass01}), (\ref{eqclass02}) and (\ref{eqclass03}) are satisfied, which means that $x$, ${\cal A}$, ${\cal B}$ and ${\cal A \cap B}$ constitute a quantum representation of the item $X$ and the concepts $A$, $B$ and their conjunction `$A$ and $B$'.}

\bigskip
\noindent
Proof: See Appendix C.

\bigskip
\noindent
We prove a similar theorem in relation with the classical items with respect to the disjunction of concepts.

\bigskip
\noindent
{\bf Theorem 8:} {\it Suppose we have an item $X$ with membership weights $\mu(A)$, $\mu(B)$ and $\mu(A\ {\rm or}\ B)$ with respect to the concepts $A$, $B$ and the disjunction `$A$ or $B$'. Suppose that inequalities \ref{disjunctionineq01}, \ref{disjunctionineq02} and \ref{disjunctionineq03} are satisfied, such that $X$ is a classical disjunction item. If we choose $x=(x_{AB},x_{AB'}, x_{A'B},x_{A'B'})$ such that
\begin{eqnarray} \label{disjunctioneqclass05}
&x_{AB}=\pm\sqrt{\mu(A)+\mu(B)-\mu(A\ {\rm or}\ B)}=\pm\sqrt{k_d} \\ \label{disjunctioneqclass06}
&x_{AB'}=\pm\sqrt{\mu(A\ {\rm or}\ B)-\mu(B)} \\ \label{disjunctioneqclass07}
&x_{A'B}=\pm\sqrt{\mu(A\ {\rm or}\ B)-\mu(A)} \\ \label{disjunctioneqclass08}
&x_{A'B'}=\pm\sqrt{1-\mu(A\ {\rm or}\ B)}
\end{eqnarray}
then $x$ is a unit vector of $\real^4$, and (\ref{eqclass01}), (\ref{eqclass02}) and (\ref{eqclass04}) are satisfied, which means that $x$, ${\cal A}$, ${\cal B}$ and ${\cal A}+{\cal B}$ constitute a quantum representation of the item $X$ and the concepts $A$, $B$ and their disjunction `$A$ or $B$'.}

\bigskip
\noindent
Proof: See Appendix D.

\bigskip
\noindent
The foregoing theorems 7 and 8 prove that we can represent all classical items in $\real^4$ the way we explained. Let us do this explicitly for the item {\it Sailboat} with respect to the concepts {\it Machine} and {\it Vehicle} and their conjunction. We have $\mu(A)=0.5641$, $\mu(B)=0.8$ and $\mu(A\ {\rm and}\ B)=0.4211$. Taking into account (\ref{eqclass05}), (\ref{eqclass06}), (\ref{eqclass07}) and (\ref{eqclass08}) we have $x_{AB}=\sqrt{0.4211}=\pm0.6489$, $x_{AB'}=\sqrt{0.5641-0.4211}=0.3782$, $x_{A'B}=\sqrt{0.8-0.4211}=0.6156$ and $x_{A'B'}=\sqrt{1-0.5641-0.8+0.4211}=0.2386$. This means that the vector $x_{{\rm {\it Sailboat}}}=(0.6489, 0.3782, 0.6156, 0.2386)$ can be chosen to represents the item {\it Sailboat} in $\real^4$. In an analogous way we calculate the vectors representing the other classical items with respect to the concepts {\it Machine} and {\it Vehicle} and their conjunction. This gives us the following $x_{{\rm {\it Raft}}}=(0.4472, 0.0716,$ $0.7246, 0.5195)$, $x_{{\rm {\it Backpack}}}=(0, 0, 0, 1)$, $x_{{\rm {\it Automobile}}}=(1, 0, 0, 0)$ and $x_{{\rm {\it Bus}}}=(1, 0, 0, 0)$.

It is easy to see why non classical items cannot be represented in this way in $\real^4$. Indeed, for the case of the conjunction, and a non classical item of the $\Delta$-type, the right-hand side of equation (\ref{eqtheorem102}) or the right-hand side of equation (\ref{eqtheorem103}) is negative, which means that (\ref{eqclass06}) or (\ref{eqclass07}) is not satisfied. For the case of the conjunction, and a non classical item of the $k$-type, the right-hand side of equation (\ref{eqtheorem104}) is negative, which means that (\ref{eqclass08}) is not satisfied. For the case of the disjunction, and for a non classical item of the $\Delta$-type, the right-hand side of equation (\ref{eqtheorem202}) or the right-hand side of equation (\ref{eqtheorem203}) is negative, which means that (\ref{disjunctioneqclass06}) or (\ref{disjunctioneqclass07}) is not satisfied. For the case of the disjunction, and a non classical item of the $k$-type, the right-hand side of equation (\ref{eqtheorem201}) is negative, which means that (\ref{disjunctioneqclass05}) is not satisfied.

\subsection{Modeling $k$-Type Non Classical Items} \label{modelingKtype}
The essential difference between the quantum representation in $\real^4$ of the classical items and an ordinary set theoretic probability representation, like the one we worked out explicitly in proving {\bf Theorem 1}, is that concepts are represented by subspaces of a vector space instead of by subsets of the sample space. In this section we show that the $k$-type non classicality can be represented in an $\real^4$ quantum formalism by choosing the subspaces representing the concepts in a different way than we did for the classical items in subsection \ref{quantumrule}. We will also show in subsection \ref{modelingDeltatype} that the $\Delta$-type of non classicality (the Guppy effect) cannot be represented by the quantum formalism by just choosing the subspaces in a different way. Another aspect of the quantum formalism is needed for the $\Delta$-type classicality, namely the aspect linked to the emergence of new states and subspaces due to superposition. Referring to our general quantum modeling scheme put forward in section \ref{generalscheme}, it are the $\Delta$-type non classical items that need a `one particle way' description, while the $k$-type of non classicality can be accounted by within only a `two particle way'. This means that the $k$-type of non classicality finds its origin in the non Boolean nature of the set of closed subspaces of the Hilbert space, a structure widely studied in a research field named `quantum logic'. Hence the $k$-type of non classicality is of a `quantum logic nature'. The $\Delta$-type of non classicality arises in a different way, due to the effect of interference and how this effect is linked to the emergence of new states and subspaces in quantum mechanics. Let us construct a `quantum logic' representation for the $k$-type of non classicality. We choose the subspaces in the following way
\begin{eqnarray} \label{eqKtype01}
&{\cal A}=\{(x_1,x_2\cos({\pi \over 4}+{\theta \over 2}), x_2\sin({\pi \over 4}+{\theta \over 2}), 0)\ \vert\ x_1, x_2 \in \real\} \\ \label{eqKtype02}
&{\cal B}=\{(x_1, x_2\cos({\pi \over 4}-{\theta \over 2}), x_2\sin({\pi \over 4}-{\theta \over 2}), 0)\ \vert\ x_1, x_2 \in \real\}
\end{eqnarray}
to represent concepts $A$ and $B$, for a given angle $\theta \in [0, 180^\circ]$. For $\theta=90^\circ={\pi \over 2}$ the choice of ${\cal A}$ and ${\cal B}$ reduces to the choice we made for a classical type item, and for $\theta=0$ we have that ${\cal A}$ and ${\cal B}$ coincide. For other values of $\theta$, however, we have a different geometric situation, in between these two extremes. 

Let us calculate the weights in this new geometrical situation applying the quantum rule we put forward in \ref{quantumrule}. The notations $e_{AB}, e_{AB'}, e_{A'B}, e_{A'B'}$ we used earlier is because $e_{AB}$ is a vector contained in ${\cal A} \cap {\cal B}$, $e_{AB'}$ a vector contained in ${\cal A} \cap {\cal B}'$, $e_{A'B}$ a vector contained in ${\cal A}' \cap {\cal B}$, and $e_{A'B'}$ a vector contained in ${\cal A}' \cap {\cal B}'$. With this new choice of the subspaces ${\cal A}$ and ${\cal B}$ this is no longer true, which is the reason that we introduce a new notation for this canonical base. Hence we denote now $e_{AB}=(1, 0, 0, 0)$, $e_2=(0, 1, 0, 0)$, $e_3=(0, 0, 1, 0)$ and $e_{A'B'}=(0, 0, 0, 1)$ and for a vector $x$ representing the item $X$ we denote the components with respect to this canonical base as follows $x=(x_{AB}, x_2, x_3, x_{A'B'})$. With respect to the chosen subspaces ${\cal A}$ and ${\cal B}$ it is important to consider two other orthonormal bases of $\real^4$ which are the following. A first basis is $e_{AB}=(1, 0, 0, 0)$, $e_A=(0,\sin({\pi \over 4}+{\theta \over 2}), \cos({\pi \over 4}+{\theta \over 2}), 0)$, $e_{A'}=(0, -\cos({\pi \over 4}+{\theta \over 2}), \sin({\pi \over 4}+{\theta \over 2}), 0)$ and $e_{A'B'}=(0, 0, 0, 1)$. And a second basis is $e_{AB}=(1, 0, 0, 0)$,  $e_B=(0, \sin({\pi \over 4}-{\theta \over 2}), \cos({\pi \over 4}-{\theta \over 2}), 0)$, $e_{B'}=(0, \cos({\pi \over 4}-{\theta \over 2}), -\sin({\pi \over 4}-{\theta \over 2}), 0)$ and $e_{A'B'}=(0, 0, 0, 1)$. Consider the vector $x=(x_{AB}, x_2, x_3, x_{A'B'})$ of $\real^4$ of unit length, which hence means that we have $x_{AB}^2+x_2^2+x_3^2+x_{A'B'}^2=1$. The orthogonal projection of this vector $x$ on the subspace ${\cal A \cap B}$, which is the subspace generated by the vector $e_{AB}$, is given by $x_{AB}$. Following the quantum rule we get 
$\mu(A\ {\rm and}\ B)=x_{AB}^2$ and hence
\begin{equation} \label{eqKtype10}
x_{AB}=\pm\sqrt{\mu(A\ {\rm and}\ B)}
\end{equation}
The subspace ${\cal A}$ is the space generated by the orthonormal set of vectors $e_{AB}$ and $e_A$, and hence $\mu(A)$ is the square of the orthogonal projection on this space, which gives $\mu(A) = x_{AB}^2+(x \cdot e_A)^2$. The subspace ${\cal B}$ is generated by the orthonormal set of vectors $e_{AB}$ and $e_B$, and hence $\mu(B)$ is the square of the orthogonal projection on this space, which gives $\mu(B)=x_{AB}^2+(x \cdot e_B)^2$. It follows that $\mu(A)=x_{AB}^2+(x_2\sin({\pi \over 4}+{\theta \over 2})+x_3\cos({\pi \over 4}+{\theta \over 2}))^2$ and hence $\mu(A)-\mu(A\ {\rm and}\ B)=(x_2\sin({\pi \over 4}+{\theta \over 2})+x_3\cos({\pi \over 4}+{\theta \over 2}))^2$ and putting $a=\mu(A)-\mu(A\ {\rm and}\ B)$ we get
\begin{equation}
\label{eqKtype12}
x_2\sin({\pi \over 4}+{\theta \over 2})+x_3\cos({\pi \over 4}+{\theta \over 2})=\pm\sqrt{a}
\end{equation}
In an analogous way we calculate
\begin{equation}
\label{eqKtype13}
x_2\sin({\pi \over 4}-{\theta \over 2})+x_3\cos({\pi \over 4}-{\theta \over 2})=\pm\sqrt{b}\end{equation}
where we have put $b=\mu(B)-\mu(A\ {\rm and}\ B)$. From (\ref{eqKtype12}) and (\ref{eqKtype13}) it is possible to calculate $x_2$ and $x_3$ in function of $a$, $b$ and $\theta$ and this gives
\begin{equation} \label{eqKtype16}
x_2={\pm\sqrt{a}\cos({\pi \over 4}-{\theta \over 2})\mp\sqrt{b}\cos({\pi \over 4}+{\theta \over 2}) \over \sin\theta} \quad x_3={\mp\sqrt{a}\sin({\pi \over 4}-{\theta \over 2})\pm\sqrt{b}\sin({\pi \over 4}+{\theta \over 2}) \over \sin\theta}
\end{equation}
Let us calculate $x_{A'B'}$. A straightforward calculation using (\ref{eqKtype16}) gives
\begin{equation} \label{eqKtype22}
x_2^2+x_3^2={a+b-2(\pm\sqrt{a})(\pm\sqrt{b})\cos\theta \over \sin^2\theta} \quad x_{A'B'}=\pm\sqrt{1-x_{AB}^2-x_2^2-x_3^2}=\pm\sqrt{q_{conj}(A, B, \theta)}
\end{equation}
where we have introduced for a pair of concepts $A$ and $B$ a quantum logic factor $q_{conj}(A, B, \theta)$ as follows
\begin{equation} \label{eqKtype23}
q_{conj}(A, B, \theta)=1-\mu(A\ {\rm and}\ B)-{{a+b-2(\pm\sqrt{a})(\pm\sqrt{b}})\cos\theta \over \sin^2\theta}
\end{equation}
This means that the item $X$ with membership weights $\mu(A)$, $\mu(B)$ and $\mu(A\ {\rm and}\ B)$ with respect to concepts $A$, $B$ and their conjunction `$A$ and $B$', can be represented by means of the vector $x=(x_{AB}, x_2, x_3, x_{A'B'})$ if and only if $0 \le q_{conj}(A, B, \theta)$.
The quantum logic factor plays the role for $k$-type of non Kolmogorovian items that was earlier plaid by the Kolmogorovian factor. The value of this quantum logic factor depends on the angle $\theta$, characterizing the angle between the two subspaces ${\cal A}$ and ${\cal B}$ representing the concepts $A$ and $B$. We have 
$q_{conj}(A, B, {\pi \over 2})=k_c$, which proves that $q$ is really a generalization of $k$.
Let us investigate the requirement that the quantum logic factor needs to be non negative in detail. After some calculation we can show that
\begin{equation}
q_{conj}(A, B, \theta)=-{1 \over \sin^2\theta}((1-\mu(A\ {\rm and}\ B))\cos^2\theta-2(\pm\sqrt{a})(\pm\sqrt{b})\cos\theta-k)
\end{equation}
This is a quadratic equation in $\cos\theta$ which discriminant $D=(1-\mu(A))(1-\mu(B))$ is never negative, which means that two (or one if the discriminant is zero) roots exist. The roots are
\begin{equation} \label{eqKtype80}
\cos\theta^{\pm}={(\pm\sqrt{a})(\pm\sqrt{b})\pm\sqrt{D} \over 1-\mu(A\ {\rm and}\ B)}
\end{equation}
For values of $\theta$ between $\theta^{-}$ and $\theta^{+}$, hence where $\cos\theta$ lies in between these two roots $\cos\theta^{-}$ and $\cos\theta^{+}$, we have that $q_{conj}(A, B, \theta)$ is non negative, and hence a quantum representation of the data exists.

Let us apply this to the $k$-type non classical items with respect to the pair of concepts {\it Machine} and {\it Vehicle} and their conjunction `{\it Machine and Vehicle}'. For the item {\it Horse Cart} we have $\mu(A)=0.3846$, $\mu(B)=0.95$ and $\mu(A\ {\rm and}\ B)=0.2895$. Hence $a=0.0951$, $b=0.6605$ and $k=-0.0451$. From (\ref{eqKtype23}) follows that we have $q=1-0.2895-{1 \over \sin^2\theta}(0.0951+0.6605-2(\pm0.3084)(\pm0.8127)\cos\theta)$ and from (\ref{eqKtype80}) follows that the values of $\theta$ where $q$ is non negative are those for which $\cos\theta$ lies between the following two roots $\cos\theta^{\pm}={1 \over 1-0.2895}((\pm0.3084)(\pm0.8127)\pm\sqrt{D})$
where we have $D=(1-\mu(A))(1-\mu(B))=(1-0.3846)(1-0.95)=0.0308$. If we choose $+0.3084$ and $+0.8127$ or $-0.3084$ and $-0.8127$ for $\pm\sqrt{a}$ and $\pm\sqrt{b}$ we find $\cos\theta=+0.1059$ and $\cos\theta=+0.5996$ as roots. If we choose $+0.3084$ and $-0.8127$ or $-0.3084$ and $+0.8127$ for $\pm\sqrt{a}$ and $\pm\sqrt{b}$ we find $\cos\theta=-0.5996$ and $\cos\theta=-0.1059$ as roots. For the choices of $\theta$ this gives. If we choose $+0.3084$ and $+0.8127$ or $-0.3084$ and $-0.8127$ for $\pm\sqrt{a}$ and $\pm\sqrt{b}$ we have $\theta \in [53.1553^\circ,83.9225^\circ]$. If we choose $+0.3084$ and $-0.8127$ or $-0.3084$ and $+0.8127$ for $\pm\sqrt{a}$ and $\pm\sqrt{b}$ we have $\theta \in [96.0775^\circ,126.8447^\circ]$. Hence, $[53.1553^\circ,83.9225^\circ]$ and $[96.0775^\circ,126.8447^\circ]$ are the intervals for $\theta$ where a solution for the data related to the item {\it Horsecart} exists, i.e. where we can represent this item by means of a vector $x_{Horsecart}$ within our quantum model in $\real^4$.

For the second $k$-type non classical item with respect to the concepts {\it Machine} and {\it Vehicle} and their conjunction {\it Machine and Vehicle}, namely $Dishwasher$, we have $\mu(A)=1$, $\mu(B)=0.025$, $\mu(A\ {\rm and}\ B)=0$, and hence $a=1$, $b=0.025$ and $k=-0.025$. We have $D=0$, which means that the intervals reduce to points, given by (\ref{eqKtype80}), and hence $\cos\theta^{\pm}=(\pm1)(\pm0.1581)$. If we choose $+1$ and $+0.1581$ or $-1$ and $-0.1581$ for $\pm\sqrt{a}$ and $\pm\sqrt{b}$ we get $\cos\theta=0.1581$, and hence $\theta=80.9026^\circ$, and if we choose $+1$ and $-0.1581$ or $-1$ and $+0.1581$ for $\pm\sqrt{a}$ and $\pm\sqrt{b}$ we get $\cos\theta=-0.1581$, and hence $\theta=99.0974^\circ$. Since these values of $\theta$ fall within the respective intervals we found for the item {\it Horsecart}, we have a solution for both items for a choice of $\theta=80.9026^\circ$ or $\theta=99.0974^\circ$. Let us construct vectors $x$ for both items corresponding to this solution. Take first $\theta=80.9026^\circ$, and consider the item {\it Horsecart}. From (\ref{eqKtype10}) it follows that $x_{AB}=\pm\sqrt{0.2895}=\pm0.5380$. From (\ref{eqKtype16}) we get $x_2=\pm0.2461$ and $x_3=\pm0.7957$. And from (\ref{eqKtype22}) we get $x_{A'B'}=\pm0.1296$. Hence the vector $x_{\rm {\it Horse Cart}}=(0.5380, 0.2461, 0.7957, 0.1296)$ represents the item {\it Horse Cart} in $\real^4$ for a choice of $\theta=80.9026^\circ$.

Interestingly, we can also represent the classical items in this new representation, which shows that the new representation, with $\theta=80.9026^\circ$ or $\theta=99.0974^\circ$, is more general than the foregoing representation for classical items, as worked out in subsection \ref{quantumrule}, since we can represent the classical items plus the non classical items of the $k$-type.
Let us calculate the vectors that represent the classical items in the new representation for the situation where $\theta=80.9026^\circ$. For example, for the item {\it Sailboat} we have $\mu(A)=0.5641$, $\mu(B)=0.8$ and $\mu(A\ {\rm and}\ B)=0.4211$, and hence $a=0.1430$ and $b=0.3789$. Using the same equations (\ref{eqKtype10}), (\ref{eqKtype16}) and (\ref{eqKtype22}) that served to calculate the vectors in the case of the non classical items {\it Horse Cart} and {\it Dishwasher} of $k$-type, we get $x_{AB}=\pm0.6489$, $x_2=\pm0.3324$, $x_3=\pm0.5911$ and $x_{A'B'}=\pm0.3451$ which means that the vector $x_{\rm {\it Sailboat}}=(0.6489, 0.3324, 0.5911, 0.3451)$ is a possible representation of the item {\it Sailboat} in $\real^4$. Equally so we find for the remaining classical items, {\it Backpack}, {\it Raft}, {\it Automobile} and {\it Bus}, that the vectors $x_{\rm {\it Backpack}}=(0, 0, 0, 1)$, $x_{\rm {\it Raft}}=(0.4472, 0.0141, 0.7257, 0.5226)$, $x_{\rm {\it Automobile}}=(1, 0, 0, 0)$ and $x_{\rm {\it Bus}}=(1, 0, 0, 0)$ represent these items in $\real^4$.
In an analogous way we can also represent the $k$-type non classical items in the case of the disjunction.

\subsection{Modeling $\Delta$-type Non-Classical Items} \label{modelingDeltatype}
We will not be able to model $\Delta$-type non-classical items in the way we did with $k$-type non-classical items in section \ref{modelingKtype}. The reason for this is fundamental, as we will show in this subsection. In this respect, we can prove the following theorem.

\bigskip
\noindent
{\it {\bf Theorem 9:} Consider two concepts $A$ and $B$, and suppose we represent them by means of subspaces ${\cal A}$ and ${\cal B}$ of $\real^n$, such that the conjunction of $A$ and $B$ is represented by subspace ${\cal A \cap B}$, and the disjunction of $A$ and $B$ is represented by subspace ${\cal A} + {\cal B}$. Suppose that $\mu(A)$, $\mu(B)$, $\mu(A\ {\rm and}\ B)$ and $\mu(A\ {\rm or}\ B)$ are the membership weights of item $X$ with respect to concepts $A$ and $B$, their conjunction `$A$ and $B$' and their disjunction `$A$ or $B$'. Then $\mu(A)$, $\mu(B)$ and $\mu(A\ {\rm and}\ B)$ satisfy inequalities (\ref{ineq01}) and (\ref{ineq02}), which means that $\Delta_c \le 0$, and hence $X$, is not a $\Delta$-type non-classical item for the conjunction.  Equally, $\mu(A)$, $\mu(B)$ and $\mu(A\ {\rm or}\ B)$ satisfy inequalities (\ref{disjunctionineq01}) and (\ref{disjunctionineq02}), which means that $\Delta_d \le 0$, and hence $X$, is not a $\Delta$-type non-classical item for the disjunction.}

\bigskip
\noindent
Proof: See Appendix E.

\bigskip
\noindent This theorem shows that if we want to represent concepts $A$ and $B$ by means of subspaces ${\cal A}$ and ${\cal B}$ of $\real^n$, such that the conjunction of $A$ and $B$ is represented by means of subspace ${\cal A \cap B}$ and the disjunction by means of subspace ${\cal A} + {\cal B}$, inequalities (\ref{ineq01}), (\ref{ineq02}), (\ref{disjunctionineq01}) and (\ref{disjunctionineq02}) are satisfied. This means that we cannot model items for which these inequalities are not satisfied, i.e. $\Delta$-type non-classical items. It confirms that it is the $\Delta$-type of non-classicality that needs the `one-particle way' quantum modeling. We propose now a construction for this `one-particle way' for this real vector space quantum model taking into the analysis of section \ref{generalscheme}. {\it Whenever two concepts $A$ and $B$ are combined, be it to form a conjunction or a disjunction, `in superposition' with the specific combination, hence in our case conjunction or disjunction, there is a process of `new concept formation'.} Consequently, the sentence `$A$ and $B$' is a superposition of a conjunction of two concepts $A$ and $B$ and a new concept $C$ formed (or emerged) out of the process of combining concept $A$ and concept $B$. The situation is similar for disjunction, the sentence `$A$ or $B$' is a superposition of a disjunction of two concepts $A$ and $B$ and a new concept $C$ formed (or emerged) out of the process of combining of concept $A$ and concept $B$.

We construct the emergent new concept representation in another 4-dimensional real vector space $\real^4$. This means that a superposition of both representations is formulated in the direct sum of both vector spaces $\real^4\oplus\real^4=\real^8$. 
We denote the canonical base of this 8-dimensional real vector space as follows: $e_{AB}=(1, 0, 0, 0, 0, 0, 0, 0)$, $e_{AB'}=(0, 1, 0, 0, 0, 0, 0, 0)$, $e_{A'B}=(0, 0, 1, 0, 0, 0, 0, 0)$, $e_{A'B'}=(0, 0, 0, 1, 0, 0, 0, 0)$, $e_5=(0, 0, 0, 0, 1, 0, 0, 0)$, $e_6=(0, 0, 0, 0, 0, 1, 0, 0)$, $e_7=(0, 0, 0, 0, 0, 0, 1, 0)$ and $e_8=(0, 0, 0, 0, 0, 0, 0, 1)$. This means that a vector $x$ has the following expansion with respect to this canonical base $x=(x_{AB},x_{AB'},x_{A'B},x_{A'B'},x_5,x_6,x_7,x_8)$.

We consider two other orthonormal bases of $\real^8$, namely a base given by vectors $e_{AB}=(1, 0, 0, 0, 0, 0,$ $0, 0)$, $e_{AB'}=(0, 1, 0, 0, 0, 0, 0, 0)$, $e_{A'B}=(0, 0, 1, 0, 0, 0, 0, 0)$, $e_{A'B'}=(0, 0, 0, 1, 0, 0, 0, 0)$, $e_5=(0, 0, 0, 0,$ $1, 0, 0, 0)$, $e_{A}=(0, 0, 0, 0, 0,\sin({\pi \over 4}+{\theta \over 2}), \cos({\pi \over 4}+{\theta \over 2}), 0)$, $e_{A'}=(0, 0, 0, 0, 0, -\cos({\pi \over 4}+{\theta \over 2}), \sin({\pi \over 4}+{\theta \over 2}), 0)$, $e_8=(0, 0, 0, 0, 0, 0, 0, 1)$. With respect to this basis, vector $x$ can be written as $x=x_{AB}e_{AB}+x_{AB'}e_{AB'}+x_{A'B}e_{A'B}+x_{A'B'}e_{A'B'}+x_5e_5+x_{A}e_{A}+x_{A'}e_{A'}+x_8e_8$, and a third base given by vectors $e_{AB}=(1, 0, 0, 0, 0, 0, 0, 0)$, $e_{AB'}=(0, 1, 0, 0, 0, 0, 0, 0)$, $e_{A'B}=(0, 0, 1, 0, 0, 0, 0, 0)$, $e_{A'B'}=(0, 0, 0, 1, 0, 0, 0, 0)$, $e_5=(0, 0, 0, 0, 1, 0, 0, 0)$, $e_{B}=(0, 0, 0, 0, 0,\sin({\pi \over 4}-{\theta \over 2}), \cos({\pi \over 4}-{\theta \over 2}), 0)$, $e_{B'}=(0, 0, 0, 0, 0, \cos({\pi \over 4}-{\theta \over 2}), -\sin({\pi \over 4}-{\theta \over 2}), 0)$, $e_8=(0, 0, 0, 0, 0, 0, 0, 1)$. Vector $x$ with respect to this basis is written as $x=x_{AB}e_{AB}+x_{AB'}e_{AB'}+x_{A'B}e_{A'B}+x_{A'B'}e_{A'B'}+x_5e_5+x_{B}e_{B}+x_{B'}e_{B'}+x_8e_8$.

Subspaces ${\cal A}$ and ${\cal B}$, representing concepts $A$ and $B$, respectively, are defined as follows:
\begin{eqnarray}
&{\cal A}=\{x_{AB}e_{AB}+x_{AB'}e_{AB'}+x_5e_5+x_{A}e_{A}\ \vert\ x_{AB}, x_{AB'}, x_5, x_{A} \in \real\} \\
&{\cal B}=\{x_{AB}e_{AB}+x_{A'B}e_{A'B}+x_5e_5+x_{B}e_{B}\ \vert\ x_{AB}, x_{A'B}, x_5, x_{B} \in \real\}
\end{eqnarray}
Following the quantum rule, we get
\begin{eqnarray} \label{eqweights01}
&\mu(A)=x_{AB}^2+x_{AB'}^2+x_5^2+x_{A}^2 \\ \label{eqweights02}
&\mu(B)=x_{AB}^2+x_{A'B}^2+x_5^2+x_{B}^2
\end{eqnarray}
To introduce subspace ${\cal C}$ representing the emerging new concept, we consider a fourth orthonormal base consisting of vectors $\{e_{AB}, e_{AB'}, e_{A'B}, e_{A'B'}, f_5, f_6, f_7, f_8 \}$,
and vector $x$ with respect to this base is written as $x=x_{AB}e_{AB}+x_{AB'}e_{AB'}+x_{A'B}e_{A'B}+x_{A'B'}e_{A'B'}+c_5f_5+c_6f_6+c_7f_7+c_8f_8$.
Subspace ${\cal C}$ is defined as follows: $ 
{\cal C}=\{c_5f_5+c_6f_6\ \vert\ c_5, c_6 \in \real\}$. Applying the quantum rule for the weights of conjunction and disjunction yields:
\begin{eqnarray} \label{eqweightconj}
&\mu(A\ {\rm and}\ B)=x_{AB}^2+c_5^2+c_6^2 \\
\label{eqweightdisj}
&\mu(A\ {\rm or}\ B)=x_{AB}^2+x_{AB'}^2+x_{A'B}^2+c_5^2+c_6^2
\end{eqnarray}
The conjunction and disjunction values of subspace ${\cal C}$ need not be identical. In fact, the following shows us they are not.
We further have that the length of $x$ equals 1, which gives that
\begin{equation} \label{eqweights04}
x_{AB}^2+x_{AB'}^2+x_{A'B}^2+x_{A'B'}^2+x_5^2+x_6^2+x_7^2+x_8^2=1
\end{equation}
We also have
\begin{eqnarray} \label{eqxA}
&x_{A}=x \cdot e_A=x_6\sin({\pi \over 4}+{\theta \over 2})+x_7\cos({\pi \over 4}+{\theta \over 2}) \\ \label{eqxA'}
&x_{A'}=x \cdot e_{A'}=-x_6\cos({\pi \over 4}+{\theta \over 2})+x_7\sin({\pi \over 4}+{\theta \over 2}) \\ \label{eqxB}
&x_B=x \cdot e_B=x_6\sin({\pi \over 4}-{\theta \over 2})+x_7\cos({\pi \over 4}-{\theta \over 2}) \\ \label{eqxB'}
&x_{B'}=x \cdot e_{B'}=x_6\cos({\pi \over 4}-{\theta \over 2})-x_7\sin({\pi \over 4}-{\theta \over 2}) \\ \label{eqc5678}
&c_5=x \cdot f_5 \quad c_6=x \cdot f_6 \quad c_7=x \cdot f_7 \quad c_8=x \cdot f_8
\end{eqnarray}
The following equalities can easily be derived from (\ref{eqxA}), (\ref{eqxA'}), (\ref{eqxB}), (\ref{eqxB'}) and (\ref{eqc5678})
\begin{eqnarray} \label{eqequality01}
&x_6^2+x_7^2=x_{A}^2+x_{A'}^2=x_{B}^2+x_{B'}^2 \\ \label{eqequality02}
&x_5^2+x_6^2+x_7^2+x_8^2=c_5^2+c_6^2+c_7^2+c_8^2
\end{eqnarray}
Equations (\ref{eqweights01}), (\ref{eqweights02}), (\ref{eqweightconj}), (\ref{eqweightdisj}), (\ref{eqweights04}), (\ref{eqxA}), (\ref{eqxA'}), (\ref{eqxB}), (\ref{eqxB'}) and (\ref{eqc5678}) are the basic equations that determine the constraints that need to be satisfied for a model within the 8-dimensional real vector space to reproduce the experimental data of Hampton (1988a,b). Let us work on them to obtain the form we need to construct the quantum model for the conjunction and disjunction data.

We will first look at the situation of the conjunction.
We isolate the classical components $x_{AB}$, $x_{AB'}$, $x_{A'B}$ and $x_{A'B'}$ of vector $x$ that represents item $X$, and express these classical components in function of the experimental data, i.e. membership weights $\mu(A)$, $\mu(B)$ and $\mu(A\ {\rm and}\ B)$, and quantum components $x_5$, $x_6$, $x_7$, $x_8$, $c_5$ and $c_6$. From (\ref{eqweightconj}) we thus get the value of $x_{AB}$, namely
\begin{equation} \label{eqweights05}
x_{AB}^2=\mu(A\ {\rm and}\ B)-c_5^2-c_6^2
\end{equation}
If we substitute this in (\ref{eqweights01}) and (\ref{eqweights02}), we get
\begin{eqnarray} \label{eqweights06}
&x_{AB'}^2=\mu(A)-\mu(A\ {\rm and}\ B)+c_5^2+c_6^2-x_5^2-x_{A}^2 \\ \label{eqweights07}
&x_{A'B}^2=\mu(B)-\mu(A\ {\rm and}\ B)+c_5^2+c_6^2-x_5^2-x_{B}^2
\end{eqnarray}
and if we substitute (\ref{eqweights05}), (\ref{eqweights06}) and (\ref{eqweights07}) in (\ref{eqweights04}) we get
\begin{equation} \label{eqweights08}
x_{A'B'}^2=1-\mu(A)-\mu(B)+\mu(A\ {\rm and}\ B)-c_5^2-c_6^2+x_{A}^2+x_{B}^2+x_5^2-x_6^2-x_7^2-x_8^2
\end{equation}
The set of equations (\ref{eqweights05}), (\ref{eqweights06}), (\ref{eqweights07}) and (\ref{eqweights08}) is equivalent to the set of basic equations (\ref{eqweights01}), (\ref{eqweights02}), (\ref{eqweightconj}) and (\ref{eqweights04}), so that, to find a solution for the modeling, we can concentrate on this set of equations. If we can find $x_5, x_6, x_7, x_8, c_5$ and $c_6$ such that $0 \le x_{AB}^2 \le 1$, $0 \le x_{AB'}^2 \le 1$, $0 \le x_{A'B}^2 \le 1$ and $0 \le x_{A'B'}^2 \le 1$, as expressed in equations (\ref{eqweights05}), (\ref{eqweights06}), (\ref{eqweights07}) and (\ref{eqweights08}), we have our solution. In the next subsection we construct such a solution.

We will now calculate the situation of the disjunction by an analogous reasoning. We subtract (\ref{eqweights01}) from (\ref{eqweightdisj}) to get
$\mu(A\ {\rm or }\ B)-\mu(A)=x_{A'B}^2+c_5^2+c_6^2-x_5^2-x_{A}^2$, and hence
\begin{equation} \label{eqdisjunctionweights05}
x_{A'B}^2=\mu(A\ {\rm or }\ B)-\mu(A)-c_5^2-c_6^2+x_5^2+x_{A}^2
\end{equation}
By analogy, subtracting (\ref{eqweights02}) from (\ref{eqweightdisj}), we get 
$\mu(A\ {\rm or }\ B)-\mu(B)=x_{AB'}^2+c_5^2+c_6^2-x_5^2-x_{B}^2$, and hence
\begin{equation} \label{eqdisjunctionweights06}
x_{AB'}^2=\mu(A\ {\rm or }\ B)-\mu(B)-c_5^2-c_6^2+x_5^2+x_{B}^2
\end{equation}
Subtracting (\ref{eqweights01}) and (\ref{eqweights02}) from (\ref{eqweightdisj}), we get $\mu(A\ {\rm or }\ B)-\mu(A)-\mu(B)=-x_{AB}^2+c_5^2+c_6^2-2x_5^2-x_{A}^2-x_{B}^2$,
and hence
\begin{equation} \label{eqdisjunctionweights07}
x_{AB}^2=\mu(A)+\mu(B)-\mu(A\ {\rm or }\ B)+c_5^2+c_6^2-2x_5^2-x_{A}^2-x_{B}^2
\end{equation}
Subtracting (\ref{eqweights01}) from (\ref{eqweights04}), we get $1-\mu(A)=x_{A'B}^2+x_{A'B'}^2+x_6^2+x_7^2+x_8^2-x_{A}^2$,
and hence
\begin{equation} \label{eqdisjunctionweights08}
x_{A'B'}^2=1-\mu(A)-x_{A'B}^2-x_6^2-x_7^2-x_8^2+x_{A}^2
=1-\mu(A\ {\rm or }\ B)+c_5^2+c_6^2-x_5^2-x_6^2-x_7^2-x_8^2
\end{equation}
The set of equations (\ref{eqdisjunctionweights05}), (\ref{eqdisjunctionweights06}), (\ref{eqdisjunctionweights07}) and (\ref{eqdisjunctionweights08}) is equivalent to the set of basic equations (\ref{eqweights01}), (\ref{eqweights02}), (\ref{eqweightdisj}) and (\ref{eqweights04}), so that, to find a solution for the modeling, we can concentrate on this set of equations. If we can find $x_5, x_6, x_7, x_8, c_5$ and $c_6$ such that $0 \le x_{AB}^2 \le 1$, $0 \le x_{AB'}^2 \le 1$, $0 \le x_{A'B}^2 \le 1$ and $0 \le x_{A'B'}^2 \le 1$, as expressed in equations (\ref{eqdisjunctionweights05}), (\ref{eqdisjunctionweights06}), (\ref{eqdisjunctionweights07}) and (\ref{eqdisjunctionweights08}), we have our solution. In section \ref{disjunctiondatasolution} we construct such a solution.

We should note that finding a solution is not merely a matter of fitting parameters. For each pair of concepts $A$ and $B$ and their conjunction `$A$ and $B$' we need to determine a fixed plane ${\cal C}$ such that for each one of the items $X$ considered with respect to this pair of concepts in Hampton (1988a), we can determine a vector $x=(x_{AB},x_{AB'},x_{A'B},x_{A'B'},x_5,x_6,x_7,x_8)$, such that equations (\ref{eqweights05}), (\ref{eqweights06}), (\ref{eqweights07}) and (\ref{eqweights08}) are satisfied, where $\mu(A)$, $\mu(B)$ and $\mu(A\ {\rm and}\ B)$ are the membership weights of item $X$ with respect to concept $A$, concept $B$ and conjunction `$A$ and $B$', respectively, experimentally determined by Hampton (1988a). This means that $c_5$ and $c_6$ are determined in function of $x_5$, $x_6$, $x_7$ and $x_8$ by means of the choice of this fixed plane ${\cal C}$. This leaves us with 8 parameters, $x_{AB},x_{AB'},x_{A'B},x_{A'B'},x_5,x_6,x_7,x_8$, and four equations, (\ref{eqweights05}), (\ref{eqweights06}), (\ref{eqweights07}) and (\ref{eqweights08}), to be satisfied, i.e. four degrees of freedom to determine three quantities $\mu(A)$, $\mu(B)$ and $\mu(A\ {\rm and}\ B)$. This shows that the problem can be solved, and even contains one additional degree of freedom. This free parameter will prove of key importance to the interpretation of the result in terms of the two different processes of thought in, i.e. classical logical thought and quantum conceptual thought.

\section{Solving the Modeling of the Disjunction and Conjunction Data} \label{disjunctiondatasolution}
In this section we will put forward a solution for the modeling of the conjunction and disjunction data of Hampton (1988a,b), proposing subspaces that describe the aspect `new concept' for conjunction `$A$ and $B$' and disjunction `$A$ or $B$'. We will first focus on the disjunction data and, as we will see, we can use the previously worked out disjunction data to model the conjunction data.
 
\subsection{Choice of the Subspace Representing the Emergent Concept} \label{choicesubspaceC}
We will use the machinery of orthogonal transformations to determine the 2-dimensional subspace ${\cal C}$ representing the new concept of disjunction `$A$ or $B$' of two concepts $A$ and $B$. We intend to make explicit the orthogonal transformation of basis $\{e_{AB}, e_{AB'}, e_{A'B}, e_{A'B'}, e_5, e_6, e_7,$ $ e_8\}$ into basis $\{e_{AB}, e_{AB'}, e_{A'B}, e_{A'B'}, f_5,$ $f_6, f_7, f_8\}$, because this makes it easier to have a geometrical interpretation, and, as we will see, it also enables us to characterize the subspace by means of one angle $\phi$. First we rotate over an angle of $90^\circ$ in the plane formed by $e_6$ and $e_8$. Such a rotation is represented by the orthogonal matrix
\begin{equation} \label{eqmatrixbasetransformation}
R_{68}(90^o)=\left(
\begin{array}{cccccccc}
1 & 0 & 0 & 0 & 0 & 0 & 0 & 0 \\
0 & 1 & 0 & 0 & 0 & 0 & 0 & 0 \\
0 & 0 & 1 & 0 & 0 & 0 & 0 & 0 \\
0 & 0 & 0 & 1 & 0 & 0 & 0 & 0 \\
0 & 0 & 0 & 0 & 1 & 0 & 0 & 0 \\
0 & 0 & 0 & 0 & 0 & 0 & 0 & -1 \\
0 & 0 & 0 & 0 & 0 & 0 & 1 & 0 \\
0 & 0 & 0 & 0 & 0 & 1 & 0 & 0
\end{array}
\right)
\end{equation}
Indeed, this matrix leaves $e_{AB}, e_{AB'}, e_{A'B}, e_{A'B'}, e_5$ and $e_7$ unchanged, and transforms $e_6$ into $e_8$ and $e_8$ into $-e_6$. This means that we have $f_5=e_5$, $f_6=e_8$, $f_7=e_7$ and $f_8=-e_6$ after the rotation, and that the subspace ${\cal C}$ would be ${\cal C}=\{(0,0,0,0,x_5,0,0,x_8)\ \vert\ x_5, x_8 \in \real\}$
if this was the only rotation considered. We need another rotation to characterize subspace ${\cal C}$. Let us specify this second rotation. We consider vector $e_{bisec}={1 \over \sqrt{2}}(e_6+e_7)$, which is the vector on the bisector of the plane formed by $e_6$ and $e_7$, and its orthogonal $e_{bisec}^\perp={1 \over \sqrt{2}}(-e_6+e_7)$. We now rotate over an angle $\phi$ in the plane formed by $e_5$ and $e_{bisec}$. Let us construct the matrix of this rotation. First we consider the matrix that describes the rotation over $45^\circ$ in the plane formed by $e_6$ and $e_7$ and denote it as $R_{67}(45^o)$. This matrix rotates $e_6$ into $e_{bisec}$ and $e_7$ into $e_{bisec}^\perp$. We also consider the matrix that rotates over an angle $\phi$ in the plane formed by $e_5$ and $e_6$, and denote it as $R_{56}(\phi)$.
We can calculate the matrix that describes the rotation over $\phi$ in the plane formed by $e_5$ and $e_{bisec}$ by multiplying three matrices. First the one that rotates backwards over $45^\circ$ in the plane formed by $e_6$ and $e_7$, i.e. $R^\tau_{67}(45^\circ)$, then the one that rotates over $\phi$ in the plane formed by $e_5$ and $e_6$, i.e. $R_{56}(\phi)$, and then the one that rotates over $45^\circ$ in the plane formed by $e_6$ and $e_7$, i.e. $R_{67}(45^\circ)$. If we multiply these three matrices $R_{5,bisec}(\phi)=R_{67}(45^\circ)R_{56}(\phi)R^\tau_{67}(45^\circ)$, and then multiply this matrix $R_{5,bisec}(\phi)$ by the original matrix of base transformation $R_{68}(90^\circ)$ given in (\ref{eqmatrixbasetransformation}), we obtain the final matrix of the base transformation and the equations for $f_5$, $f_6$, $f_7$ and $f_8$
\begin{equation}
R_{5,bisec}(\phi)R_{68}(90^\circ)=\left(
\begin{array}{cccccccc}
1 & 0 & 0 & 0 & 0 & 0 & 0 & 0 \\
0 & 1 & 0 & 0 & 0 & 0 & 0 & 0 \\
0 & 0 & 1 & 0 & 0 & 0 & 0 & 0 \\
0 & 0 & 0 & 1 & 0 & 0 & 0 & 0 \\
0 & 0 & 0 & 0 & \cos\phi & 0 & -{1 \over \sqrt{2}}\sin\phi & {1 \over \sqrt{2}}\sin\phi \\
0 & 0 & 0 & 0 & {1 \over \sqrt{2}}\sin\phi & 0 & {1 \over 2}(\cos\phi-1) & -{1 \over 2}(\cos\phi+1) \\
0 & 0 & 0 & 0 & {1 \over \sqrt{2}}\sin\phi & 0 & {1 \over 2}(\cos\phi+1) & -{1 \over 2}(\cos\phi-1) \\
0 & 0 & 0 & 0 & 0 & 1 & 0 & 0
\end{array}
\right)
\end{equation}
\begin{eqnarray}
&f_5=(0, 0, 0, 0, \cos\phi, {1 \over \sqrt{2}}\sin\phi, {1 \over \sqrt{2}}\sin\phi, 0)=e_5\cos\phi + {e_6 \over \sqrt{2}}\sin\phi+ {e_7 \over \sqrt{2}}\sin\phi \\
&f_6=(0, 0, 0, 0, 0, 0, 0, 1)=e_8 \\
&f_7=(0, 0, 0, 0,-{1 \over \sqrt{2}}\sin\phi, {1 \over 2}(\cos\phi-1), {1 \over 2}(\cos\phi+1), 0) \nonumber \\
&=-{e_5 \over \sqrt{2}}\sin\phi+{e_6 \over 2}(\cos\phi-1)+{e_7 \over 2}(\cos\phi+1) \\
&f_8=(0, 0, 0, 0, {1 \over \sqrt{2}}\sin\phi,-{1 \over 2}(\cos\phi-1),-{1 \over 2}(\cos\phi+1), 0) \nonumber \\
&={e_5 \over \sqrt{2}}\sin\phi-{e_6 \over 2}(\cos\phi-1)-{e_7 \over 2}(\cos\phi+1) 
\end{eqnarray}
The transposed matrix of this one gives us the coordinates with respect to the new base in function of the coordinates with respect to the old base and the corresponding equations. Hence we have
\begin{equation}
\left(
\begin{array}{c}
x_{AB} \\
x_{AB'} \\
x_{A'B} \\
x_{A'B'} \\
c_5  \\
c_6  \\
c_7  \\
c_8 
\end{array}
\right)=\left(
\begin{array}{cccccccc}
1 & 0 & 0 & 0 & 0 & 0 & 0 & 0 \\
0 & 1 & 0 & 0 & 0 & 0 & 0 & 0 \\
0 & 0 & 1 & 0 & 0 & 0 & 0 & 0 \\
0 & 0 & 0 & 1 & 0 & 0 & 0 & 0 \\
0 & 0 & 0 & 0 & \cos\phi & {1 \over \sqrt{2}}\sin\phi & {1 \over \sqrt{2}}\sin\phi & 0 \\
0 & 0 & 0 & 0 & 0 & 0 & 0 & 1 \\
0 & 0 & 0 & 0 & -{1 \over \sqrt{2}}\sin\phi & {1 \over 2}(\cos\phi-1) & {1 \over 2}(\cos\phi+1) & 0 \\
0 & 0 & 0 & 0 & {1 \over \sqrt{2}}\sin\phi & -{1 \over 2}(\cos\phi+1) & -{1 \over 2}(\cos\phi-1) & 0
\end{array}
\right) \cdot \left(
\begin{array}{c}
x_{AB} \\
x_{AB'} \\
x_{A'B} \\
x_{A'B'} \\
x_5  \\
x_6  \\
x_7  \\
x_8 
\end{array}
\right)
\end{equation}
\begin{equation}
c_5=x_5\cos\phi+{x_6+x_7 \over \sqrt{2}}\sin\phi \quad {\rm and} \quad c_6=x_8
\end{equation}

\subsection{Determining the basic equations and the condition of modularity}

The set of basic equations (\ref{eqweights01}), (\ref{eqweights02}), (\ref{eqweightdisj}) and (\ref{eqweights04}) that determine the model now are
\begin{eqnarray} \label{eqdisj01A}
&\mu(A)=x_{AB}^2+x_{AB'}^2+x_5^2+x_{A}^2 \\ \label{eqdisj02A}
&\mu(B)=x_{AB}^2+x_{A'B}^2+x_5^2+x_{B}^2 \\ \label{eqdisj03A}
&\mu(A\ {\rm or}\ B)=x_{AB}^2+x_{AB'}^2+x_{A'B}^2+(x_5\cos\phi+{x_6+x_7 \over \sqrt{2}}\sin\phi)^2+x_8^2 \\ \label{eqdisj04A}
&1=x_{AB}^2+x_{AB'}^2+x_{A'B}^2+x_{A'B'}^2+x_5^2+x_6^2+x_7^2+x_8^2
\end{eqnarray}
and the derived equations (\ref{eqdisjunctionweights05}), (\ref{eqdisjunctionweights06}), (\ref{eqdisjunctionweights07}) and (\ref{eqdisjunctionweights08}) are
\begin{eqnarray} \label{eqdisj05A}
&x_{AB}^2=\mu(A)+\mu(B)-\mu(A\ {\rm or}\ B)+(x_5\cos\phi+{x_6+x_7 \over \sqrt{2}}\sin\phi)^2-2x_5^2+x_8^2-x_{B}^2-x_{A}^2 \\ \label{eqdisj06A}
&x_{AB'}^2=\mu(A\ {\rm or}\ B)-\mu(B)-(x_5\cos\phi+{x_6+x_7 \over \sqrt{2}}\sin\phi)^2-x_8^2+x_5^2+x_{B}^2 \\ \label{eqdisj07A}
&x_{A'B}^2=\mu(A\ {\rm or}\ B)-\mu(A)-(x_5\cos\phi+{x_6+x_7 \over \sqrt{2}}\sin\phi)^2-x_8^2+x_5^2+x_{A}^2 \\ \label{eqdisj08A}
&x_{A'B'}^2=1-\mu(A\ {\rm or}\ B)+(x_5\cos\phi+{x_6+x_7 \over \sqrt{2}}\sin\phi)^2-x_5^2-x_6^2-x_7^2
\end{eqnarray}
We want to diminish the number of variables and substitute $x_6$ and $x_7$ in function of $x_{A}$ and $x_{B}$. Using (\ref{eqxA}) and (\ref{eqxB}), we calculate $x_6$ and $x_7$ in function of $x_{A}$ and $x_{B}$. We get
\begin{eqnarray}
&x_{A}\cos({\pi \over 4}-{\theta \over 2})=x_6\sin({\pi \over 4}+{\theta \over 2})\cos({\pi \over 4}-{\theta \over 2})+x_7\cos({\pi \over 4}+{\theta \over 2})\cos({\pi \over 4}-{\theta \over 2}) \\
&x_{B}\cos({\pi \over 4}+{\theta \over 2})=x_6\sin({\pi \over 4}-{\theta \over 2})\cos({\pi \over 4}+{\theta \over 2})+x_7\cos({\pi \over 4}-{\theta \over 2})\cos({\pi \over 4}+{\theta \over 2})
\end{eqnarray}
\begin{equation}
x_{A}\cos({\pi \over 4}-{\theta \over 2})-x_{B}\cos({\pi \over 4}+{\theta \over 2})=x_6(\sin({\pi \over 4}+{\theta \over 2})\cos({\pi \over 4}-{\theta \over 2})-\sin({\pi \over 4}-{\theta \over 2})\cos({\pi \over 4}+{\theta \over 2}))=x_6\sin\theta
\end{equation}
hence, for $\theta\not=0$ and $\theta\not=\pi$, we get (\ref{eqx6}), and an analogous calculation for $\theta\not=0$ and $\theta\not=\pi$ gives us (\ref{eqx7})
\begin{eqnarray} \label{eqx6}
&x_6={x_{A}\cos({\pi \over 4}-{\theta \over 2})-x_{B}\cos({\pi \over 4}+{\theta \over 2}) \over \sin\theta} \\  \label{eqx7}
&x_7={x_{B}\sin({\pi \over 4}+{\theta \over 2})-x_{A}\sin({\pi \over 4}-{\theta \over 2}) \over \sin\theta}
\end{eqnarray}
After a straightforward calculation we get
\begin{eqnarray}
&x_6^2+x_7^2={x_{A}^2+x_{B}^2-2x_{A}x_{B}\cos\theta \over \sin^2\theta} \\
&{x_6+x_7 \over \sqrt{2}}={1 \over 2\cos{\theta \over 2}}(x_{A}+x_{B})
\end{eqnarray}
where we have used
\begin{eqnarray}
&\cos({\pi \over 4}-{\theta \over 2})-\sin({\pi \over 4}-{\theta \over 2})={1 \over \sqrt{2}}(\cos{\theta \over 2}-\sin{\theta \over 2}-\cos{\theta \over 2}-\sin{\theta \over 2})=\sqrt{2}\sin{\theta \over 2} \\
&\sin({\pi \over 4}+{\theta \over 2})-\cos({\pi \over 4}+{\theta \over 2})={1 \over \sqrt{2}}(\cos{\theta \over 2}+\sin{\theta \over 2}-\cos{\theta \over 2}+\sin{\theta \over 2})=\sqrt{2}\sin{\theta \over 2}
\end{eqnarray}
The four basic equations (\ref{eqweights01}), (\ref{eqweights02}), (\ref{eqweightdisj}) and (\ref{eqweights04}) that determine the model now are 
\begin{eqnarray} \label{eqdisj01B}
&\mu(A)=x_{AB}^2+x_{AB'}^2+x_5^2+x_{A}^2 \\ \label{eqdisj02B}
&\mu(B)=x_{AB}^2+x_{A'B}^2+x_5^2+x_{B}^2 \\ \label{eqdisj03B}
&\mu(A\ {\rm or}\ B)=x_{AB}^2+x_{AB'}^2+x_{A'B}^2+(x_5\cos\phi+{\sin\phi \over 2\cos{\theta \over 2}}(x_{A}+x_{B}))^2+x_8^2 \\ \label{eqdisj04B}
&1=x_{AB}^2+x_{AB'}^2+x_{A'B}^2+x_{A'B'}^2+x_5^2+{x_{A}^2+x_{B}^2-2x_{A}x_{B}\cos\theta \over \sin^2\theta}+x_8^2
\end{eqnarray}
and the derived equations (\ref{eqdisjunctionweights05}), (\ref{eqdisjunctionweights06}), (\ref{eqdisjunctionweights07}) and (\ref{eqdisjunctionweights08}) now are
\begin{eqnarray} \label{eqdisj05B}
&x_{AB}^2=\mu(A)+\mu(B)-\mu(A\ {\rm or}\ B)+(x_5\cos\phi+{\sin\phi \over 2\cos{\theta \over 2}}(x_{A}+x_{B}))^2+x_8^2-2x_5^2-x_{B}^2-x_{A}^2 \\ \label{eqdisj06B}
&x_{AB'}^2=\mu(A\ {\rm or}\ B)-\mu(B)-(x_5\cos\phi+{\sin\phi \over 2\cos{\theta \over 2}}(x_{A}+x_{B}))^2-x_8^2+x_5^2+x_{B}^2 \\ \label{eqdisj07B}
&x_{A'B}^2=\mu(A\ {\rm or}\ B)-\mu(A)-(x_5\cos\phi+{\sin\phi \over 2\cos{\theta \over 2}}(x_{A}+x_{B}))^2-x_8^2+x_5^2+x_{A}^2 \\ \label{eqdisj08B}
&x_{A'B'}^2=1-\mu(A\ {\rm or}\ B)+(x_5\cos\phi+{\sin\phi \over 2\cos{\theta \over 2}}(x_{A}+x_{B}))^2-x_5^2-{x_{A}^2+x_{B}^2-sx_{A}x_{B}\cos\theta \over \sin^2\theta}
\end{eqnarray}  
This is a complicated set of equations to solve. We will use a combination of a graphical method to determine the neighborhood of a solution, and a convergence method to determine the exact solution. We consider the $(x_{A}, x_{B})$ plane of the two variables $x_{A}$ and $x_{B}$, and try to find values of $x_5, x_{A}, x_{B}, x_8$, such that the expressions to the right of the equality sign in equations (\ref{eqdisj05B}), (\ref{eqdisj05B}), (\ref{eqdisj05B}) and (\ref{eqdisj05B}) are positive or zero. In this case, we can take the square root of the values of these expressions and attribute the values of these square roots to $x_{AB}$, $x_{AB'}$, $x_{A'B}$ and $x_{A'B'}$, respectively, and as a result we have a global solution, a value for the vector $x=(x_{AB}, x_{AB'}, x_{A'B}, x_{A'B'}, x_5, x_6, x_7, x_8)$ after having calculated the values of $x_6$ and $x_7$ using (\ref{eqx6}) and (\ref{eqx7}).

Let us give a concrete example to illustrate this technique. We consider the item {\it Discus Throwing} for the pair of concepts {\it Hobbies} and {\it Games} and their disjunction {\it Hobbies or Games}. The membership weights measured in Hampton (1988a) are $\mu(A)=1$, $\mu(B)=0.75$ and $\mu(A\ {\rm or}\ B)=0.75$. Figure 1 represents, for the four equations (\ref{eqdisj05B}), (\ref{eqdisj05B}), (\ref{eqdisj05B}) and (\ref{eqdisj05B}), the regions where the right-hand side of each equation is positive. We have taken $\theta=108.4354^\circ$ and $\phi=12^\circ$. These regions are formed by ellipses for equations 1 and 4. We have written $eq\ 1$ and $eq\ 4$ inside both ellipses, which are the regions where the right-hand side of equations 1 and 4 are positive. The regions are formed by hyperbola for equations 2 and 3. We have written $eq\ 2$ and $eq\ 3$ at the sides of the hyperbola where the right-hand side of equations 2 and 3 are positive. Figure 1 shows that there is only one point, denoted as $S$, which is located in a positive region for the two ellipses and for the two hyperbola. For Figure 1 we have taken $x_5=0.725697377973203$ and $x_8=0$, so that Figure 1 shows that the solution, namely the location of point $S$, must be close to a point with coordinates $x_{A}=-0.45$ and $x_{B}=0.15$. With these values $x_5=0.725697377973203$, $x_{A}=-0.45$, $x_{B}=0.15$ $x_8=0$, we start an approximation process for the equations (\ref{eqdisj05B}), (\ref{eqdisj05B}), (\ref{eqdisj05B}) and (\ref{eqdisj05B}), subject to the constraint that the right-hand sides of all equations need to be positive or zero. In this way, following several iterations, we can find a solution whose error is smaller than $10^{-15}$. For example, in the case of the item {\it Discus Throwing}, we find $x_{AB}=0.450673881687196$, $x_{AB'}=0.260196686908723$, $x_{A'B}=0$, $x_{A'B'}=0$, $x_5=0.725697377973203$, $x_{A}=-0.450060053870078$, $x_{B}=0.142324867707058$, $x_8=0$.  Making use of (\ref{eqx6}) and (\ref{eqx7}), we find $x_6=-0.444248347107600$ and $x_7=0.072093399016160$. We have fully written out these values with an error margin smaller than $10^{-15}$ here to indicate our level of approximation for the iteration process. However, to avoid using numbers of this length, we will generally write only four decimals. Hence, the vector representing the item {\it Discus Throwing} is
$x_{\rm{\it Discus Throwing}}=(0.4507, 0.2602, 0, 0, 0.7257, -0.4442, 0.0721, 0)$
\begin{figure}[h]
\centerline {\includegraphics[width=8.5cm]{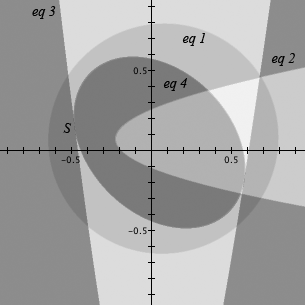}}
\caption{A graphical representation in the $(x_{A}, x_{B})$ plane of the `regions of solutions' of equations (\ref{eqdisj05B}), (\ref{eqdisj05B}), (\ref{eqdisj05B}) and (\ref{eqdisj05B}) for the item {\it Discus Throwing} with respect to the pair of concepts {\it Hobbies} and {\it Games} and their disjunction {\it Hobbies or Games}. Hampton (1988b) measured membership weights $\mu(A)=1$, $\mu(B)=0.75$ and $\mu(A\ {\rm or}\ B)=0.75$. We have $\theta=108.4354^\circ$ and $\phi=12^\circ$. The solution regions of equations 1 and 4 are ellipses, which we have denoted as $eq\ 1$ and $eq\ 4$. The solution regions of equations 2 and 3 are hyperbola, which we have denoted as $eq\ 2$ and $eq\ 3$. Point $S$ is the only point at the intersection of all four regions of solutions.}
\end{figure}
It would be beyond the scope of this article to explain in detail how we have calculated the angles for the different pairs of concepts. What we can say is that in most cases there were several options for the two angles $\theta$ and $\phi$. For two pairs of concepts, however, more specifically the pairs {\it Instruments} and {\it Tools} and {\it Sportswear} and {\it Sports Equipment}, there proved to be only a very narrow window for both angles allowing to find solutions for all items.

Note that if we have a solution, we can calculate the total classical weight $\mu_c(total)$  and the total quantum weight $\mu_q(total)$ corresponding to this solution, and also the `classical weights', $\mu_c(A)$, $\mu_c(B)$ and $\mu_c(A\ {\rm or}\ B)$, and the `quantum weights', $\mu_q(A)$, $\mu_q(B)$ and $\mu_q(A\ {\rm or}\ B)$, corresponding to this solution, as well as the `relative classical weights', $\mu_c^r(A)$, $\mu_c^r(B)$ and $\mu_c^r(A\ {\rm or}\ B)$, and the `relative quantum weights', $\mu_q^r(A)$, $\mu_q^r(B)$ and $\mu_q^r(A\ {\rm or}\ B)$, corresponding to this solution. These values describe the classical and quantum aspects of the considered item with respect to the considered pairs of concepts and their disjunction. These are given by the following equations:
\begin{eqnarray} \label{eqmutotal}
&\mu_c(total)=x_{AB}^2+x_{AB'}^2+x_{A'B}^2+x_{A'B'}^2 \quad \mu_q(total)=x_5^2+x_6^2+x_7^2+x_8^2 \\ \label{eqmucq}
&\mu_c(A)=x_{AB}^2+x_{AB'}^2 \quad \mu_q(A)=x_5^2+x_{A}^2 \quad \mu_c(B)=x_{AB}^2+x_{A'B}^2 \quad \mu_q(B)=x_5^2+x_{B}^2\\ \label{eqmucqAorB}
&\mu_c(A\ {\rm or}\ B)=x_{AB}^2+x_{AB'}^2+x_{A'B}^2 \quad \mu_q(A\ {\rm or}\ B)=(x_5\cos\phi+{\sin\phi \over 2\cos{\theta \over 2}}(x_{A}+x_{B}))^2+x_8^2 \\ \label{eqmucr}
&\mu_c^r(A)={\mu_c(A) \over \mu_c(total)} \quad \mu_c^r(B)={\mu_c(B) \over \mu_c(total)} \quad \mu_c^r(A\ {\rm or}\ B)={\mu_c(A\ {\rm or}\ B) \over \mu_c(total)} \\ \label{eqmuqr}
&\mu_q^r(A)={\mu_q(A) \over \mu_q(total)} \quad \mu_q^r(B)={\mu_q(B) \over \mu_q(total)} \quad \mu_q^r(A\ {\rm or}\ B)={\mu_q(A\ {\rm or}\ B) \over \mu_q(total)}
\end{eqnarray}  
To further clarify the meaning of these new quantities, let us calculate them for the case of the solution we found for the item {\it Discus Throwing} with respect to the pair of concepts {\it Hobbies} and {\it Games} and their disjunction {\it Hobbies or Games}. We find $\mu_c(total)=0.2708$, $\mu_q(total)=0.7292$, $\mu_c(A)=0.2708$, $\mu_q(A)=0.7292$, $\mu_c(B)=0.2031$, $\mu_q(B)=0.5469$, $\mu_c(A\ {\rm or}\ B)=0.2708$, $\mu_q(A\ {\rm or}\ B)=0.4292$, $\mu_c^r(A)=1$, $\mu_c^r(B)=0.75$, $\mu_c^r(A\ {\rm or}\ B)=1$, $\mu_q^r(A)=1$, $\mu_q^r(B)=0.75$ and $\mu_q^r(A\ {\rm or}\ B)=0.5886$.

A closer examination of the `relative classical and quantum weights' is the most revealing for it shows us that the relative weights of the item {\it Discus Throwing} equal the originally measured weights for both concepts $A$ and $B$. The relative classical weight of the item with respect to the disjunction equals 1, which is the maximum of the weights with respect to both concepts. This means that the relative classical weight satisfies the maximum rule of fuzzy set theory. The relative quantum weight with respect to the disjunction is 0.5886, which is smaller than the original weight of 0.7 with respect to the disjunction. This means that the relative quantum weight `confirms' the underextension measured in the experiment. In fact, the measured underextension is a result of the quantum aspects of the item, and the underextension for the relative quantum weights is stronger than the underextension measured. Mixed with the `maximum rule satisfying relative classical weight' it results in the underextension measured in the experiment.

A more correct interpretation would be to acknowledge the effect of `quantum superposition'. The classical aspect, expressed by the relative classical weights, behaves in line with the maximum rule of fuzzy set for the disjunction, while the quantum aspect gives rise to underextension.

Before we proceed, we will introduce an extra condition, which we will call the `modularity condition'. This condition is that the classical weights, i.e. the weights that would result if only the classical aspects of the item were involved, must be proportional to the measured weights for both concepts $A$ and $B$. The reason for this condition is that on measuring $\mu(A)$ and $\mu(B)$, there is no combination of concepts yet. This implies that the classical weight involving a combination should be proportional to the weights that do not involve a combination. In this way the classical weights will be the same, subject to a proportionality factor, if the concepts under consideration are combined with other concepts. The modularity condition can be expressed as follows:
\begin{equation} \label{modularity}
\mu(A) \cdot \mu_c(B) = \mu(B) \cdot \mu_c(A)
\end{equation}
Note that the condition makes that, for $\mu(A)\not=0$ and $\mu(B)\not=0$, we have
\begin{equation}
{\mu_c^r(A) \over \mu(A)}={\mu_c(A) \over \mu_c(total) \cdot \mu(A)}={\mu_c(B) \over \mu_c(total) \cdot \mu(B)}={\mu_c^r(B) \over \mu(B)}
\end{equation}
which means that the `relative classical weights' are proportional to the measured weights.

For the item {\it Discus Throwing}, this factor of proportionality is equal to 1, and as a consequence the relative classical weight of {\it Discus Throwing} for both concepts is equal to the measured weight. This is true of many items, as we can see in Tables 3 and 4, where we have calculated the relative classical and relative quantum weights for all items with respect to all pairs of concepts in Hampton(1988a,b). For some items the modularity factor is different from 1. Table 3 presents the values of the relative classical and quantum weights with respect to all items and pairs of concepts and their disjunctions as referred to in Hampton (1988b), and also the components of the vectors representing these items.

\subsection{Working out the Conjunction Data}

To model the conjunction data, we can directly use the modeling of the disjunction data worked out in the previous sections. Indeed, suppose that we consider a pair of concepts $A$ and $B$ and their conjunction `$A$ and $B$', and an item $X$, with $\mu(A)$, $\mu(B)$, $\mu(A\ {\rm and}\ B)$ being the membership weights of item $X$ with respect to $A$, $B$ and `$A$ and $B$'. If we take equations (\ref{eqweights01}), (\ref{eqweights02}), (\ref{eqweightconj}) and (\ref{eqweightdisj}), which determine the model for the disjunction and the conjunction, we can see that if we consider two hypothetical concepts $A'$, $B'$ and their disjunction `$A'$ or $B'$' such that $\mu(A')=1-\mu(A)$, $\mu(B')=1-\mu(B)$ and $\mu(A'\ {\rm or}\ B')=1-\mu(A\ {\rm and}\ B)$ are the membership weights of an item $X$ with respect to $A'$, $B'$ and their disjunction `$A'$ or $B'$', the same vector $x$ that models the item $X$ for the concepts $A'$, $B'$ and `$A'$ or $B'$', also models this item $X$ for the concepts $A$, $B$ and `$A$ and $B$'.

Here is a concrete example to clarify what we mean. Let us consider the item {\it TV} with respect to the pair of concepts {\it Furniture} and {\it Household Appliances} and their conjunction `{\it Furniture and Household Appliances}'. We have $\mu(A)=0.7$, $\mu(B)=0.9$ and $\mu(A\ {\rm and}\ B)=0.925$. Hence we introduce $A'$, with $\mu(A')=1-\mu(A)=0.3$, $B'$, with $\mu(B')=1-\mu(B)=0.1$, and `$A'$ or $B'$' with $\mu(A' {\rm or}\ B')=1-\mu(A\ {\rm and}\ B)=0.075$. Suppose we construct the vector $x$, as we have done in previous sections, modeling item $X$ with respect to $A'$, $B'$ and `$A'$ or $B'$'. Replacing $A$ with $A'$ and $B$ with $B'$ in the equations (\ref{eqweights01}), (\ref{eqweights02}) and (\ref{eqweightdisj}), we obtain
\begin{eqnarray}
&\mu(A')=x_{A'B'}^2+x_{A'B}^2+x_8^2+x_{A'}^2 \\
&\mu(B')=x_{A'B'}^2+x_{AB'}^2+x_8^2+x_{B'}^2 \\
&\mu(A'\ {\rm or}\ B')=x_{A'B'}^2+x_{A'B}^2+x_{AB'}^2+c_7^2+c_8^2 \\
&1=x_{A'B'}^2+x_{A'B}^2+x_{AB'}^2+x_{AB}^2+x_8^2+x_7^2+x_6^2+x_5^2
\end{eqnarray}
If we now apply (\ref{eqequality01}), it follows from this that
\begin{equation} \label{eqmuA}
\mu(A)=1-\mu(A')=x_{AB}^2+x_{AB'}^2+x_6^2+x_7^2-x_{A'}^2
=x_{AB}^2+x_{AB'}^2+x_{A}^2
\end{equation}
In an analogous way, again making use of (\ref{eqequality01}), it follows that
\begin{equation} \label{eqmuB}
\mu(B)=x_{AB}^2+x_{A'B}^2+x_{B}^2
\end{equation}
and, making use of (\ref{eqequality02}), we also get
\begin{equation} \label{eqmuAandB}
\mu(A\ {\rm and}\ B)=1-\mu(A'\ {\rm or}\ B')=x_{AB}^2+x_8^2+x_7^2+x_6^2+x_5^2-c_7^2-c_8^2 
=x_{AB}^2+c_5^2+c_6^2
\end{equation}
If we compare equations (\ref{eqweights01}), (\ref{eqweights02}) and (\ref{eqweightconj}) with equations (\ref{eqmuA}), (\ref{eqmuB}) and (\ref{eqmuAandB}), we can see that $x$ models the conjunction data $\mu(A)$, $\mu(B)$ and $\mu(A\ {\rm and}\ B)$ of item $X$. In Table 4 we have calculated for each item $X$ and for each pair of concepts $A$, $B$ and their conjunction $A$ and $B$ the vector $x_{X}$ - element of $\real^8$ - that represents this item $X$. We have also calculated the `relative classical and quantum weights' for each item $X$ with respect to the different pairs of concepts and their conjunction measured in Hampton (1988a). Analogous to our discussion of the disjunction, we introduce the following equations for the conjunction:
\begin{eqnarray} \label{eqmucqAandB}
&\mu_c(A\ {\rm and}\ B)=x_{AB}^2 \quad \mu_q(A\ {\rm and}\ B)=c_5^2+c_6^2 \\ \label{eqmucqrconj}
&\mu_c^r(A\ {\rm and}\ B)={\mu_c(A\ {\rm and}\ B) \over \mu_c(total)} \quad \mu_q^r(A\ {\rm and}\ B)={\mu_q(A\ {\rm and}\ B) \over \mu_q(total)}
\end{eqnarray} 
To illustrate this, let us consider the item {\it Library} with respect to the pair of concepts {\it Building} and {\it Dwelling} and their conjunction {\it Building and Dwelling}. Hampton (1988a) measured $\mu(A)=0.95$, $\mu(B)=0.175$ and $\mu(A\ {\rm and}\ B)=0.3077$, which is a case of `overextension'. Table 4 shows that vector $x$ representing this item is given by $x_{\rm{\it Library}}=(0.3809, 0.8015, 0, 0.2036, 0, 0.3927, 0.0919, 0.0923)$, and for the relative classical and relative quantum weights we find $\mu_c^r(A)=0.95$, $\mu_c^r(B)=0.175$, $\mu_c^r(A\ {\rm and}\ B)=0.175$, $\mu_q^r(A)=0.95$, $\mu_q^r(B)=0.175$ and $\mu_q^r(A\ {\rm and}\ B)=0.9503$. The relative classical weights satisfy the minimum rule of fuzzy set theory. On the other hand, the relative quantum weights involve a far greater overextension than that of the measured values of the membership weights in Hampton (1988a). The reason for this is that these values are determined by a combination of classical and quantum aspects of the item {\it Library}. The weights of the classical aspects that satisfy the minimum rule are responsible for keeping the overextension caused by the quantum aspects restricted to the moderate overextension apparent in the weights found in Hampton (1988a). In the next section we will elaborate on the hypothesis that we introduced in subsection \ref{QuantumFieldTheory} about the `superposed layers of the human thought'. 

\subsection{Classical Logical and Quantum Conceptual Thought}

In section \ref{QuantumFieldTheory} we introduced the idea of the two `superposed layers of human thought'. In this section we will give several examples to illustrate these layers, taking advantage of the explicit nature of the worked out superposition and the introduction of the corresponding relative classical and relative quantum weights.

We consider the pair of concepts {\it Spices} and {\it Herbs} and their disjunction {\it Spices or Herbs}. In Table 3 we can see that different types of items have been tested by Hampton (1988b) in relation with this pair of concepts. Let us first consider one of the $k$-type non-classical items, for example {\it MSG}. Total classical weight is $0.6950$ and total quantum weight is $0.3050$. Hence the item's behavior is `more classical than quantum', or `classical logical thought' is more dominant than `quantum conceptual thought'. The relative classical weights are $\mu_c^r(A)=0.15$, $\mu_c^r(B)=0.1$ and $\mu_c^r(A\ {\rm or}\ B)=0.25$, compared to the measured weights of Hampton (1988b), which are $\mu(A)=0.15$, $\mu(B)=0.1$, $\mu(A\ {\rm or}\ B)=0.425$. We can see that the weights with respect to the individual concepts $A$ and $B$ are equal. However, for the weight with respect to the disjunction, we see that the relative classical weight is less. This must be so, because {\it MSG} is a $k$-type non-classical item, which means that $\mu(A)+\mu(B) < \mu(A\ {\rm or}\ B)$. Note that the relative classical weight is exactly equal to $\mu(A)+\mu(B)$, which means that the `classical part' of the item has done the best it can in coping with the $k$-type non-classicality, indeed $\mu_c^r(A\ {\rm or}\ B)=0.25=\mu_c^r(A)+\mu_c^r(B)$ is as great as it can in order to remain classical. The relative quantum weights are $\mu_q^r(A)=0.15$, $\mu_q^r(B)=0.1$ and $\mu_q^r(A\ {\rm or}\ B)=0.8239$. Hence most of all $\mu_q^r(A\ {\rm or}\ B)$ is very great. This means that the subjects considered item {\it MSG} to be very much characteristic of {\it Spices or Herbs}. In other words, they considered {\it MSG} to be one of those items that typically make one doubt about whether they are {\it Spices} or whether they are {\it Herbs}, while at the same time they found {\it MSG} to be not very characteristic of {\it Spices} alone, or {\it Herbs} alone. This is reflected by $\mu(A)=\mu_c^r(A)=\mu_q^r(A)=0.15$ and $\mu(A)=\mu_c^2(A)=\mu_q^r(A)=0.1$.

We find a similar pattern for the other $k$-type non-classical items {\it Saccharin}, {\it Sugar}, {\it Vinegar} and {\it Lemon Juice}. For {\it Vinegar} the value found is even greater: $\mu_q^r(A\ {\rm or}\ B)=0.9000$. The most extreme case is that of {\it Sugar}, which subjects qualified as `does not belong to {\it Spices}' ($\mu(A)=0$), and `does not belong to {\it Herbs}' ($\mu(B)=0$), but `does belong moderately to {\it Spices or Herbs}' ($\mu(A\ {\rm or}\ B)=0.2$). If we look at the elements of its quantum representation in Table 3, we can see the following. We have $\mu_c^2(A)=\mu_q^r(A)=0$ and $\mu_c^2(B)=\mu_q^r(B)=0$, which is not a surprise. Also, we have $\mu_c^r(A\ {\rm or}\ B)=0$, which means that the subjects indeed did not assign any weight to the membership of the disjunction as far as this is determined by `classical logical thought'. And Table 3 shows that $\mu_q^r(A\ {\rm or}\ B)=1$. Consequently, as far as `quantum conceptual thought' is concerned, subjects decided that the membership weight of the item {\it Sugar} with respect to the new concept {\it Spices or Herbs} equals 1. To make this more concrete, we can imagine the following situation. Asked whether the item {\it Sugar} is a member of {\it Spices or Herbs}, a subject may be reasoning as follows: `It is clear that sugar is not a spice and nor is it a herb (which means that following the rules of classical logic it is not a `spice or herb'), but then again, sugar is a substance that one may readily doubt about whether it is a spice or whether it is a herb. Hence, in this sense, it is a member of `{\it Spices or Herbs}'. As said, such thought processes do not happen serially, but in parallel, or rather, `in superposition'.

Let us now turn to the $\Delta$-type non-classical items, for example the item {\it Salt}. Subjects found it to be much of a member of {\it Spices} ($\mu(A)=0.75$), but not very much a member of {\it Herbs} ($\mu(B)=0.1$), while the membership weight with respect to the disjunction is considerable ($\mu(A\ {\rm or}\ B)=0.6$). `Classical logical thought' gives rise to values for the relative classical weights that are the following $\mu_c^r(A)=0.75$, $\mu_c^r(B)=0.1$, $\mu_c^r(A\ {\rm or}\ B)=0.75$. These values satisfy the maximum rule of fuzzy set theory, and hence are the ones one would intuitively expect for the disjunction. For the quantum relative weights we find $\mu_q^r(A)=0.75$, $\mu_q^r(B)=0.1$, $\mu_q^r(A\ {\rm or}\ B)=0.2955$. The underextension here is very significant, much greater than for the measured values. It means that the subjects consider {\it Salt} an item that raises hardly any doubts as to whether it is a member of {\it Spices} or a member of {\it Herbs}, even though it is not a very strong member of either of the two. 

A similar pattern appears for the other $\Delta$-type non-classical items with respect to this pair of concepts, {\it Curry}, {\it Oregano}, {\it Chili Pepper}, {\it Mustard}, {\it Turmeric}, {\it Vanilla}, {\it Chires} and {\it Root Ginger}. Note that many of these $\Delta$-type non-classical items have a great total quantum weight, for example {\it Mustard} has $\mu_q(total)=0.6448$. This means the `new concept' has a great impact, also for these $\Delta$-type non-classical items, and even more for these than for the previously considered $k$-type non-classical items. But the presence of the `new concept' can mainly be explained by the fact that the subjects do not regard these items as characteristic of the new concept `$A$ or $B$'. So whereas for a $k$-type non-classical item, subjects will argue more or less by saying `that this is an item that indeed raises doubts about whether it is a member of {\it Spices} or a member of {\it Herbs}, for $\Delta$-type non-classical items they will also -- and very much so -- take into account the new concept `$A$ or $B$', and decide, unlike in the case of the $k$-type non-classical items, that these $\Delta$-type non-classical items are `not' characteristic of `$A$ or $B$' and hence `not' the type of items that raise doubts about whether they are members of $A$ or members of $B$.

What about the classical items? Well, our model shows that they are not really classical. We could have decided to model each of the classical items purely within the classical part, i.e. the subspace generated by the vectors $\{e_{AB}, e_{AB'}, e_{A'B}, e_{A'B'}\}$ of $\real^8$. The membership weights for classical items satisfy the inequalities characterizing classical data, so that this modeling would have been possible, as made explicitly clear in subsection \ref{quantumrule}. However, in view of the results yielded by our quantum model, it is much more plausible to assume that a classical item is classical only in appearance and that in reality it is a superposition of a classical item with a greater value for the disjunction than the value measured and superposed with a $\Delta$-type non-classical item with underextension for the disjunction. This is how we have modeled the classical items, taking for the value of the `relative classical disjunction' the medium of the measured value and the maximum value. Here is an example. Let us consider the item {\it Poppyseeds} with respect to the pair of concepts {\it Spices} and {\it Herbs} and their disjunction {\it Spices or Herbs}. The measured values of the membership weights are $\mu(A)=0.4$, $\mu(B)=0.4$ and $\mu(A\ {\rm or}\ B)=0.4$. The maximum value of the disjunction is the sum of $\mu(A)$ and $\mu(B)$, hence $0.8$. Hence we have chosen to take the relative classical weight of the disjunction equal to the medium of the maximum value $0.8$ and the measured value $0.4$, i.e. $\mu_c^r(A\ {\rm or}\ B)=0.6$. As a consequence we find $\mu_c^r(A)=0.4$, $\mu_c^r(B)=0.4$, $\mu_c^r(A\ {\rm or}\ B)=0.6$, $\mu_q^r(A)=0.4$, $\mu_q^r(B)=0.4$, $\mu_q^r(A\ {\rm or}\ B)=0.1270$. This gives a very small value for the membership weight with respect to the new concept `{\it Spices or Herbs}'. Subjects find {\it Poppyseeds} not at all characteristic of items that raise doubts as to whether they are {\it Spices} or {\it Herbs}.

We have so far compared `classical logical thought' and `quantum conceptual thought' only for the situation of disjunction. We will therefore now discuss an example of conjunction. We will first consider a $\Delta$-type non-classical item, for example the item {\it Dogsled} with respect to the pair of concepts {\it Machine} and {\it Vehicle} and their conjunction {\it Machine and Vehicle}. The measured membership weights are $\mu(A)=0.1795$, $\mu(B)=0.925$ and $\mu(A\ {\rm and}\ B)=0.275$. This points to a moderate overextension for the conjunction. For the relative classical membership weights, we find $\mu(A)_c^r=0.1795$, $\mu(B)_c^r=0.925$ and $\mu(A\ {\rm and}\ B)_c^r=0.1795$. This means that the minimum rule of fuzzy set theory is satisfied for the conjunction weight. For the relative quantum membership weights, we find $\mu(A)_q^r=0.1795$, $\mu(B)_q^r=0.925$ and $\mu(A\ {\rm and}\ B)_q^r=0.9829$. With respect to the new concept {\it Machine and Vehicle}, and with the subjects now reflecting in terms of `quantum conceptual thought', {\it Dogsled} is found to be only very weakly a member of {\it Machine} (0.1795) and very strongly a member of {\it Vehicle} (0.925). However, {\it Dogsled} is even more strongly a member of the new concept `{\it Machine and Vehicle}' (0.9829).

A similar pattern is found for the other $\Delta$ type items, {\it Bicycle}, {\it Roadroller},  {\it Elevator}, {\it Course Liner}, {\it Skateboard}, {\it Bulldozer}, {\it Lawn Mover} and {\it Ski Lift}. We already said that the proportionality factor related to the modularity condition is not always 1, let us therefore consider a relevant example. For the item {\it Course Liner} we have $\mu(A)=0.875$, $\mu(B)=0.875$, $\mu(A\ {\rm and}\ B)=0.95$, $\mu(A)_c^r=0.9424$, $\mu(B)_c^r=0.9424$, $\mu(A\ {\rm and}\ B)_c^r=0.9424$, $\mu(A)_q^r=0.4306$, $\mu(B)_q^r=0.4306$ and $\mu(A\ {\rm and}\ B)_q^r=1$. This means that subjects find {\it Course Liner} to be as much a {\it Machine} as a {\it Vehicle}, but the relative classical weights are much greater (0.9424) than the relative quantum weights (0.4306); together they middle out to the measured weights (0.875). The relative quantum weight of a member of the new concept {\it Machine and Vehicle} is 1.

\section{Fundamentals of Concept Formation and Combination}

The above might suggest that the process of thought we have discussed is very specific. In this section we will show that it is quite the contrary, it is fundamental and directly related to the process of concept formation.

\subsection{General Concept Formation} \label{conceptformation}

Often a concept can be presented as the disjunction -- the `or' -- of a set of other concepts. Let us consider the following example. {\it Animal} can be {\it Dog} or {\it Cat} or {\it Horse} or {\it Rabbit} or $\ldots$ followed by a long list of all the usually known animals. {\it Barking} is characteristic of {\it Dog}, so that it is fair to state that its weight with respect to {\it Dog} would be close to 1. With respect to {\it Animal}, however, {\it Barking} is not very characteristic, so that its weight would be rather small. This indicates the effect of underextension, exactly what appears in the Hampton (1988b) experiments with respect to the disjunction. In other words, if disjunction means `the formation of a new concept', then underextension is a natural effect, and also a fundamental effect.

It is also true, however, that a concept can often be presented as a conjunction -- the `and' -- of a set of other concepts. For example, {\it Dog} is {\it Has four legs} and {\it Likes to Bark} and {\it Has fur} and {\it Likes to Swim} and $\dots$ followed by a long list of characteristics of a {\it Dog}. {\it Humans Friend} is characteristic of {\it Dog}, hence we could attribute it a weight close to 1 with respect to {\it Dog}. However, {\it Humans Friend} is not very characteristic of {\it Has Four Legs} or {\it Likes to Swim} etc $\dots$, so that weights for these concepts would be rather small. This shows that if the conjunction is used to form a new concept, overextension is a natural effect, and also a fundamental effect.

These examples illustrate that the effects we have been modeling in our quantum modeling scheme are natural effects related to the capacity of the human mind to form new concepts. Moreover, we can put forward an interesting line of reasoning with regard to {\it Fodor's Puzzle of Concept Acquisition} (Fodor 1975; Margolis and Laurence, 2002). As we demonstrated in the foregoing, in our theory, conjunction and disjunction comprise two aspects which we referred to as `two-particle way' and `one-particle way'. Usually, certainly if things are considered from a logical perspective, only the `two-particle way' is identified as existing, and hence conjunction and disjunction appear as just `logical combinations that do not generate new conceptual knowledge' -- one of Fodor's arguments. In our quantum modeling scheme, `superposition' introduces a new emergent state, coming about each time concepts combine. Consequently, `new concept formation' is considered to be as important an event as the classical logical combination of concepts, and it can be readily modeled using the mathematical formalism of quantum mechanics.

\subsection{The Modeling of Large Collections of Combinations of Concepts} \label{largecollection}
In this subsection we will show how our quantum modeling scheme gives rise to a mathematical formalism for the description of large collections of combinations of concepts. The structure of Fock space again plays a fundamental role. We will not repeat here the details of this construction, which can be found in Aerts (2007b), but only outline its basic elements and suggest the potential of this approach.

Let us consider the situation of a combination of $n$ concepts and in parallel consider a concrete example, namely the sentence `The cat eats the food while the child plays in the garden', which is a combination of 12 concepts. For a combination of $n$ concepts, the quantum superposition state that corresponds to `considering the combination as one new concept' is the state ${1 \over \sqrt{n}}\sum_{i=1}^n|A_n\rangle$ where $|A_i\rangle$ is the state of concept number $i$, and $\langle A_i|A_i\rangle=1$ while $\langle A_i|A_j\rangle=0$ for $i\not=j$, all this defined in a Hilbert space ${\cal H}$. Hence the state which describes the example sentence as `one new concept' is given by ${1 \over \sqrt{12}}\sum_{i=1}^{12}|A_{12}\rangle$. Note that some concepts of such a large combination carry much more meaning than others with respect to a given context. This difference is accounted for by introducing a more general superposition state, i.e. a weighted linear combination instead of a sum, as demonstrated in Aerts (2007b), but we will not consider this situation here. The state where the $n$ concepts are considered individually, and where only their logical combinations are taken into account, is the state $|A_1\rangle \otimes \ldots \otimes|A_i\rangle \otimes \ldots \otimes |A_n\rangle$ element of the $n$-times tensor product ${\cal H}\otimes\ldots\otimes{\cal H}$. Hence, for the example sentence, this is the state $|A_1\rangle \otimes \ldots \otimes|A_i\rangle \otimes \ldots \otimes |A_{12}\rangle$. These are the two end-elements of the direct sum of Hilbert spaces that Fock space is, the first containing the description of the whole combination as one new concept, and the second containing the description of the combination as a classical combination of concepts. In between there are parts of the complete state that contain the effects due to the fact that `also parts of the combination can be considered as whole and new concepts'. In the example sentence, `The cat eats the food' and `While the child plays in the garden' are good examples of parts that will be assigned considerable weight for being considered `new concepts of their own'. How can Fock space provide a description for all these possibilities? Let us demonstrate this using the same example. The first part of the sentence consists of a combination of five concepts, so that the state ${1 \over \sqrt{5}}\sum_{i=1}^{5}|A_{5}\rangle$ is the state that describes this part of the sentence as a whole new concept. The second part of the sentence consists of the seven remaining concepts. Hence the state ${1 \over \sqrt{7}}\sum_{i=1}^{6}|A_{12}\rangle$ describes this part of sentence as a whole new concept. We now have a situation with `two concepts', i.e. one for each of both parts of the sentence. This situation is described by ${1 \over \sqrt{5}}\sum_{i=1}^{5}|A_{5}\rangle\otimes{1 \over \sqrt{7}}\sum_{i=1}^{6}|A_{12}\rangle$ element of the two times tensor product ${\cal H}\otimes{\cal H}$. This is how the Fock space model of our quantum modeling scheme realizes each possibility of considering certain subsets of the combination of $n$ concepts as individual concepts.

Fock space can become very extensive as in quantum field theory when it describes the micro-world, so that it is important to consider possibilities of approximation. The guiding idea of such approximate modeling is linked to what we have understood so far with respect to our general modeling scheme. Considering the example sentence, we can ask the question `how many instantiations are there primarily involved in the scenery of this sentence?'. Clearly, the concepts {\it Cat}, {\it Food}, {\it Child} and {\it Garden} give rise to instantiations within possible situations described by the sentence. This means that a `four-entity way', with the state described by a vector in a 4-times tensor product ${\cal H}\otimes{\cal H}\otimes{\cal H}\otimes{\cal H}$, will be a possible first-order approximate model for this sentence. If the scenery described by the sentence considers also the concepts {\it Eat} and {\it Play} as giving rise to instantiations, we get an extra element for an eventually better approximation. By adding {\it Eat} and {\it Play} to the set of concepts that determine the basic dimension of the tensor product component, we take into account `ways of eating', which gives rise to states for the concept {\it Eat}, and `ways of playing', which gives rise to states for the concept {\it Play}. This would result in a 6-times tensor product space description. The direct sum of both is an element of Fock space.

In principle, our quantum modeling scheme, with the construction of the state in Fock space as outlined above, extends to very large parts of combinations of concepts, pieces of text, documents, collections of documents, books etc \ldots. We will have to investigate the validity of this modeling by performing quantitative experiments and measuring weights related to typicalities of items and applicability of features, as we did in Aerts and Gabora (2005a,b) to prove the presence of contextual influence. Basing ourselves on the concrete modeling of the data found in Hampton (1988a,b), we are confident about the outcome.

\scriptsize
\setlongtables 
\begin{longtable}{|l@{}lll@{}lll@{}lll@{}l|} 
\caption{For each item $X$ and for each pair of concepts $A$, $B$ and their disjunction $A$ or $B$ we calculate the vector $x$, element of $\real^8$, that represents this item $X$. We also calculate the `relative classical' and `relative quantum weights' as defined in (\ref{eqmucr}) and (\ref{eqmuqr}), and the total quantum weight $\mu_q$ defined in (\ref{eqmutotal}). The total classical weight equals 1 minus the total quantum weight. Under each item we specify its {\it type}, $\Delta$ means $\Delta$-type non-classical, k means $k$-type non-classical, and c means classical.} \\
\hline
{\it Item} & $\mu(A)$ & $\mu(B)$ & $\mu(A\ {\rm or}\ B)$ & $\mu_c^r(A)$ & $\mu_c^r(B)$ & $\mu_c^r(A\ {\rm or}\ B)$ & $\mu_q^r(A)$ & $\mu_q^r(B)$ & $\mu_q^r(A\ {\rm or}\ B)$ & $\mu_q$   \\
{\it type} & $x_{AB}$ & $x_{AB'}$ & $x_{A'B}$ & $x_{A'B'}$ & $x_5$ & $x_6$ & $x_7$ & $x_8$ & $x_{A}$ & $x_{B}$ 
  \\
\endfirsthead 
\hline 
{\it Item} & $\mu(A)$ & $\mu(B)$ & $\mu(A\ {\rm or}\ B)$ & $\mu_c^r(A)$ & $\mu_c^r(B)$ & $\mu_c^r(A\ {\rm or}\ B)$ & $\mu_q(A)$ & $\mu_q(B)$ & $\mu_q(A\ {\rm or}\ B)$ & $\mu_q$  \\
{\it type} & $x_{AB}$ & $x_{AB'}$ & $x_{A'B}$ & $x_{A'B'}$ & $x_5$ & $x_6$ & $x_7$ & $x_8$ & $x_{A}$ & $x_{B}$ 
  \\
\hline \hline
\endhead
\hline
\hline
\multicolumn{11}{|l|}{\it $A$=Home Furnishing, $B$=Furniture, $\theta=131.1440^\circ$, $\phi=0^\circ$} \\
\hline
\hline
{\it
Mantelpiece
} &
0.8
&
0.4
&
0.75
&
0.8
&
0.4
&
0.8
&
0.8
&
0.4
&
0.4414
&
0.1394
\\
$\Delta$ &
0.5867
&
0.5867
&
0
&
0.4149
&
0.1918
&
-0.2756
&
0.0437
&
0.1574
&
-0.1378
&
0.2734
\\ \hline
{\it
Window Seat
} &
0.9
&
0.9
&
0.8
&
0.9
&
0.9
&
0.9
&
0.9
&
0.9
&
0.7236
&
0.5667
\\
$\Delta$ &
0.6244
&
0
&
0
&
0.2081
&
0.6166
&
-0.2799
&
0.2799
&
0.1729
&
-0.3604
&
0.3604
\\ \hline
{\it
Painting
} &
0.9
&
0.5
&
0.85
&
0.9
&
0.5
&
0.9
&
0.9
&
0.5
&
0.4680
&
0.1157
\\
$\Delta$ &
0.6649
&
0.5947
&
0
&
0.2974
&
0.2216
&
-0.2480
&
0.0070
&
0.0712
&
-0.2347
&
0.0937
\\ \hline
{\it
Light Fixture
} &
0.8
&
0.4
&
0.775
&
0.8
&
0.4
&
0.8
&
0.8
&
0.4
&
0.4512
&
0.0717
\\
\hline
\hline
\end{longtable}
\normalsize

\scriptsize
\setlongtables 
\begin{longtable}{|l@{}lll@{}lll@{}lll@{}l|} 
\caption{For each item $X$ and for each pair of concepts $A$, $B$ and their conjunction $A$ and $B$ we calculate the vector $x$, element of $\real^8$, that represents this item $X$. We also calculate the `relative classical' and `relative quantum weights' as defined in (\ref{eqmucr}), (\ref{eqmuqr}) and (\ref{eqmucqrconj}), and the total quantum weight $\mu_q$ defined in (\ref{eqmutotal}). The total classical weight equals 1 minus the total quantum weight. Under each item we specify its {\it type}, $\Delta$ means $\Delta$-type non-classical, k means $k$-type non-classical, and c means classical.} \\
\hline
{\it Item} & $\mu(A)$ & $\mu(B)$ & $\mu(A\ {\rm and}\ B)$ & $\mu_c^r(A)$ & $\mu_c^r(B)$ & $\mu_c^r(A\ {\rm and}\ B)$ & $\mu_q^r(A)$ & $\mu_q^r(B)$ & $\mu_q^r(A\ {\rm and}\ B)$ & $\mu_q$   \\
{\it type} & $x_{AB}$ & $x_{AB'}$ & $x_{A'B}$ & $x_{A'B'}$ & $x_5$ & $x_6$ & $x_7$ & $x_8$ & $x_{A}$ & $x_{B}$ 
  \\
\endfirsthead 
\hline 
{\it Item} & $\mu(A)$ & $\mu(B)$ & $\mu(A\ {\rm and}\ B)$ & $\mu_c^r(A)$ & $\mu_c^r(B)$ & $\mu_c^r(A\ {\rm and}\ B)$ & $\mu_q(A)$ & $\mu_q(B)$ & $\mu_q(A\ {\rm and}\ B)$ & $\mu_q$  \\
{\it type} & $x_{AB}$ & $x_{AB'}$ & $x_{A'B}$ & $x_{A'B'}$ & $x_5$ & $x_6$ & $x_7$ & $x_8$ & $x_{A}$ & $x_{B}$ 
  \\
\hline \hline
\endhead
\hline
\hline
\multicolumn{11}{|l|}{\it $A$=Furniture, $B$=Household Appliances, $\theta=90^\circ$, $\phi=0^\circ$} \\
\hline
\hline
{\it
Filing Cabinet
} &
0.9744
&
0.3077
&
0.5263
&
0.9744
&
0.3077
&
0.3077
&
0.9744
&
0.3077
&
0.7179
&
0.5329
\\
$\Delta$ &
0.3791
&
0.5580
&
0
&
0.1094
&
0.3877
&
0.6074
&
0.1169
&
0.0006
&
0.3689
&
0.0137
\\ \hline
{\it
Clothes Washer
} &
0.15
&
1
&
0.725
&
0.15
&
1
&
0.15
&
0.15
&
1
&
0.85
&
0.8214
\\
$\Delta$ &
0.1637
&
0
&
0.3896
&
0
&
0.3510
&
0
&
0.8356
&
0
&
0
&
0.6982
\\ \hline
{\it
Vacuum Cleaner
} &
0.075
&
1
&
0.3846
&
0.075
&
1
&
0.075
&
0.075
&
1
&
0.9250
&
0.3643
\\
$\Delta$ &
0.2184
&
0
&
0.7669
&
0
&
0.1653
&
0
&
-0.5805
&
0
&
0
&
0.3369
\\ \hline
{\it
Hifi
} &
0.5789
&
0.7895
&
0.7895
&
0.6923
&
0.8462
&
0.6923
&
0.3333
&
0.6667
&
1
&
0.3158
\\
\hline
\hline
\end{longtable}
\normalsize

\section*{References}
\begin{description}
\setlength{\itemsep}{-2mm}
\item Aerts, D. (2002). Being and change: foundations of a realistic operational formalism. In D. Aerts, M. Czachor and T. Durt (Eds.), {\it Probing the Structure of Quantum Mechanics: Nonlinearity, Nonlocality, Probability and Axiomatics} (pp. 71-110). Singapore: World Scientific. Archive reference and link: http://uk.arxiv.org/abs/quant-ph/0205164.

\item Aerts, D. (2007a). Quantum interference and superposition in cognition: Development of a theory for the disjunction of concepts. Archive address and link: http://arxiv.org/abs/0705.0975.

\item Aerts, D. (2007b). General quantum modeling of combining concepts: A quantum field model in Fock space. Archive address and link: http://arxiv.org/abs/0705.1740.

\item Aerts, D., \& Aerts, S. (1994). Applications of quantum statistics in psychological studies of decision processes. {\it Foundations of Science, 1}, 85-97. Reprinted in B. Van Fraassen (1997), (Eds.), {\it Topics in the Foundation of Statistics} (pp. 111-122). Dordrecht: Kluwer Academic.

\item Aerts, S. and Aerts, D. (2008). When can a data set be described by quantum theory? In P. Bruza, W. Lawless, K. van Rijsbergen, D. Sofge, B. Coecke and S. Clark (Eds.), {\it Proceedings of the Second Quantum Interaction Symposium, Oxford 2008}, pp. 27-33. London: College Publications.

\item Aerts, D., Aerts, S. \& Gabora, L. (2009). Experimental evidence for quantum structure in cognition. In Bruza P.D., Sofge D., Lawless, W., Van Rijsbergen, C.J., Klusch, M. (Eds.). {\it Proceedings of QI 2009-Third International Symposium on Quantum Interaction, Lecture Notes in Computer Science}. Berlin: Springer.

\item Aerts, D. \& D'Hooghe, B. (2009). Classical logical versus quantum conceptual thought: Examples in economics, decision theory and concept theory. In Bruza P.D., Sofge D., Lawless, W., Van Rijsbergen, C.J., Klusch, M. (Eds.). {\it Proceedings of QI 2009-Third International Symposium on Quantum Interaction, Lecture Notes in Computer Science}. Berlin: Springer.

\item Aerts, D., \& Czachor, M. (2004). Quantum aspects of semantic analysis and symbolic artificial intelligence. {\it Journal of Physics A, Mathematical and Theoretical, 37}, L123-L132. 

\item Aerts, D., Czachor, M. and D'Hooghe, B. (2006). Towards a quantum evolutionary scheme: violating Bell's inequalities in language. In N. Gontier, J. P. Van Bendegem and D. Aerts (Eds.), {\it Evolutionary Epistemology, Language and Culture - A non adaptationist systems theoretical approach}. Dordrecht: Springer.

\item Aerts, D., \& Gabora, L. (2005a). A theory of concepts and their combinations II: A Hilbert space representation. {\it Kybernetes, 34}, 192-221.

\item Aerts, D., \& Gabora, L. (2005b). A theory of concepts and their combinations I: The structure of the sets of contexts and properties. {\it Kybernetes, 34}, 167-191.

\item Allais, M. (1953). Le comportement de l'homme rationnel devant le risque: critique des postulats et axiomes de l'\'ecole Am\'ericaine. {\it Econometrica, 21}, 503-546.

\item Baaquie, B. E. (2004). {\it Quantum Finance: Path Integrals and Hamiltonians for Options and Interest Rates}. Cambridge UK: Cambridge University Press.

\item Bagassi, M., Macchi, L. (2007). The `vanishing' of the disjunction effect by sensible procrastination. {\it Mind \& Society 6}, 41Ð52.

\item Barsalou, L. (1987). The instability of graded structure: Implications of the nature of concepts. In U. Neisser (Eds.), {\it Concepts and Conceptual Development: Ecological and Intellectual factors in Categorization}. Cambridge: Cambridge University Press.

\item Berry, M. W., Dumais, S. T., and O'Brien, G. W. (1995). Using linear algebra for intelligent information retrieval. {\it SIAM Review, 37}, 573-595.

\item Bruner, J. (1990). {\it Acts of Meaning}. Cambridge, MA: Harvard University Press.

\item Bruza, P. D. and Cole, R. J. (2005). Quantum logic of semantic space: An exploratory investigation of context effects in practical reasoning. In S. Artemov, H. Barringer, A. S. d'Avila Garcez, L.C. Lamb, J. Woods (Eds.) {\it We Will Show Them: Essays in Honour of Dov Gabbay}. College Publications.

\item Bruza, P.D., Kitto, K., Nelson, D. and McEvoy, K. (2008). Entangling words and meaning. {\it Proceedings of the Second Quantum Interaction Symposium}, University of Oxford.

\item Bruza, P.D., Kitto, K., Nelson D. and McEvoy,  C. (2009). Extracting spooky-activation-at-a-distance from considerations of entanglement, In Bruza P.D., Sofge D., Lawless, W., Van Rijsbergen, C.J., Klusch, M. (Eds.). {\it Proceedings of QI 2009-Third International Symposium on Quantum Interaction, Lecture Notes in Computer Science}. Berlin: Springer.

\item Bruza, P.D., Widdows, D. \& Woods J.H. (in press). A Quantum logic of down below. In {\it Handbook of Quantum Logic and Quantum Structures}. Volume 2. Elsevier.

\item Busemeyer, J. R., Wang, Z., \& Townsend, J. T. (2006). Quantum dynamics of human decision making. {\it Journal of Mathematical Psychology, 50}, 220-241. 

\item Busemeyer, J. R., Matthew, M., Wang, Z. (2006)  A Quantum Information Processing Theory Explanation of Disjunction Effects. {\it Proceedings of the Cognitive Science Society}.

\item Deerwester, S., Dumais, S. T. \& Harshman, R. (1990). Indexing by Latent Semantic Analysis. {\it Journal of the Society for Information Science, 41}, 391-407. 

\item Dirac, P. A. M. (1958). {\it Quantum mechanics}, 4th ed. London: Oxford University Press.

\item Ellsberg, D. (1961). Risk, Ambiguity, and the Savage Axioms. {\it Quarterly Journal of Economics, 75}, 643-669.

\item Flender, C., Kitto, K. \& Bruza, P. D. (2009). Beyond ontology in information systems. In Bruza, P., Sofge, D., Lawless, W., Rijsbergen, K., Klusch, M. (Eds.). Proceedings 
of the Third Quantum Interaction Symposium. Volume 5494 of Lecture Notes in 
Artificial Intelligence, Springer.

\item Fodor, J. (1975). {\it The Language of Thought}. New York: Thomas Crowell.

\item Franco, R. (2007). Quantum mechanics, Bayes' theorem and the conjunction fallacy. Archive address and link: http://arxiv.org/abs/0708.3948 

\item Freud, S. (1899). {\it Die Traumdeutung}. Berlin: Fischer-Taschenbuch.

\item Gabora, L., \& Aerts, D. (2002a). Contextualizing concepts. In {\it Proceedings of the 15th International FLAIRS Conference. Special track: Categorization and Concept Representation: Models and Implications}, Pensacola Florida, May 14-17, American Association for Artificial Intelligence (pp. 148-152). 

\item Gabora, L., \& Aerts, D. (2002b). Contextualizing concepts using a mathematical generalization of the quantum formalism. {\it Journal of Experimental and Theoretical Artificial Intelligence, 14}, 327-358. Preprint at http://arXiv.org/abs/quant-ph/0205161 

\item Gabora, L., Rosch, E., Aerts, D. (2008). Toward an ecological theory of concepts. {\it Eco- 
logical Psychology 20}, 84Ð116.

\item G\"ardenfors, P. (2004). {\it Conceptual Spaces: The Geometry of Thought}. Boston, USA: MIT-Press.

\item Hampton, J. A. (1987). Inheritance of attributes in natural concept conjunctions. {\it Memory \& Cognition, 15}, 55-71. 

\item Hampton, J. A. (1988a). Overextension of conjunctive concepts: Evidence for a unitary model for concept typicality and class inclusion. {\it Journal of Experimental Psychology: Learning, Memory, and Cognition, 14}, 12-32.

\item Hampton, J. A. (1988b). Disjunction of natural concepts. {\it Memory \& Cognition, 16}, 579-591.

\item Hampton, J. A. (1991). The combination of prototype concepts. In P. Schwanenflugel (Ed.), {\it The Psychology of Word Meanings}. Hillsdale, NJ: Erlbaum.

\item Hampton, J. A. (1993). Prototype models of concept representation. In I. Van Mechelen, J. Hampton, R. S. Michalski, \& P. Theuns (Eds.), {\it Categories and Concepts: Theoretical Views and Inductive Data Analysis} (pp. 67-95). London, UK: Academic Press.

\item Hampton, J.A. (1996). Conjunctions of visually-based categories: overextension and compensation. {\it Journal of Experimental Psychology: Learning, Memory and Cognition, 22}, 378-396.

\item Hampton, J. A. (1997a). Conceptual combination: Conjunction and negation of natural concepts. {\it Memory \& Cognition, 25}, 888-909.

\item Hampton, J. A. (1997b). Conceptual combination. In K. Lamberts \& D. Shanks (Eds.), {\it Knowledge, Concepts, and Categories} (pp. 133-159). Hove: Psychology Press. 

\item Haven, E. (2005). Pilot-wave theory and financial option pricing. {\it International Journal of Theoretical Physics, 44}, 1957-1962.

\item Hettel, T., Flender, C., Barros, A.: Scaling Choreography Modelling for B2B Value- 
Chain Analysis. In: Proceeding of the 6th International Conference on Business 
Process Management (BPM 2008), 1-4 September 2008, Milan, Italy. (2008)

\item James, W. (1910). {\it Some Problems of Philosophy}. Cambridge, MA: Harvard University Press.

\item Komatsu, L. K. (1992). Recent views of conceptual structure. {\it Psychological Bulletin, 112}, 500-526. 

\item Kolmogorov, A. N. (1977). {\it Grundbegriffe der Wahrscheinlichkeitsrechnung}. Reprint der Erstauflage Berlin 1933. Berlin: Springer.

\item Khrennikov, A. (2008). A model of quantum-like decision-making with applications to psychology and cognitive science. Archive address and link: http://uk.arxiv.org/abs/0711.1366

\item Khrennikov, A. (2009). Design of an experiment to test quantum probabilistic behavior of the financial market. http://uk.arxiv.org/abs/0902.1922

\item Kunda, Z., Miller, D. T., \& Claire, T. (1990). Combining social concepts: The role of causal reasoning. {\it Cognitive Science, 14}, 551-577. 

\item Landauer, T. K., Foltz, P. W. and Laham, D. (1998). Introduction to Latent Semantic Analysis. Discourse Processes 25, 259-284.

\item Lund, K \& Burgess, C. (1996). Producing high-dimensional semantic spaces from lexical co-occurrence. {\it Behavior Research Methods, Instruments and Computers, 28}, 203-208.

\item Margolis, E. and Laurence, S. (2002). Radical concept nativism. {\it Cognition, 86}, 25-55.

\item Nelson, D. L. \&  McEvoy, C. L. (2007). Entangled associative structures and context. In P. Bruza, W. Lawless, K. van Rijsbergen, \& D. Sofge (Eds.) {\it Proceedings of the Association for the Advancement of Artificial Intelligence (AAAI) Spring Symposium 8: Quantum Interaction}, March 26-28, 2007, Stanford University.

\item Osherson , D. N. \& Smith, E. E. (1981). On the adequacy of prototype theory as a theory of concepts. {\it Cognition, 9}, 35-58. 

\item Osherson, D. N. \& Smith, E. E. (1982). Gradedness and conceptual combination. {\it Cognition, 12}, 299-318. 

\item Piaget, J. (1990). {\it Le Langage et la Pens\'ee Chez l'Enfant}. Paris: Delachaux et Niestl\'e.

\item Pitowsky, I. (1989). {\it Quantum Probability, Quantum Logic. Lecture Notes in Physics 321}. Heidelberg:
Springer.

\item Rips, L J. (1995). The current status of research on concept combination. {\it Mind \& Language, 10}, 72-104. 

\item Rosch, E. (1973a). Natural categories. {\it Cognitive Psychology, 4}, 328.

\item Rosch, E. (1973b). On the internal structure of perceptual and semantic categories. In T. E. Moore (Ed.), {\it Cognitive Development and the Acquisition of Language}. New York: Academic Press. 

\item Savage, L.J. (1944). {\it The Foundations of Statistics}. New-York: Wiley.

\item Schaden, M. (2002). Quantum finance: A quantum approach to stock price fluctuations. {\it Physica A, 316}, 511-538. 

\item Smith, E. E., \& Medin, D. L. (1981). {\it Categories and Concepts}. Cambridge, MA.: Harvard University Press.

\item Smith, E. E., \& Osherson, D. N. (1984). Conceptual combination with prototype concepts. {\it Cognitive Science, 8}, 357-361. 

\item Smith, E. E., Osherson, D. N., Rips, L. J., \& Keane, M. (1988). Combining prototypes: A selective modification model. {\it Cognitive Science, 12}, 485-527. 

\item Smolensky, P. (1990). Tensor product variable binding and the representation of symbolic structures in connectionist systems. {\it Artificial Intelligence, 46}, 159-216.

\item Springer, K., \& Murphy, G. L. (1992). Feature availability in conceptual combination. {\it Psychological Science, 3}, 111-117. 

\item Storms, G., De Boeck, P., Van Mechelen, I., \& Geeraerts, D. (1993). Dominance and non-commutativity effects in concept conjunctions: Extensional or intensional basis? {\it Memory \& Cognition, 21}, 752-762.

\item Storms, G., de Boeck, P., Hampton, J.A., \& van Mechelen, I. (1999). Predicting conjunction typicalities by component typicalities. {\it Psychonomic Bulletin and Review, 6}, 677-684. 

\item Tversky, A. \& Kahneman, D. (1982). Judgments of and by representativeness. In D. Kahneman, P. Slovic
\& A. Tversky (Eds.), {\it Judgment under uncertainty: Heuristics and biases}. Cambridge, UK: Cambridge
University Press.

\item Tversky, A. \& Shafir, E. (1992). The disjunction effect in choice under uncertainty. {\it Psychological Science, 3}, 305-309.

\item Van Rijsbergen, K. (2004). {\it The Geometry of Information Retrieval}. Cambridge UK: Cambridge University Press. 

\item Widdows, D. (2003). Orthogonal negation in vector spaces for modelling word-meanings and document retrieval. In {\it Proceedings of the 41st Annual Meeting of the Association for Computational Linguistics} (pp. 136-143). Sapporo, Japan, July 7-12. 

\item Widdows, D. (2006). {\it Geometry and Meaning}. CSLI Publications: University of Chicago Press. 

\item Widdows, D., \& Peters, S. (2003). Word vectors and quantum logic: Experiments with negation and disjunction. In {\it Mathematics of Language 8} (pp. 141-154). Indiana: Bloomington. 

\item Zadeh, L. (1965). Fuzzy sets. {\it Information \& Control, 8}, 338-353.

\end{description}

\appendix

\section{Appendix: Proof of Theorem 1:} If $\mu(A), \mu(B)$ and $\mu(A\ {\rm and}\ B)$ are classical conjunction data, there exists a Kolmogorovian probability space $(\Omega,\sigma(\Omega),P)$ and events $E_A, E_B \in \sigma(\Omega)$ such that $P(E_A) = \mu(A)$, $P(E_B) = \mu(B)$ and $P(E_A \cap E_B) = \mu(A\ {\rm and}\ B)$. From the general properties of a Kolmogorovian probability space it follows that we have $0 \le P(E_A \cap E_B) \le P(E_A) \le 1$ and $0 \le P(E_A \cap E_B) \le P(E_B) \le 1$, which proves that inequalities (\ref{ineq01}) and (\ref{ineq02}) are satisfied. From the same general properties of a Kolmogorovian probability space it also follows that we have $P(E_A \cup E_B) = P(E_A) + P(E_B) - P(E_A \cap E_B)$, and since $P(E_A \cup E_B) \le 1$ we also have $P(E_A) + P(E_B) - P(E_A \cap E_B) \le 1$. This proves that inequality (\ref{ineq03}) is satisfied. We have now proved that for classical conjunction data $\mu(A), \mu(B)$ and $\mu(A\ {\rm and}\ B)$ the three inequalities are satisfied. Now suppose that we have an item $X$ such that for its membership weights $\mu(A), \mu(B), \mu(A\ {\rm and}\ B)$ with respect to concepts $A$ and $B$ and their conjunction `$A$ and $B$', inequalities (\ref{ineq01}), (\ref{ineq02}) and (\ref{ineq03}) are satisfied. We will prove that as a consequence $\mu(A), \mu(B)$ and $\mu(A\ {\rm and}\ B)$ are classical conjunction data. To this end, we will explicitly construct a Kolmogorovian probability space that models these data. Consider the set $\Omega=\{1, 2, 3, 4\}$ and $\sigma(\Omega) = {\cal P}(\Omega)$, the set of all subsets of $\Omega$. We define
\begin{eqnarray} \label{eqtheorem101}
&P(\{1\}) = \mu(A\ {\rm and}\ B) \\ \label{eqtheorem102}
&P(\{2\}) = \mu(A) - \mu(A\ {\rm and}\ B) \\ \label{eqtheorem103}
&P(\{3\}) = \mu(B) - \mu(A\ {\rm and}\ B) \\ \label{eqtheorem104}
&P(\{4\}) = 1-\mu(A)-\mu(B)+\mu(A\ {\rm and}\ B)
\end{eqnarray}
and further for an arbitrary subset $S \subseteq \{1,2,3,4\}$ we define
\begin{equation} \label{defarbitrarysubset}
P(S) = \sum_{a\in S}P(\{a\})
\end{equation}
Let us prove that $P: \sigma(\Omega) \rightarrow [0,1]$ is a probability measure. For this purpose, we need to prove that $P(S) \in [0,1]$ for an arbitrary subset $S \subseteq \Omega$, and that the `sum formula' for a probability measure is satisfied to comply with (\ref{KolmogorovianMeasure}). The sum formula for a probability measure is satisfied because of definition (\ref{defarbitrarysubset}). What remains to be proved is that $P(S) \in [0,1]$ for an arbitrary subset $S \subseteq \Omega$. $P(\{1\}), P(\{2\}), P(\{3\})$ and $P(\{4\})$ are contained in $[0,1]$ as a direct consequence of inequalities (\ref{ineq01}), (\ref{ineq02}) and (\ref{ineq03}). Further, we have $P(\{1,2\}) = \mu(A), P(\{1,3\})=\mu(B), P(\{3,4\}) = 1-\mu(A), P(\{2,4\}) = 1-\mu(B), P(\{2,3,4\}) = 1-\mu(A\ {\rm and}\ B)$ and $P(\{1,2,3\}) = \mu(A)+\mu(B)-\mu(A\ {\rm and}\ B)$, and all these are contained in $[0,1]$ as a consequence of inequalities (\ref{ineq01}), (\ref{ineq02}) and (\ref{ineq03}). Consider $P(\{2,3\}) = \mu(A)+\mu(B)-2\mu(A\ {\rm and}\ B)$. From inequality (\ref{ineq03}) it follows that $\mu(A)+\mu(B)-2\mu(A\ {\rm and}\ B) \le \mu(A)+\mu(B)-\mu(A\ {\rm and}\ B) \le 1$. Further, we have, following inequalities (\ref{ineq01}) and (\ref{ineq02}), $\mu(A\ {\rm and}\ B) \le \mu(A)$ and $\mu(A\ {\rm and}\ B) \le \mu(B)$ and hence $2\mu(A\ {\rm and}\ B) \le \mu(A) + \mu(B)$. From this it follows that $0 \le \mu(A)+\mu(B)-2\mu(A\ {\rm and}\ B)$. Hence we have proved that $P(\{2,3\}) = \mu(A)+\mu(B)-2\mu(A\ {\rm and}\ B) \in [0,1]$. We have $P(\{1,4\}) = 1-\mu(A)-\mu(B)+2\mu(A\ {\rm and}\ B) = 1 - P(\{2,3\})$ and hence $P(\{1,4\}) \in [0,1]$. We have $P(\{1,2,4\})=1-\mu(B)+\mu(A\ {\rm and}\ B) = 1-P(\{3\}) \in [0,1]$ and 
$P(\{1,3,4\})=1-\mu(A)+\mu(A\ {\rm and}\ B) = 1-P(\{2\}) \in [0,1]$. The last subset to control is $\Omega$ itself. We have $P(\Omega)=P(\{1\}) + P(\{2\}) + P(\{3\}) + P(\{4\}) = 1$. We have verified all subsets $S \subseteq \Omega$, and hence proved that $P$ is a probability measure. Since $P(\{1\}) = \mu(A\ {\rm and}\ B)$, $P(\{1,2\})=\mu(A)$ and $P(\{1,3\}=\mu(B)$, we have modeled the data $\mu(A)$, $\mu(B)$ and $\mu(A\ {\rm and}\ B)$ by means of a Kolmogorovian probability space, and hence they are classical conjunction data.

\section{Appendix: Proof of Theorem 4}
If $\mu(A), \mu(B)$ and $\mu(A\ {\rm or}\ B)$ are classical disjunction data, there exists a Kolmogorovian probability space $(\Omega, \sigma(\Omega), P)$ and events $E_A, E_B \in \sigma(\Omega)$ such that $P(E_A) = \mu(A)$, $P(E_B) = \mu(B)$ and $P(E_A \cup E_B) = \mu(A\ {\rm or}\ B)$. We have $0 \le P(E_A) \le P(E_A \cup E_B) \le 1$ and $0 \le P(E_B) \le P(E_A \cup E_B) \le 1$, which proves that inequalities (\ref{disjunctionineq01}) and (\ref{disjunctionineq02}) are satisfied. We have $P(E_A \cap E_B) = P(E_A) + P(E_B) - P(E_A \cup E_B)$, and since $0 \le P(E_A \cap E_B)$ we also have $0 \le P(E_A) + P(E_B) - P(E_A \cup E_B)$. This proves that inequality (\ref{disjunctionineq03}) is satisfied. Hence we have proved that for classical disjunction data $\mu(A), \mu(B)$ and $\mu(A\ {\rm or}\ B)$ the three inequalities are satisfied. Suppose now that we have an item $X$ such that for its membership weights $\mu(A), \mu(B), \mu(A\ {\rm or}\ B)$ with respect to concepts $A$ and $B$ and their disjunction `$A$ or $B$' inequalities (\ref{disjunctionineq01}), (\ref{disjunctionineq02}) and (\ref{disjunctionineq03}) are satisfied. We can now prove that as a consequence $\mu(A), \mu(B)$ and $\mu(A\ {\rm or}\ B)$ are classical disjunction data. Consider the set $\Omega=\{1, 2, 3, 4\}$ and $\sigma(\Omega) = {\cal P}(\Omega)$, the set of all subsets of $\Omega$. 
We define
\begin{eqnarray} \label{eqtheorem201}
&P(\{1\}) = \mu(A)+\mu(B)-\mu(A\ {\rm or}\ B) \\ \label{eqtheorem202}
&P(\{2\}) = \mu(A\ {\rm or}\ B)-\mu(B) \\ \label{eqtheorem203}
&P(\{3\}) = \mu(A\ {\rm or}\ B)-\mu(A) \\ \label{eqtheorem204}
&P(\{4\}) = 1-\mu(A\ {\rm or}\ B)
\end{eqnarray}
and further for an arbitrary subset $S \subseteq \{1,2,3,4\}$ we define
\begin{equation} \label{disjunctiondefarbitrarysubset}
P(S) = \sum_{a\in S}P(\{a\})
\end{equation}
Let us show that $P: \sigma(\Omega) \rightarrow [0,1]$ is a probability measure. We need to prove that $P(S) \in [0,1]$ for an arbitrary subset $S \subseteq \Omega$, and that the `sum formula' for a probability measure is satisfied to comply with (\ref{KolmogorovianMeasure}). The sum formula for a probability measure is satisfied because of definition (\ref{disjunctiondefarbitrarysubset}). What remains to be proved is that $P(S) \in [0,1]$ for an arbitrary subset $S \subseteq \Omega$. $P(\{1\}), P(\{2\}), P(\{3\})$ and $P(\{4\})$ are contained in $[0,1]$ as a direct consequence of inequalities (\ref{disjunctionineq01}), (\ref{disjunctionineq02}) and (\ref{disjunctionineq03}). Further, we have $P(\{1,2\}) = \mu(A), P(\{1,3\})=\mu(B), P(\{3,4\}) = 1-\mu(A), P(\{2,4\}) = 1-\mu(B), P(\{2,3,4\}) = 1-\mu(A)-\mu(B)+\mu(A\ {\rm or}\ B)$ and $P(\{1,2,3\}) = \mu(A\ {\rm or}\ B)$, and all these are contained in $[0,1]$ as a consequence of inequalities (\ref{disjunctionineq01}), (\ref{disjunctionineq02}) and (\ref{disjunctionineq03}). Consider $P(\{2,3\}) = 2\mu(A\ {\rm or}\ B)-\mu(A)-\mu(B)$. From (\ref{disjunctionineq01}) and (\ref{disjunctionineq02}) it follows that $0 \le \mu(A\ {\rm or}\ B) - \mu(A)$ and $0 \le \mu(A\ {\rm or}\ B) - \mu(B)$, and this gives $0 \le 2\mu(A\ {\rm or}\ B) - \mu(A)-\mu(B)$. From (\ref{disjunctionineq03}) it follows that $\mu(A\ {\rm or}\ B)-\mu(A)-\mu(B) \le 0$, and hence $2\mu(A\ {\rm or}\ B)-\mu(A)-\mu(B) \le \mu(A\ {\rm or}\ B) \le 1$. This proves that $P(\{2,3\}) = 2\mu(A\ {\rm or}\ B)-\mu(A)-\mu(B) \in [0.1]$. We have $P(\{1,4\}) = 1+\mu(A)+\mu(B)-2\mu(A\ {\rm or}\ B) = 1 - P(\{2,3\})$ and hence $P(\{1,4\}) \in [0,1]$. We have $P(\{1,2,4\})=1+\mu(A)-\mu(A\ {\rm or}\ B) = 1-P(\{3\}) \in [0,1]$ and $P(\{1,3,4\})=1+\mu(B)-\mu(A\ {\rm or}\ B) = 1-P(\{2\}) \in [0,1]$. The last subset to control is $\Omega$ itself. We have $P(\Omega)=P(\{1\}) + P(\{2\}) + P(\{3\}) + P(\{4\}) = 1$. We have verified all subsets $S \subseteq \Omega$, and hence proved that $P$ is a probability measure. Since $P(\{1,2,3\}) = \mu(A\ {\rm or}\ B)$, $P(\{1,2\})=\mu(A)$ and $P(\{1,3\}=\mu(B)$, we have modeled the data $\mu(A)$, $\mu(B)$ and $\mu(A\ {\rm or}\ B)$ by means of a Kolmogorovian probability space, and hence they are classical disjunction data.

\section{Appendix: Proof of Theorem 7}
From {\bf Theorem 1} it follows that $x_{AB}$, $x_{AB'}$, $x_{A'B}$ and $x_{A'B'}$ are well-defined. Indeed, taking into account (\ref{eqtheorem101}), (\ref{eqtheorem102}), (\ref{eqtheorem103}) and (\ref{eqtheorem104}), we can see that $x_{AB}=\pm\sqrt{P(\{1\})}$, $x_{AB'}=\pm\sqrt{P(\{2\})}$, $x_{A'B}=\pm\sqrt{P(\{3\})}$ and $x_{A'B'}=\pm\sqrt{P(\{P(4)\})}$. Since $P(\{1\}), P(\{2\}), P(\{3\}), P(\{4\}) \in [0,1]$, which we proved in {\bf Theorem 1}, $x_{AB}$, $x_{AB'}$, $x_{A'B}$ and $x_{A'B'}$ are well-defined. We have $x_{AB}^2=\mu(A\ {\rm and}\ B)$, which proves that (\ref{eqclass03}) is satisfied. We have $x_{AB}^2+x_{AB'}^2=\mu(A)-\mu(A\ {\rm and}\ B)+\mu(A\ {\rm and}\ B)=\mu(A)$ and $x_{AB}^2+x_{A'B}^2=\mu(B)-\mu(A\ {\rm and}\ B)+\mu(A\ {\rm and}\ B)=\mu(B)$, which proves that (\ref{eqclass01}) and (\ref{eqclass02}) are satisfied. We have $x_{AB}^2+x_{AB'}^2+x_{A'B}^2+x_{A'B'}^2=\mu(A\ {\rm and}\ B)+\mu(A)-\mu(A\ {\rm and}\ B)+\mu(B)-\mu(A\ {\rm and}\ B)+1-\mu(A)-\mu(B)+\mu(A\ {\rm and}\ B)=1$, which proves that $x$ is a unit vector of $\real^4$.

\section{Appendix: Proof of Theorem 8}
From {\bf Theorem 4} it follows that $x_{AB}$, $x_{AB'}$, $x_{A'B}$ and $x_{A'B'}$ are well-defined. Indeed, taking into account (\ref{eqtheorem201}), (\ref{eqtheorem202}), (\ref{eqtheorem203}) and (\ref{eqtheorem204}) we can see that $x_{AB}=\pm\sqrt{P(\{1\})}$, $x_{AB'}=\pm\sqrt{P(\{2\})}$, $x_{A'B}=\pm\sqrt{P(\{3\})}$ and $x_{A'B'}=\pm\sqrt{P(\{P(4)\})}$. Since $P(\{1\})$, $P(\{2\})$, $P(\{3\})$, $P(\{4\}) \in [0,1]$, which we proved in {\bf Theorem 4}, $x_{AB}$, $x_{AB'}$, $x_{A'B}$ and $x_{A'B'}$ are well-defined. We have $x_{AB}^2+x_{AB'}^2=\mu(A)+\mu(B)-\mu(A\ {\rm or}\ B)+\mu(A\ {\rm or}\ B)-\mu(B)=\mu(A)$ and $x_{AB}^2+x_{A'B}^2=\mu(A)+\mu(B)-\mu(A\ {\rm or}\ B)+\mu(A\ {\rm or}\ B)-\mu(A)=\mu(B)$, which proves that (\ref{eqclass01}) and (\ref{eqclass02}) are satisfied. We have $x_{AB}^2+x_{AB'}^2+x_{A'B}^2=\mu(A)+\mu(B)-\mu(A\ {\rm or}\ B)+\mu(A\ {\rm or}\ B)-\mu(B)+\mu(A\ {\rm or}\ B)-\mu(A)=\mu(A\ {\rm or}\ B)$, which proves that (\ref{eqclass04}) is satisfied. Further, we have $x_{AB}^2+x_{AB'}^2+x_{A'B}^2+x_{A'B'}^2=\mu(A\ {\rm and}\ B)+1-\mu(A\ {\rm and}\ B)=1$, which proves that $x$ is a unit vector of $\real^4$.

\section{Appendix: Proof of Theorem 9}
Suppose that item $X$ is represented by vector $x$ of $\real^n$. From the quantum rule formulated in section \ref{quantumrule} it follows that $\mu(A)=P_{\cal A}(x)$, $\mu(B)=P_{\cal B}(x)$, $\mu(A\ {\rm and}\ B)=P_{\cal A \cap B}(x)$ and $\mu(A\ {\rm or}\ B)=P_{\cal A+\cal B}(x)$, where $P_{\cal A}$, $P_{\cal B}$, $P_{\cal A \cap B}$ and $P_{\cal A+\cal B}$ are the orthogonal projections on  subspaces ${\cal A}$, ${\cal B}$, ${\cal A \cap B}$ and ${\cal A+\cal B}$, respectively. We have ${\cal A \cap B} \subseteq {\cal A}$ and ${\cal A \cap B} \subseteq {\cal B}$, from which it follows that $\|P_{\cal A \cap B}(x)\|^2 \le \|P_{\cal A}(x)\|^2$ and $\|P_{\cal A \cap B}(x)\|^2 \le \|P_{\cal B}(x)\|^2$ and hence $\mu(A\ {\rm and}\ B) \le \mu(A)$ and $\mu(A\ {\rm and}\ B) \le \mu(B)$, which shows that inequalities (\ref{disjunctionineq01}) and (\ref{disjunctionineq02}) are satisfied, that $\Delta_c \le 0$, and that $X$ is not a $\Delta$-type non-classical item for the conjunction. We also have ${\cal A} \subseteq {\cal A+\cal B}$ and ${\cal B} \subseteq {\cal A+\cal B}$, from which it follows that $\|P_{\cal A}(x)\|^2 \le \|P_{\cal A+\cal B}(x)\|^2$ and $\|P_{\cal B}(x)\|^2 \le \|P_{\cal A+\cal B}(x)\|^2$ and hence $\mu(A) \le \mu(A\ {\rm or}\ B)$ and $\mu(B) \le \mu(A\ {\rm or}\ B)$, which shows that inequalities (\ref{disjunctionineq01}) and (\ref{disjunctionineq02}) are satisfied, that $\Delta_d \le 0$, and that $X$ is not a $\Delta$-type non-classical item for the disjunction.

\end{document}